\documentclass[11pt]{article}

\usepackage[margin=1in]{geometry}
\usepackage{amsmath,amssymb,amsthm,mathtools,bm}
\usepackage{enumitem}
\usepackage[expansion=false]{microtype}
\usepackage{xcolor}
\usepackage[normalem]{ulem}
\usepackage{pdflscape}
\usepackage{float}
\usepackage{comment}
\usepackage{natbib}
\newtheorem{assumption}{Assumption}
\newtheorem{theorem}{Theorem}
\newtheorem{lemma}{Lemma}
\newtheorem{remark}{Remark}
\newtheorem{proposition}{Proposition}
\newtheoremstyle{plainnopunct}
  {\topsep}{\topsep}{\itshape}{}{\bfseries}{}{0.5em}{}
\theoremstyle{plainnopunct}
\newtheorem{propositionnopunct}[proposition]{Proposition}
\theoremstyle{plain}

\newcommand{\E}{\mathbb{E}}
\newcommand{\Pp}{\mathbb{P}}
\newcommand{\R}{\mathbb{R}}
\newcommand{\Var}{\operatorname{Var}}
\newcommand{\Cov}{\operatorname{Cov}}
\newcommand{\op}{o_{\Pp}}
\newcommand{\Op}{O_{\Pp}}
\newcommand{\norm}[1]{\left\lVert #1\right\rVert}
\newcommand{\trans}{^{\mathsf T}}

\begin{document}

\title{Estimating treatment-effect heterogeneity across sites, in multi-site randomized experiments with few units per site.\thanks{We are very grateful to Manuel Arellano, Dmitry Arkhangelsky, Xavier D'Haultfoeuille, Peng Ding, Roberto Galbiati, Grégory Jolivet, David McKenzie, Julien Monardo, Timo Schenk, Liyang Sun, Christine Valente, Yanos Zylberberg, and seminar participants at the 2025 Encounters in Econometric Theory Workshop, CEMFI, KU Leuven, the University of Bristol, and the World Bank for their helpful comments. Cl\'{e}ment de Chaisemartin was funded by the European Union (ERC, REALLYCREDIBLE,GA N°101043899). \\
\textbf{AI disclosure statement:} Versions 1–3 of the arXiv manuscript were written without AI assistance. The current version was prepared with the assistance of ChatGPT for the theory and Claude for the empirical application and simulations. When we developed the paper's new identification results, not present in the earlier versions, ChatGPT helped us as follows. ChatGPT helped us prove Equation~\eqref{eq:corr-third}, Proposition~\ref{prop:rct-unbiased}, and Proposition~\ref{prop:late-prediction}, following detailed prompts and after those results had been conjectured by us. Following a lose prompt, ChatGPT conjectured and prove Point~2 of Proposition~\ref{prop:late-variance-identification}. We checked all AI-generated arguments and take responsibility for the final manuscript.\\
\textbf{Estimated CO\textsubscript{2}e emissions generated by AI use:}
Our interactions with ChatGPT in preparing this paper involved approximately 220.1 million input tokens—of which 205.9 million were cached—and 0.92 million output tokens, corresponding to an estimated 1.92 kg CO\textsubscript{2}e under GPT's reference scenario; this estimate covers only the operational emissions associated with the marginal electricity consumed in processing our prompts and excludes emissions from training the models and their predecessors. Our interactions with Claude in preparing this paper involved approximately 94.6 million input tokens, of which 87.7 million were cached, and 0.35 million output tokens, corresponding to an estimated 0.83 kg CO2e under the same reference scenario.}}

  \author{Cl\'{e}ment de Chaisemartin\textcircled{r}\hspace{.05cm}\thanks{Department of Economics, Sciences Po} \and
    Antoine Deeb\thanks{Development Impact, World Bank}}

\maketitle

\begin{abstract}
Multi-site randomized experiments, which are common in economics and medicine, can reveal how treatment effects vary across settings. However, differences between site-specific estimates reflect both treatment-effect heterogeneity and estimation noise. We build on established measurement-error corrections and develop new ones to study this heterogeneity in experiments with many sites and few units per site. In particular, our results can be used to estimate the variance and centered third moment of treatment effects across sites, relate site effects to untreated outcomes, conduct mediation analyses, assess the roles of provider effectiveness and site responsiveness, and study heterogeneity in local average treatment effects (LATEs) under imperfect compliance.
We apply our methods to counseling programs delivered by public and private providers in about 200 French employment offices.
Average employment gains of about 2 percentage points conceal cross-site standard deviations of intention-to-treat effects of 8 to 9 percentage points. We also find substantial LATE heterogeneity.
Effects are negatively associated with untreated employment rates. The two programs also tend to be more effective in the same sites, consistent with a role for heterogeneous site responsiveness.
\end{abstract}

\vfill

\newpage

``Vérité au-deçà des Pyrénées, erreur au-delà.''\\
\textit{``Truth on this side of the Pyrenees, error on the other.''}\\
---Blaise Pascal, \textit{Pensées}

\section{Introduction}\label{sec:introduction}

In multi-site randomized controlled trials (RCTs), on top of learning whether the treatment works on average, researchers may want to know how much treatment effects vary across sites, whether effects are larger where untreated outcomes are worse, and whether sites' effects on a mediator predict their effects on the final outcome. When compliance with random assignment is imperfect, researchers may also want to know whether differences in intention-to-treat effects reflect only differences in take-up rates or also differences in local average treatment effects (LATEs). Answering these questions is difficult when each site contains few units. Due to estimation error, site-specific estimates can differ substantially even when the underlying effects are similar. Moreover, estimation error can create misleading relationships between estimated effects and other estimated site characteristics. Accordingly, while experiments with many sites and few units per site are common in economics\footnote{From 2014 to 2016, 12 multi-site RCTs were published in \textit{AEJ: Applied Economics}, with a median of 46 sites and a median of 15 randomization units per site (see Table~\ref{table:exps} in the appendix).} and medicine,\footnote{In medicine, such experiments are usually called multi-center RCTs. Reviewing 206 multi-center RCTs published in four leading medical journals, \citet{kahan2015many} report medians of 20 centers and 24 patients per center; 30\% of the trials included more than 50 centers.} researchers rarely systematically investigate effect heterogeneity across sites, and those that do so generally use parametric random-effects models.\footnote{In Table~\ref{table:exps}, the only paper that systematically investigates effect heterogeneity across sites is \citet{walters2015inputs}, who does so using a parametric random-coefficient model. In medicine, \cite{hemming2021extending} note that analyses of treatment-effect heterogeneity across centers remain uncommon. Exceptions exist \citep[see, e.g.][]{lingsma2011between}, though they usually rely on parametric random-effects models \citep{hemming2021extending}.}

We develop methods for studying treatment-effect heterogeneity in multi-site RCTs with many sites and few units per site. Section~\ref{sec:framework} develops the statistical framework. It introduces classical measurement-error corrections, which subtract the contribution of estimation noise from observed cross-site variances and covariances \citep{fuller1987measurement}, and also proposes a novel corrected estimator of the centered third moment of site effects. Section~\ref{sec:multi-site-rcts} then shows that under i.i.d. sampling within sites, random assignment provides the unbiased estimators needed to implement these corrections.

The remainder of Section~\ref{sec:multi-site-rcts} shows how these results can be used to conduct six types of heterogeneity analyses. First, to assess whether effects vary across sites, we estimate the variance of average treatment effects (ATEs) across
sites. Second, to assess whether heterogeneity is asymmetric, we estimate ATEs' centered third moment, which measures whether unusually large treatment effects are concentrated above or below the average.

Third, to assess if effects are larger where untreated outcomes are worse, we
regress sites' ATEs on their average untreated outcome. This regression may inform targeting to new sites outside the RCT, if the relationship
between treatment effects and untreated outcomes carries over to the new sites, and if outcomes
among previous cohorts provide a useful proxy for new sites' untreated outcomes. A central difficulty is that the control-group mean enters positively in the estimated untreated outcome and negatively in the estimated ATE. Our regression estimator corrects for the resulting negative covariance between the estimators, and for measurement error in the regressor.

Fourth, we consider imperfect compliance. Site-specific local average treatment effects (LATEs) are ratios and generally lack unbiased estimators, so the usual measurement-error corrections cannot be applied directly to estimated LATEs. Nevertheless, we identify and estimate a variance of LATEs that weights sites proportionally
to their squared first stages. This target depends only on estimable second moments of intention-to-treat (ITT) effects and first stages. We also identify and estimate a lower bound for the complier-weighted variance. These results allow researchers to test whether differences in take-up alone explain ITT heterogeneity. We also identify and estimate the covariance between LATEs and first stages under complier weights. This reveals whether sites with larger shares of compliers also tend to have compliers who benefit more from treatment; a positive covariance is consistent with Roy selection
across sites, whereby take-up is higher where compliers' gains are larger.  Finally, we identify and estimate weighted regressions relating site-specific LATEs on different outcomes.

Fifth, we study predictive mediation regressions: do sites with larger treatment effects on a mediator also have larger effects on the final outcome? With a binary mediator, under monotonicity and a strong cross-site independence assumption, the regression's intercept identifies the average direct effect, while its slope identifies the average across sites of the mediator's effect among each site's mediator compliers. We derive a testable implication of homogeneous direct and mediator effects across sites, a sufficient condition for the cross-site independence assumption.

Sixth, when different providers deliver two similar treatments within the same sites, we use the variance of the difference between their site-specific effects to assess the roles of provider effectiveness and site responsiveness. We give conditions under which a constant difference across sites rules out heterogeneous provider effectiveness as a source of effect heterogeneity. More generally, this variance can help assess the roles of sites' responsiveness and providers' effectiveness, although it does not separately identify their contributions.

Section~\ref{sec:application} applies the \texttt{multisite} Stata package, which computes the estimators proposed in this paper, to the experiment of \citet{behaghel2014private}. That experiment compared publicly and privately provided intensive counseling for job seekers in French public-employment offices. Average effects on employment after six months are 2.4 percentage points for public counseling and 1.9 percentage points for private counseling. These averages conceal substantial heterogeneity: the estimated standard deviations of site-specific ITTs are approximately 9.2 and 8.5 percentage points, respectively. We also find substantial LATE heterogeneity. Both programs' estimated effects are negatively associated with employment rates without treatment, with particularly precise evidence for private counseling. This association is concentrated in the component of employment rate without treatment not predicted by job seekers' observed characteristics. The programs also tend to be more effective in the same sites, despite being delivered by different providers. This positive association is consistent with a role for site responsiveness.

Finally, a natural question is how the methods proposed in this paper compare with those proposed by \citet{cddf2025} and \citet{athey2019generalized} to investigate treatment-effect heterogeneity. In Appendix~\ref{app:blp}, we show that, under a Gaussian model, our test of homogeneous effects based on the corrected variance estimator has strictly greater local power than a test based on the best linear predictor (BLP) approach of \citet{cddf2025}, with the full set of site indicators as predictors and where the training is done by least squares. Intuitively, our test is more powerful because it uses the entire sample to estimate site-specific treatment effects and corrects their dispersion for estimation noise, whereas the BLP approach relies on sample splitting, which reduces the precision with which treatment effects are estimated. Our power-comparison result applies to a specific implementation of the BLP approach, but more general approaches are less easily amenable to a power comparison. Results in our empirical application are consistent with a power advantage of our method, even when compared with the machine-learning implementation of BLP proposed by \citet{cddf2025}, and with the generalized random forest approach of \citet{athey2019generalized}. An important advantage of those methods, however, is their broader applicability: they can be used to investigate treatment-effect heterogeneity with respect to general covariates, whereas our methods can only be used to investigate heterogeneity across sites, or more generally across the values of a discrete covariate with finite support.

\subsection*{Related literature and paper outline}

The corrected-moment estimators we use build directly on the measurement-error literature. Subtracting estimation-error variance from the observed variance of estimated effects is familiar from classical measurement-error models \citep{fuller1987measurement}, random-effects meta-analysis \citep{dersimonian1986meta}, and empirical work on industry wage effects \citep{krueger1988efficiency}, teacher effects \citep{aaronson2007teachers}, and employers' propensities to discriminate \citep{kline2022systemic}. A particularly close connection is to the leave-out variance-component estimators of \citet{kline2020leave}. Specializing their random-coefficient model to a binary regressor yields a leave-one-out variance estimator for a group-specific slope that coincides with the conventional variance estimator for a difference in means used below. Similarly, the regression estimator we consider specializes the corrected measurement-error estimator in Equation~(3.1.7) of \citet{fuller1987measurement}. Related estimators have been used with survey averages \citep{deaton1985panel} and teachers' value added \citep{rose2022effects} as regressors. Another related approach regresses outcomes on empirical-Bayes posterior means of estimated group effects \citep[see, e.g.,][]{kline2022systemic}. \citet{chen2025eb} show that this approach generally requires additional restrictions that corrected-moment estimators do not require.

To our knowledge, our centered-third-moment estimator is new. The closest
precedents are \citet{rimoldini2014skewness}, whose correction assumes independent Gaussian measurement errors with known variances, and \citet{benmoshe2026unbiased}, who derive unbiased estimators of higher central moments in unbalanced multilevel models with mutually independent additive random effects and within-level i.i.d.\ errors. Our framework instead allows the sampling-error distribution to vary with the site-specific conditional mean.

Our paper's main focus is to apply corrected-moment estimators to multi-site randomized experiments. Proposition~\ref{prop:rct-unbiased} establishes that under our sampling and random-assignment assumptions, one can form unbiased estimators of site-level effects, and of their variances, covariances, and third central moments, as required by the corrections. The subsequent propositions develop the six applications described above, including results for LATE heterogeneity that cannot be obtained by directly applying the correction to site-specific LATE estimators.
These applications connect to work on LATE heterogeneity in multi-site RCTs
\citep{walters2015inputs,adusumilli2024heterogeneity,
angrist2023implementation}, mediation analysis
\citep{imai2010identification,daniels1997meta,dunn2015evaluation,
reardon2013assumptions,vanderweele2013threeway,kwon2026testing}, and the
contribution of implementation to effect heterogeneity
\citep{bloom2003linking,hong2024organizational,angrist2023implementation}. We discuss these connections in the corresponding
subsections of Section~\ref{sec:multi-site-rcts}.

The remainder of the paper is organized as follows. Section~\ref{sec:framework} presents the general framework and corrected-moment estimators. Section~\ref{sec:multi-site-rcts} develops their applications to multi-site RCTs. Section~\ref{sec:application} applies the methods to public and private counseling programs for French job seekers. Appendix~\ref{app:inference} presents inference results. Appendix~\ref{app:simulations} reports simulations. Appendix~\ref{app:blp} compares tests of treatment-effect heterogeneity based on our corrected-variance estimator and on a version of the BLP approach of \citet{cddf2025}. Appendix~\ref{sec:proofs} contains the proofs, including those of Appendix~\ref{app:blp}. Appendix~\ref{sec:survey} presents a survey of multi-site RCTs published in economics.

\section{General framework and estimators}\label{sec:framework}

In this section, we assume we observe unbiased noisy realizations of a univariate and a multivariate conditional expectation. In such settings, we propose estimators of the variance of the univariate conditional expectation, of its third central moment, and of the coefficient from the regression of the univariate on the multivariate conditional expectation. The following section will apply these results to multi-site RCTs.

\subsection{Setup and target parameters}\label{subsec:framework-setup}

Let $(X,Y,S)$ be random variables, where $Y$ and $S$ are scalar and
$X\in\R^K$. Throughout, vectors are
understood to be column vectors. For any random variable or vector $R$ such that the expectation below is well defined, let
\[
  \mu_R(s)=\E[R\mid S=s].
\]
Let $W(S)$ be such that $W(S)>0$ and $0<\E[W(S)]<\infty$. $W(S)$ is a weighting variable that gives greater weight to some values of $S$ than others. For instance, $S$ may represent the identifier of a cluster of units, and $W(S)$ could represent the cluster's number of units. Let $w(S)=W(S)/\E[W(S)]$. For any random variables $R_1$ and $R_2$, we let
\begin{align*}
&\E_w[R_1]=\E[w(S)R_1]\\
&\Var_w[R_1]=\E[w(S)(R_1-\E_w[R_1])^2]\\
&\Cov_w(R_1,R_2)=\E[w(S)(R_1-\E_w[R_1])(R_2-\E_w[R_2])].
\end{align*}
Our three target parameters are
\begin{align*}
  \sigma^2:=\Var_w(\mu_Y(S)),
  \qquad
  \beta&:=\Var_w(\mu_X(S))^{-1}\Cov_w(\mu_X(S),\mu_Y(S)),\qquad
  \kappa_Y:=\E_w\!\left[\{\mu_Y(S)-\E_w[\mu_Y(S)]\}^3\right].
\end{align*}
$\sigma^2$ is the weighted variance of $\mu_Y(S)$. The vector $\beta$ is the slope coefficient in the population OLS regression of $\mu_Y(S)$ on a constant and $\mu_X(S)$, weighted by $w(S)$. $\kappa_Y$ is the weighted third central moment of $\mu_Y(S)$.

We observe $(S_1,\ldots,S_n)$, $n$ i.i.d. random variables with the same distribution as $S$. We also observe $(W(S_1),\ldots,W(S_n))$, the weights of observations $1$ to $n$. For $i=1,\ldots,n$, let
\[
  \widehat\mu_X(S_i)
  \qquad\text{and}\qquad
  \widehat\mu_Y(S_i)
\]
denote estimators of $\mu_X(S_i)$ and $\mu_Y(S_i)$, respectively. Let
\[
  \widehat V_{X,i}\in\R^{K\times K},
  \qquad
  \widehat V_{Y,i}\in\R,
  \qquad
  \widehat C_{XY,i}\in\R^K,
  \qquad
  \widehat C_{YV,i}\in\R,
  \qquad
  \widehat K_{Y,i}\in\R
\]
 denote estimators of $\Var(\widehat\mu_X(S_i)\mid S_i)$,
 $\Var(\widehat\mu_Y(S_i)\mid S_i)$,
 $\Cov(\widehat\mu_X(S_i),\widehat\mu_Y(S_i)\mid S_i)$,
 $\Cov(\widehat\mu_Y(S_i),\widehat V_{Y,i}\mid S_i)$ and
 $\E[\{\widehat\mu_Y(S_i)-\mu_Y(S_i)\}^3\mid S_i]$, respectively.

\begin{remark}[Predictors measured without error]\label{rem:error-free-x}
The framework allows some or all coordinates of $\mu_X(S_i)$ to be measured without error. For every such coordinate $k$, one simply has
\[
 \widehat\mu_{X,k}(S_i)=\mu_{X,k}(S_i).
\]
The corresponding row and column of $\widehat V_{X,i}$ and the corresponding coordinate of $\widehat C_{XY,i}$ are then equal to zero.
\end{remark}

\begin{assumption}[Sampling and unbiasedness]\label{ass:sampling}
For every $i=1,\ldots,n$:
\begin{enumerate}[label=(\roman*)]
  \item\label{ass:sampling-iid}
  We observe the vector
  \[
    \left(S_i,W(S_i),\widehat\mu_X(S_i),\widehat\mu_Y(S_i),
    \widehat V_{X,i},\widehat V_{Y,i},
    \widehat C_{XY,i},\widehat C_{YV,i},\widehat K_{Y,i}\right).
  \]
  Those vectors are i.i.d. across $i$.

  \item\label{ass:sampling-means}
  Almost surely, all the following conditional unbiasedness restrictions hold:
  \begin{align*}
    &\E[\widehat\mu_X(S_i)\mid S_i]=\mu_X(S_i),\qquad
    \E[\widehat\mu_Y(S_i)\mid S_i]=\mu_Y(S_i),\\
    &\E[\widehat V_{X,i}\mid S_i]
      =\Var(\widehat\mu_X(S_i)\mid S_i),\qquad
    \E[\widehat V_{Y,i}\mid S_i]
      =\Var(\widehat\mu_Y(S_i)\mid S_i),\\
    &\E[\widehat C_{XY,i}\mid S_i]
      =\Cov(\widehat\mu_X(S_i),\widehat\mu_Y(S_i)\mid S_i),\qquad\E[\widehat C_{YV,i}\mid S_i]=\Cov(\widehat\mu_Y(S_i),\widehat V_{Y,i}\mid S_i),\\
    &\E[\widehat K_{Y,i}\mid S_i]
      =\E[\{\widehat\mu_Y(S_i)-\mu_Y(S_i)\}^3\mid S_i].
  \end{align*}
\end{enumerate}
\end{assumption}
Section 3.1.1 of \citet{fuller1987measurement} considers a closely related setup in which a dependent variable and a vector of regressors are observed with measurement error, and where the variance of the measurement errors can be unbiasedly estimated. We further assume that the third centered moment of the estimation
error and its covariance with its variance estimator can be unbiasedly estimated. These additional restrictions allow us to construct the corrected estimator of $\kappa_Y$ below.


\begin{lemma}\label{lem:corrected}
Under Assumption~\ref{ass:sampling}, provided the expectations below exist,
\begin{align}
 &\E_w\!\left[\{\widehat\mu_Y(S_i)-\E_w[\mu_Y(S)]\}^2-\widehat V_{Y,i}
 \right]
 =\Var_w(\mu_Y(S)),\label{eq:corr-delta}\\
 &\E_w\!\left[
   \{\widehat\mu_X(S_i)-\E_w[\mu_X(S)]\}
   \{\widehat\mu_X(S_i)-\E_w[\mu_X(S)]\}\trans-\widehat V_{X,i}
 \right]
 =\Var_w(\mu_X(S)),\label{eq:corr-x}\\
 &\E_w\!\left[
   \{\widehat\mu_X(S_i)-\E_w[\mu_X(S)]\}
   \{\widehat\mu_Y(S_i)-\E_w[\mu_Y(S)]\}
   -\widehat C_{XY,i}
 \right]
 =\Cov_w(\mu_X(S),\mu_Y(S)),\label{eq:corr-xdelta}\\
 &\E_w\!\left[
   \{\widehat\mu_Y(S_i)-\E_w[\mu_Y(S)]\}^3
   -3\{\widehat\mu_Y(S_i)-\E_w[\mu_Y(S)]\}\widehat V_{Y,i}
   +3\widehat C_{YV,i}-\widehat K_{Y,i}
 \right]=\kappa_Y.\label{eq:corr-third}
\end{align}
\end{lemma}
Equations \eqref{eq:corr-delta}-\eqref{eq:corr-xdelta} follow from the laws of total variance and covariance, combined with Assumption 1. Equation \eqref{eq:corr-third} additionally follows by expanding a cube.

\subsection{Estimators}\label{subsec:framework-estimators}

Let
\[
 \widehat{w}(S_i)
 =\frac{W(S_i)}{\frac1n\sum_{j=1}^nW(S_{j})},
 \qquad
 \overline{\widehat\mu}_X
 =\frac1n\sum_{i=1}^n\widehat{w}(S_i)\widehat\mu_X(S_i),
 \qquad
 \overline{\widehat\mu}_Y
 =\frac1n\sum_{i=1}^n\widehat{w}(S_i)\widehat\mu_Y(S_i).
\]
We estimate $\sigma^2$ by
\begin{equation}\label{eq:sigma-hat}
 \widehat\sigma^2
 =\frac1n\sum_{i=1}^n
 \widehat{w}(S_i)\left[
   \{\widehat\mu_Y(S_i)-\overline{\widehat\mu}_Y\}^2
   -\widehat V_{Y,i}
 \right]=\frac1n\sum_{i=1}^n \widehat{w}(S_i)\{\widehat\mu_Y(S_i)-\overline{\widehat\mu}_Y\}^2 -\frac1n\sum_{i=1}^n\widehat w(S_i)\widehat V_{Y,i}.
\end{equation}
$\widehat\sigma^2$ is a corrected estimator of the variance of $\mu_Y(S)$: it subtracts the average estimated variance of $\widehat\mu_Y(S_i)$ from the variance of $\widehat\mu_Y(S_i)$ across the $n$ observations. 

Similarly, we consider the following corrected estimators of $\Var_w(\mu_X(S))$ and $\Cov_w(\mu_X(S),\mu_Y(S))$:
\begin{align}
 \widehat{\Var}(\mu_X(S))
 &=\frac1n\sum_{i=1}^n
 \widehat{w}(S_i)\left[
   \{\widehat\mu_X(S_i)-\overline{\widehat\mu}_X\}
   \{\widehat\mu_X(S_i)-\overline{\widehat\mu}_X\}\trans
   -\widehat V_{X,i}
 \right],\label{eq:varx-hat}\\
 \widehat{\Cov}(\mu_X(S),\mu_Y(S))
 &=\frac1n\sum_{i=1}^n
 \widehat{w}(S_i)\left[
   \{\widehat\mu_X(S_i)-\overline{\widehat\mu}_X\}
   \{\widehat\mu_Y(S_i)-\overline{\widehat\mu}_Y\}
   -\widehat C_{XY,i}
 \right].\label{eq:cov-hat}
\end{align}
Whenever $\widehat{\Var}(\mu_X(S))$ is nonsingular, let
\begin{equation}\label{eq:beta-hat}
 \widehat\beta
 =\widehat{\Var}(\mu_X(S))^{-1}
  \widehat{\Cov}(\mu_X(S),\mu_Y(S)).
\end{equation}
Finally, following \eqref{eq:corr-third} we estimate $\kappa_Y$ by
\begin{equation}\label{eq:kappa-hat}
 \widehat\kappa_Y=\frac1n\sum_{i=1}^n\widehat w(S_i)\left[
 \{\widehat\mu_Y(S_i)-\overline{\widehat\mu}_Y\}^3
 -3\{\widehat\mu_Y(S_i)-\overline{\widehat\mu}_Y\}\widehat V_{Y,i}
 +3\widehat C_{YV,i}-\widehat K_{Y,i}\right].
\end{equation}
When $\widehat\sigma^2>0$, one can use $\widehat\kappa_Y/(\widehat\sigma^2)^{3/2}$ to estimate the skewness of $\mu_Y(S)$.

In Appendix~\ref{app:inference}, we establish estimators' asymptotic normality and their influence functions, under an asymptotic sequence where $n$ goes to infinity but the variances of $\widehat\mu_Y(S_i)$ and $\widehat\mu_X(S_i)$ do not go to zero. In Appendix~\ref{app:simulations}, we report simulations assessing the finite-sample validity of the confidence intervals (CIs) based on these influence functions, in DGPs where site effects can have light or heavy tails, and no or a lot of skewness. The 95\%-level CI for $\beta$ has coverage close to its nominal level in every design considered, with as few as 50 sites. The 95\%-level CI for $\sigma^2$ can slightly undercover with 50 and 100 sites, but has coverage close to its nominal level with 200 sites, except in the DGP with both heavy tails and a lot of skewness. With a studentized site-level bootstrap, the CI for $\sigma^2$ has coverage close to its nominal level even with 50 or 100 sites, including in the DGP with both heavy tails and a lot of skewness. Finally, with 100 sites or more, the 95\%-level CI for $\kappa_Y$ has coverage close to its nominal level in all DGPs, except for the one with both heavy tails and a lot of skewness. The studentized site-level bootstrap does not improve coverage for that parameter, except in the DGP with both heavy tails and a lot of skewness.

\section{Application to multi-site RCTs}\label{sec:multi-site-rcts}

In this section, we show that the results of the previous section can be applied to multi-site RCTs.

Assume that $n$ sites $(S_1,...,S_n)$ are drawn with replacement from a super-population of sites. In each site, $N_i$ units are drawn with replacement from the site's super-population of units. In each site, we allocate $N_{0,i}$ units to a control group and $N_{d,i}$ units to treatment arm $d$, for $d\in \{1,...,\mathcal{D}\}$, $\mathcal{D}\geq 1$, where $N_{d,i}\geq 3$ for every $d\in \{0,...,\mathcal{D}\}$, and $\sum_{d=0}^\mathcal{D} N_{d,i}=N_i$. For each unit $u\in \{1,...,N_i\}$, let $D_{u,i}\in \{0,...,\mathcal{D}\}$ denote the treatment arm it is assigned to. For every $i\in\{1,\ldots,n\}$ and $u\in\{1,\ldots,N_i\}$, we measure a vector of outcomes $\bm{O}_{u,i}\in \mathbb{R}^{L}$, $L\geq1$, with coordinates $O_{u,i,1},...,O_{u,i,L}$. For every $d\in \{0,...,\mathcal{D}\}$, let $\bm{O}_{u,i}(d)$ denote the potential-outcome vector of unit $u$ in site $i$ under treatment $d$. The observed outcome vector satisfies $\bm O_{u,i}=\bm O_{u,i}(D_{u,i})$. Under our assumptions on the sampling of sites and units within sites, we have:
\begin{assumption}\label{hyp:iid}
\begin{enumerate}
\item The vectors $(S_i,N_i,(N_{d,i})_{d=0}^\mathcal{D},((\bm{O}_{u,i}(d))_{d=0}^\mathcal{D})_{u=1}^{N_i})$ are i.i.d. across $i$.
\item For all $i$, the vectors $(\bm{O}_{u,i}(d))_{d=0}^\mathcal{D}$ are i.i.d. across $u$ conditional on $S_i$, and we let $(\bm{O}_{i}(d))_{d=0}^\mathcal{D}$ denote a vector with the same probability distribution as $(\bm{O}_{u,i}(d))_{d=0}^\mathcal{D}$ conditional on $S_i$.
\item For every $d\in\{0,\ldots,\mathcal D\}$,
$\E[\norm{\bm O_i(d)}^{3}\mid S_i]<\infty$ almost surely.
\end{enumerate}
\end{assumption}
In multi-site RCTs, participating sites and units are not always probability samples from well-defined populations. In that case, Assumption~\ref{hyp:iid} can be interpreted as a super-population modelling device: the observed sites and units are regarded as draws from hypothetical populations of sites and units that could have participated. This interpretation does not make the sample representative of any particular external population. The i.i.d. sampling assumption within sites plays a key role: together with random assignment, it makes the usual variance estimator for a difference in means unbiased for its sampling variance around the site’s super-population ATE. With fixed sampled units and randomness arising only from treatment assignment, this variance estimator is generally conservative, so subtracting it need not recover the cross-site variance of treatment effects. The assumption that sites are an i.i.d. sample is less crucial. Versions 1--3 of this paper (arXiv:2405.17254) instead treated the observed sites as a fixed finite population. For results common to those versions and the present one, moving to i.i.d. sampling leaves the estimators considered below unchanged, but changes the interpretation of their variance estimators: under the finite-population framework, the estimators' asymptotic variances could in general only be conservatively estimated, whereas under Assumption~\ref{hyp:iid} they can be consistently estimated.

Let
\[
 \mathcal{V}_i=\left(S_i,N_i,(N_{d,i})_{d=0}^{\mathcal D},
 ((\bm O_{u,i}(d))_{d=0}^{\mathcal D})_{u=1}^{N_i}\right)
\]
collect site $i$'s identifier, its number of units, its arm sizes, and its sampled potential outcomes.
To allocate units to treatment arms, a completely randomized experiment is conducted in each site, independently across sites:
\begin{assumption}\label{hyp:strat_completely_randomized}
$N_i$ and the arm sizes $(N_{0,i},\ldots,N_{\mathcal D,i})$ are measurable with respect to $S_i$. For every $i\in\{1,\ldots,n\}$ and every assignment vector $\bm d=(d_1,\ldots,d_{N_i})\in\{0,\ldots,\mathcal{D}\}^{N_i}$ such that $\sum_{u=1}^{N_i}\mathbf 1\{d_u=d\}=N_{d,i}
 \quad\text{for every }d\in\{0,\ldots,\mathcal{D}\},$
\[
 \Pp\!\left((D_{u,i})_{u=1}^{N_i}=\bm d\mid (\mathcal{V}_{i'})_{i'=1}^n\right)
 =\binom{N_i}{N_{0,i},\ldots,N_{\mathcal{D},i}}^{-1}
 =\frac{\prod_{d=0}^{\mathcal{D}}N_{d,i}!}{N_i!}\quad\text{almost surely}.
\]
Moreover, conditional on $(\mathcal{V}_i)_{i=1}^n$, the assignment vectors $(D_{u,i})_{u=1}^{N_i}$ are independent across $i$.
\end{assumption}
Assumption~\ref{hyp:strat_completely_randomized} states that within every site, a lottery assigns a fixed number of units to each arm, as in a stratified experiment with sites as strata. Importantly, the assumptions let sites differ arbitrarily in size, in the share of units assigned to each arm, and in the effects of the treatments. Moreover, the number of units per site can be small.

Let
$\mu_{O}(S_i)=(\E[\bm{O}_{i}(d)\mid S_i])_{d=0}^{\mathcal{D}}$ denote the $L(\mathcal{D}+1)$ vector of the average outcomes $(O_{i,1},\ldots,O_{i,L})$ under treatments $(0,\ldots,\mathcal{D})$ in site $i$.
The corresponding estimator is the $L(\mathcal D+1)$ vector of
arm-specific sample means
\[
 \widehat\mu_O(S_i)
 =\left(\frac{1}{N_{d,i}}\sum_{u=1}^{N_i}
 \mathbf 1\{D_{u,i}=d\}\bm O_{u,i}\right)_{d=0}^{\mathcal D}.
\]
To estimate $\Var(\widehat\mu_O(S_i)\mid S_i)$, we use
$\widehat V_{O,i}$, an $L(\mathcal D+1)\times
L(\mathcal D+1)$ matrix whose entries are
\begin{align*}
 [\widehat V_{O,i}]_{(d,\ell),(d',\ell')}
 &=\mathbf 1\{d=d'\}\frac{1}{N_{d,i}(N_{d,i}-1)}
 \sum_{u=1}^{N_i}\mathbf 1\{D_{u,i}=d\}
 \left(O_{u,i,\ell}-\overline O_{d,i,\ell}\right)
 \left(O_{u,i,\ell'}-\overline O_{d,i,\ell'}\right),
\end{align*}
for every $(d,d')\in\{0,\ldots,\mathcal D\}^2$ and
$(\ell,\ell')\in\{1,\ldots,L\}^2$, where
\[
 \overline O_{d,i,\ell}
 =\frac{1}{N_{d,i}}\sum_{u=1}^{N_i}
 \mathbf 1\{D_{u,i}=d\}O_{u,i,\ell}.
\]
Under i.i.d. sampling within sites and completely randomized assignment, sample means from distinct arms are independent conditional on $S_i$, after averaging over both sampling and assignment. Their conditional covariances are therefore zero, explaining the block-diagonal form of $\widehat V_{O,i}$.

Finally, to estimate centered third moments,
for any non-random $L(\mathcal D+1)$ vector
$c=(c_0\trans,\ldots,c_{\mathcal D}\trans)\trans$, with $c_d\in\R^L$, let
\begin{align*}
 \widehat M_{3,O,i}(c)
 &=\sum_{d=0}^{\mathcal D}
 \frac{1}{N_{d,i}(N_{d,i}-1)(N_{d,i}-2)}
 \sum_{u=1}^{N_i}\mathbf 1\{D_{u,i}=d\}
 \left[c_d\trans\{\bm O_{u,i}-\overline{\bm O}_{d,i}\}\right]^3,
\end{align*}
where $\overline{\bm O}_{d,i}=N_{d,i}^{-1}\sum_u
\mathbf 1\{D_{u,i}=d\}\bm O_{u,i}$.
\begin{proposition}\label{prop:rct-unbiased}
Under Assumptions~\ref{hyp:iid} and
\ref{hyp:strat_completely_randomized}, for every $i$,
\begin{align}
 \E[\widehat\mu_O(S_i)\mid S_i]
 &=\mu_O(S_i),\label{eq:unbiased_estimator}\\
 \E[\widehat V_{O,i}\mid S_i]
 &=\Var(\widehat\mu_O(S_i)\mid S_i),
 \label{eq:unbiased_variance}\\
 \E[\widehat M_{3,O,i}(c)\mid S_i]
 &=\E\!\left[
 \{c\trans(\widehat\mu_O(S_i)-\mu_O(S_i))\}^3\mid S_i
 \right]\quad\text{for every non-random }c.
 \label{eq:unbiased_third_moment}
\end{align}
Let $c$ and $B$ be a non-random $L(\mathcal D+1)$ vector and a non-random $K\times L(\mathcal D+1)$ matrix, respectively, and let
\[
 Y_i=c\trans(\bm{O}_{i}(d))_{d=0}^{\mathcal{D}},
 \qquad X_i=B(\bm{O}_{i}(d))_{d=0}^{\mathcal{D}}.
\]
Define
\begin{align*}
 \widehat\mu_Y(S_i)
   &=c\trans\widehat\mu_O(S_i),
 &\widehat\mu_X(S_i)
   &=B\widehat\mu_O(S_i),\\
 \widehat V_{Y,i}
   &=c\trans\widehat V_{O,i}c,
 &\widehat V_{X,i}
   &=B\widehat V_{O,i}B\trans,\\
 \widehat C_{XY,i}
   &=B\widehat V_{O,i}c,
   &\widehat C_{YV,i}&=\widehat K_{Y,i}
 =\widehat M_{3,O,i}(c).
\end{align*}
Then,
\[
 \left(S_i,W(S_i),\widehat\mu_X(S_i),\widehat\mu_Y(S_i),
 \widehat V_{X,i},\widehat V_{Y,i},
 \widehat C_{XY,i},\widehat C_{YV,i},\widehat K_{Y,i}\right)
\]
satisfies Assumption~\ref{ass:sampling}.
\end{proposition}
Proposition~\ref{prop:rct-unbiased} shows that the corrected-moment estimators in the previous section are applicable to multi-site experiments,  whenever  $\mu_Y(S)$ and $\mu_X(S)$ are known linear transformations of arm-specific average potential outcomes. The site-level estimates and their sampling-moment corrections must be unbiased but need not be consistent as the number of sites grows.
The result follows from the usual unbiasedness of sample means, sample
covariances, and third centered moments. $\widehat K_{Y,i}$ and
$\widehat C_{YV,i}$ are equal, because the centered third moment of a sample mean equals its covariance with its estimated variance under i.i.d. sampling.

For the applications below, write
\[
 \widehat C_{d,i,\ell\ell'}
 =[\widehat V_{O,i}]_{(d,\ell),(d,\ell')},
 \qquad
 \widehat V_{d,i,\ell}=\widehat C_{d,i,\ell\ell}.
\]
These estimate the sampling covariance of the arm-$d$ means of outcomes $\ell$ and $\ell'$, and the sampling variance of the arm-$d$ mean of outcome $\ell$, respectively. With one outcome, abbreviate $\widehat V_{d,i,1}$ as $\widehat V_{d,i}$.

As an example, with one treatment arm and one outcome ($\mathcal{D}=1$, $L=1$),
	$$\widehat{\bm\mu}_O(S_i)=(\overline{O}_{0,i,1},\overline{O}_{1,i,1})\trans$$ and
	$\widehat{V}_{O,i}$ is diagonal, with the estimated variances of the two averages
	on its diagonal. Letting $\bm c=(-1,1)\trans$ and $B=(1,0)$ yields
	$\widehat{\mu}_Y(S_i)$ equal to the site's estimated ATE and
	$\widehat{\mu}_X(S_i)$ equal to its control-group mean. Then
	$\widehat{V}_{Y,i}=\bm c\trans\widehat{V}_{O,i}\bm c=\widehat V_{0,i}+\widehat V_{1,i}$ recovers the standard
	variance formula for a difference in means, and
	$\widehat{C}_{XY,i}=B\widehat{V}_{O,i}\bm c=-\widehat{V}_{0,i}$ captures the
	mechanical negative covariance between the difference in means and the control group average. These variance and covariance corrections can also be obtained by specializing the leave-out covariance estimator in the random-coefficient model in \citet{kline2020leave} to a binary regressor.
Finally, with $\bm c=(-1,1)\trans$
\begin{align*}
 \widehat C_{YV,i}=\widehat K_{Y,i}
 &=\frac{1}{N_{1,i}(N_{1,i}-1)(N_{1,i}-2)}
 \sum_{u=1}^{N_i}\mathbf 1\{D_{u,i}=1\}
 (O_{u,i,1}-\overline O_{1,i,1})^3\\
 &\quad-\frac{1}{N_{0,i}(N_{0,i}-1)(N_{0,i}-2)}
 \sum_{u=1}^{N_i}\mathbf 1\{D_{u,i}=0\}
 (O_{u,i,1}-\overline O_{0,i,1})^3.
\end{align*}

\begin{remark}[Possible choices for the weights]
One may let the weights $W(S_i)$ be equal to $N_i$, thus giving more weight to sites with more units. One could also let $W(S_i)=1$, thus giving the same weight to each site. Those are two common choices of weights, but other choices are possible.
\end{remark}

\begin{remark}[Site-level predictors measured without error]
As noted in Remark~\ref{rem:error-free-x}, $X_i$ may also include site-specific variables that do not need to be estimated and are measured without error. Examples include the unemployment rate or GDP per capita in site $i$. For those coordinates, the corresponding variance and covariance corrections are zero.
\end{remark}

\begin{remark}[Minimum number of units per arm]
The maintained requirement $N_{d,i}\geq3$ is needed only to construct
$\widehat M_{3,O,i}(c)$, and hence to estimate
$\kappa_Y$. The variance and regression estimators $\widehat\sigma^2$ and
$\widehat\beta$ require only $N_{d,i}\geq2$ in every arm.
\end{remark}

We now highlight six potentially relevant applications of Theorems \ref{thm:sigma}, \ref{thm:beta}, \ref{thm:kappa} and Proposition \ref{prop:rct-unbiased}.

\subsection{Estimating the variance and centered third moment of ATEs across sites}\label{subsec:rct-ate-variance}

In this subsection, we momentarily assume that
$\mathcal D=1$ and $L=1$: the RCT has one treatment arm, and we are interested
in the treatment effect on one outcome. Let $\sigma^2_{\mathrm{ATE}}=\Var_w(\E[O_{i,1}(1)-O_{i,1}(0)| S_i])$ denote the weighted variance across sites of $\E[O_{i,1}(1)-O_{i,1}(0)| S_i]$, the average effect of the treatment on $O_{i,1}$ in site $i$. Estimating this variance reveals whether the overall ATE summarizes effects that are similar across sites or instead masks differences across sites.
Let
\begin{align*}
 \kappa_{\mathrm{ATE}}
 =\E_w\!\left[\left\{
 \E[O_{i,1}(1)-O_{i,1}(0)\mid S_i]
 -\E_w\!\left[\E[O_{i,1}(1)-O_{i,1}(0)\mid S_i]\right]
 \right\}^3\right]
\end{align*}
denote the corresponding weighted centered third moment. Estimating
$\kappa_{\mathrm{ATE}}$ reveals whether there is some asymmetry in the distribution of treatment effects across sites, with unusually large effects concentrated above or
below the overall ATE.

To estimate $\sigma^2_{\mathrm{ATE}}$, one lets
\[
 Y_i=O_{i,1}(1)-O_{i,1}(0).
\]
The corresponding estimator of $\mu_Y(S_i)$ and variance correction are
\begin{align*}
 \widehat\mu_Y(S_i)&=\overline O_{1,i,1}-\overline O_{0,i,1},
 &\widehat V_{Y,i}&=\widehat V_{1,i}+\widehat V_{0,i}.
\end{align*}
Plugging those into \eqref{eq:sigma-hat} yields an estimator of $\sigma^2_{\mathrm{ATE}}$.

To estimate $\kappa_{\mathrm{ATE}}$, Proposition~\ref{prop:rct-unbiased} and
Theorem~\ref{thm:kappa} use the same difference-in-means estimator and variance
correction, together with
\begin{align*}
 \widehat C_{YV,i}=\widehat K_{Y,i}
 &=\frac{1}{N_{1,i}(N_{1,i}-1)(N_{1,i}-2)}
 \sum_{u=1}^{N_i}\mathbf 1\{D_{u,i}=1\}
 (O_{u,i,1}-\overline O_{1,i,1})^3\\
 &\quad-\frac{1}{N_{0,i}(N_{0,i}-1)(N_{0,i}-2)}
 \sum_{u=1}^{N_i}\mathbf 1\{D_{u,i}=0\}
 (O_{u,i,1}-\overline O_{0,i,1})^3.
\end{align*}
Plugging these quantities into \eqref{eq:kappa-hat} yields an estimator of
$\kappa_{\mathrm{ATE}}$.

\subsection{Regressing sites' ATEs on their average untreated outcome}\label{subsec:rct-ate-untreated}

In this subsection, we again assume that
$\mathcal D=1$ and $L=1$. Let $\beta_{\mathrm{ATE},0}$ denote the coefficient on
$\E[O_{i,1}(0)| S_i]$ in a regression of
$\E[O_{i,1}(1)-O_{i,1}(0)| S_i]$ on an intercept and
$\E[O_{i,1}(0)| S_i]$, weighted by $W(S_i)$. $\E[O_{i,1}(0)| S_i]$ measures how well the site would fare in the
absence of the intervention, so the regression reveals whether treatment
effects tend to be larger or smaller in sites with worse untreated outcomes.
This regression may also inform targeting outside the RCT if the relationship between treatment effects and untreated outcomes carries over to the new sites, and if outcomes among previous cohorts provide a useful proxy for new sites' untreated outcomes. Then, if $\beta_{\mathrm{ATE},0}$ is, say, negative, one may prioritize non-RCT sites with the worst outcomes among previous cohorts.

To estimate $\beta_{\mathrm{ATE},0}$, one lets
\[
 Y_i=O_{i,1}(1)-O_{i,1}(0),
 \qquad
 X_i=O_{i,1}(0).
\]
The corresponding estimators of $\mu_Y(S_i)$ and $\mu_X(S_i)$ and
variance--covariance corrections are
\begin{align*}
 \widehat\mu_Y(S_i)&=\overline O_{1,i,1}-\overline O_{0,i,1},
 &\widehat\mu_X(S_i)&=\overline O_{0,i,1},\\
 \widehat V_{Y,i}&=\widehat V_{1,i}+\widehat V_{0,i},
 &\widehat V_{X,i}&=\widehat V_{0,i},
 &\widehat C_{XY,i}&=-\widehat V_{0,i}.
\end{align*}
Plugging those into \eqref{eq:beta-hat} yields an estimator of $\beta_{\mathrm{ATE},0}$. Note that a naive regression of estimated effects on estimated untreated outcomes is generally biased. In particular, the control-group mean enters positively in
the estimator of the untreated outcome and negatively in the
difference-in-means estimator, making their estimation errors mechanically
negatively correlated. $\widehat{\beta}_{\mathrm{ATE},0}$ removes this covariance, as well as
the usual attenuation bias coming from measurement error in the estimation of $\E[O_{i,1}(0)\mid S_i]$.

\subsection{Estimating LATEs' heterogeneity in RCTs with imperfect compliance}\label{subsec:late-heterogeneity}

\subsubsection{Set-up}\label{subsubsec:late-setup}

In this subsection, we assume that $\mathcal D=1$ and $L=2$, and allow for imperfect compliance with treatment assignment. Outcome $O_{i,1}$ is treatment take-up and outcome $O_{i,2}$ is the outcome of interest. Let $O_{i,1}(d)$ denote the treatment that a generic unit in site $i$ would receive if assigned to arm $d$, and let $O_{i,2}(d,o_1)$ denote the potential value of the outcome of interest if the unit were assigned to arm $d$ and its treatment were set to $o_1$. Thus, the assignment-indexed potential outcome introduced above satisfies
\[
 O_{i,2}(d)=O_{i,2}(d,O_{i,1}(d)).
\]
Let
\begin{align*}
&\mathrm{FS}_i=O_{i,1}(1)-O_{i,1}(0),\\
&\mathrm{ITT}_i=O_{i,2}(1)-O_{i,2}(0).
\end{align*}
Then, $\mu_{\mathrm{FS}}(S_i)$ and $\mu_{\mathrm{ITT}}(S_i)$ respectively denote site $i$'s first-stage effect and its intention-to-treat (ITT) effect on the outcome of interest.

We impose the usual exclusion restriction and monotonicity conditions.
\begin{assumption}[Exclusion and monotonicity]\label{ass:late}
For every $o_1$ in the support of the received treatment,
\[
 O_{i,2}(1,o_1)=O_{i,2}(0,o_1)
 \quad\text{almost surely}.
\]
Moreover, for $d\in\{0,1\}$ $O_{i,1}(d)\in\{0,1\}$ almost surely,\footnote{This condition ensures that results apply to site-specific LATEs. Without it, \eqref{eq:late-variance-FSsq} below instead applies to site-specific average causal responses \citep{angrist1995}. More generally, results in the spirit of \eqref{eq:late-variance-FSsq} apply to site-specific parameters that can be written as ratios with positive denominators, provided the numerators and denominators admit unbiased estimators, whose variances and covariance can also be unbiasedly estimated. This may accommodate suitably defined estimands with a scalar nonbinary instrument, though researchers need to verify that their numerator and denominator estimators satisfy the stated requirements.}  $O_{i,1}(1)\geq O_{i,1}(0)$ almost surely, and $\mu_{\mathrm{FS}}(S_i)>0$ almost surely.
\end{assumption}
The exclusion restriction states that assignment affects outcomes only through the treatment received, while monotonicity rules out defiers.

Define the LATE of site $i$ on the outcome of interest as
\[
 \mu_{\mathrm{LATE}}(S_i)
 =\frac{\mu_{\mathrm{ITT}}(S_i)}{\mu_{\mathrm{FS}}(S_i)}.
\]
As shown by \cite{Imbens1994}, under Assumption~\ref{ass:late} $\mu_{\mathrm{LATE}}(S_i)$ is the average effect of the treatment on the outcome among compliers in site $i$.

Let
\[
 \mathrm{FS}=\E_w[\mu_{\mathrm{FS}}(S)],
 \qquad
 \mathrm{ITT}=\E_w[\mu_{\mathrm{ITT}}(S)]
\]
denote the weighted average first-stage and ITT effects across all sites.
The overall LATE is
\begin{equation}\label{eq:overall-late}
 \mathrm{LATE}
 =\frac{\mathrm{ITT}}{\mathrm{FS}}
 =\E_w\!\left[
   \frac{\mu_{\mathrm{FS}}(S)}{\mathrm{FS}}
   \mu_{\mathrm{LATE}}(S)
 \right].
\end{equation}
Thus, the overall LATE averages site-specific LATEs with weights proportional to $w(S)$ times sites' first stages.

For estimation, let $\widehat\mu_{\mathrm{FS}}(S_i)$ and
$\widehat\mu_{\mathrm{ITT}}(S_i)$ be the usual difference-in-means estimators.
Unbiased estimators of their variances and covariances can be obtained following Proposition~\ref{prop:rct-unbiased}. Apply \eqref{eq:cov-hat} to define
$\widehat C_{\mathrm{FS},\mathrm{ITT}}$, the corrected estimator of
$\Cov_w(\mu_{\mathrm{FS}}(S),\mu_{\mathrm{ITT}}(S))$. Also let
\[
 \widehat{\mathrm{FS}}
 =\frac1n\sum_{i=1}^n\widehat w(S_i)
       \widehat\mu_{\mathrm{FS}}(S_i),
 \qquad
 \widehat{\mathrm{ITT}}
 =\frac1n\sum_{i=1}^n\widehat w(S_i)
       \widehat\mu_{\mathrm{ITT}}(S_i),
 \qquad
 \widehat{\mathrm{LATE}}
 =\frac{\widehat{\mathrm{ITT}}}{\widehat{\mathrm{FS}}}.
\]

\subsubsection{Estimating a variance of LATEs across sites}\label{subsubsec:late-variance}

We consider two target parameters. The first is the variance of site-specific LATEs under the complier weights defining the overall LATE in \eqref{eq:overall-late}:
\begin{equation}\label{eq:late-variance}
 \sigma^2_{\mathrm{LATE}}
 =\E_w\!\left[
   \frac{\mu_{\mathrm{FS}}(S)}{\mathrm{FS}}
   \{\mu_{\mathrm{LATE}}(S)-\mathrm{LATE}\}^2
 \right].
\end{equation}
The second uses weights proportional to $w(S)$ times sites' squared first stages:
\begin{equation}\label{eq:late-variance_feasible}
 \sigma^2_{\mathrm{LATE},\mathrm{FS}^2}
 =\E_w\!\left[
   \frac{\mu_{\mathrm{FS}}(S)^2}{\E_w[\mu_{\mathrm{FS}}(S)^2]}
   \{\mu_{\mathrm{LATE}}(S)-\mathrm{LATE}_{\mathrm{FS}^2}\}^2
 \right],
\end{equation}
where
\[
 \mathrm{LATE}_{\mathrm{FS}^2}
 =\frac{\E_w[\mu_{\mathrm{FS}}(S)^2\mu_{\mathrm{LATE}}(S)]}
 {\E_w[\mu_{\mathrm{FS}}(S)^2]}.
\]
The squared-first-stage weights give greater importance to sites with stronger first stages. Such weights also arise in saturated 2SLS regressions.\footnote{Theorem~3 of \citet{angrist1995} shows that such regressions weight covariate-specific LATEs by the conditional variance of the first-stage fitted value. With a binary instrument, this variance equals the squared conditional first stage times the conditional variance of the instrument. With equally probable treatment assignment in every covariate cell, the latter factor is constant.}

If first stages are constant across sites, $\sigma^2_{\mathrm{LATE}}$ and $\sigma^2_{\mathrm{LATE},\mathrm{FS}^2}$ are both equal to
$\Var_w(\mu_{\mathrm{ITT}}(S))/\mathrm{FS}^2$, which can be consistently estimated using the corrected ITT variance and the average first stage. In the remainder of this section, we assume that first stages are heterogeneous. With heterogeneous first stages, corrected estimators cannot be applied directly to site-specific LATE estimators: $\widehat\mu_{\mathrm{ITT}}(S_i)/\widehat\mu_{\mathrm{FS}}(S_i)$ is a ratio and is generally not unbiased, and its estimated variance can be very large when the estimated first stage is small. In an application to Head Start, \citet{walters2015inputs} finds that applying the corrected variance approach to site-specific LATE estimates yields a negative corrected variance estimate.

Instead, we propose estimators that use only moments of ITTs and first stages.
\begin{proposition}[Identifying $\sigma^2_{\mathrm{LATE},\mathrm{FS}^2}$ and a lower bound for $\sigma^2_{\mathrm{LATE}}$]
\label{prop:late-variance-identification}
Suppose Assumption~\ref{ass:late} holds and the moments below exist. Then
\begin{equation}
 \sigma^2_{\mathrm{LATE},\mathrm{FS}^2}
 =\frac{\E_w[
 \{\mu_{\mathrm{ITT}}(S)-\mathrm{LATE}_{\mathrm{FS}^2}\mu_{\mathrm{FS}}(S)\}^2]}
 {\E_w[\mu_{\mathrm{FS}}(S)^2]}. \label{eq:late-variance-FSsq}
\end{equation}
If $\E_w[\mu_{\mathrm{FS}}(S)^2]<\mathrm{FS}$,
\begin{align}
 \sigma^2_{\mathrm{LATE}}
 &\geq \underline{\sigma}^2_{\mathrm{LATE}}
 :=\frac{\E_w[
 \{\mu_{\mathrm{ITT}}(S)-\mathrm{LATE}\mu_{\mathrm{FS}}(S)\}^2]}
 {\mathrm{FS}}\notag\\
 &\quad+\frac{\{\E_w[\mu_{\mathrm{FS}}(S)\mu_{\mathrm{ITT}}(S)]
 -\mathrm{LATE}\times\E_w[\mu_{\mathrm{FS}}(S)^2]\}^2}
 {\mathrm{FS}\{\mathrm{FS}-\E_w[\mu_{\mathrm{FS}}(S)^2]\}}
 \label{eq:late-variance-lower-bound}\\
 &=\frac{\E_w[\mu_{\mathrm{FS}}(S)^2]}{\mathrm{FS}}
 \left\{\sigma^2_{\mathrm{LATE},\mathrm{FS}^2}
 +\frac{\mathrm{FS}}{\mathrm{FS}-\E_w[\mu_{\mathrm{FS}}(S)^2]}
 (\mathrm{LATE}_{\mathrm{FS}^2}-\mathrm{LATE})^2\right\}.
 \notag
\end{align}
\end{proposition}
Because first stages are strictly positive, $\sigma^2_{\mathrm{LATE},\mathrm{FS}^2}$ is zero if and only if LATEs are constant across sites. Whenever the lower bound is defined, $\underline{\sigma}^2_{\mathrm{LATE}}$ has the same property. \eqref{eq:late-variance-FSsq} and \eqref{eq:late-variance-lower-bound} can therefore be used to test if differences in take-up alone explain ITT heterogeneity.

\eqref{eq:late-variance-FSsq} follows from the fact that
\[
 \mu_{\mathrm{FS}}(S)^2\{\mu_{\mathrm{LATE}}(S)-\mathrm{LATE}_{\mathrm{FS}^2}\}^2
 =\{\mu_{\mathrm{ITT}}(S)-\mathrm{LATE}_{\mathrm{FS}^2}\mu_{\mathrm{FS}}(S)\}^2.
\]
For \eqref{eq:late-variance-lower-bound}, binary take-up and monotonicity imply
$0<\mu_{\mathrm{FS}}(S)\leq1$. Since $\E_w[\mu_{\mathrm{ITT}}(S)-\mathrm{LATE}\,\mu_{\mathrm{FS}}(S)]=0$, the fact that $$1-\mu_{\mathrm{FS}}(S)=\sqrt{(1-\mu_{\mathrm{FS}}(S))\mu_{\mathrm{FS}}(S)}\sqrt{(1-\mu_{\mathrm{FS}}(S))/\mu_{\mathrm{FS}}(S)}$$ and Cauchy--Schwarz imply
\begin{align*}
 &\left\{\E_w\!\left[
 \mu_{\mathrm{FS}}(S)
 \{\mu_{\mathrm{ITT}}(S)-\mathrm{LATE}\,\mu_{\mathrm{FS}}(S)\}
 \right]\right\}^2\\
 &=\left\{\E_w\!\left[
 \{1-\mu_{\mathrm{FS}}(S)\}
 \{\mu_{\mathrm{ITT}}(S)-\mathrm{LATE}\,\mu_{\mathrm{FS}}(S)\}
 \right]\right\}^2\\
 &\leq \E_w[\mu_{\mathrm{FS}}(S)\{1-\mu_{\mathrm{FS}}(S)\}]\,
 \E_w\!\left[
 \frac{1-\mu_{\mathrm{FS}}(S)}{\mu_{\mathrm{FS}}(S)}
 \{\mu_{\mathrm{ITT}}(S)-\mathrm{LATE}\,\mu_{\mathrm{FS}}(S)\}^2
 \right]\\
 &=\{\mathrm{FS}-\E_w[\mu_{\mathrm{FS}}(S)^2]\}
 \left\{\mathrm{FS}\,\sigma^2_{\mathrm{LATE}}
 -\E_w\!\left[
 \{\mu_{\mathrm{ITT}}(S)-\mathrm{LATE}\,\mu_{\mathrm{FS}}(S)\}^2
 \right]\right\}.
\end{align*}
Rearranging yields \eqref{eq:late-variance-lower-bound}.
The equality below \eqref{eq:late-variance-lower-bound} follows by writing
$$\mu_{\mathrm{LATE}}(S)-\mathrm{LATE}=\mu_{\mathrm{LATE}}(S)-\mathrm{LATE}_{\mathrm{FS}^2}+\mathrm{LATE}_{\mathrm{FS}^2}-\mathrm{LATE},$$ expanding the square, and using $\E_w[\mu_{\mathrm{FS}}(S)^2
\{\mu_{\mathrm{LATE}}(S)-\mathrm{LATE}_{\mathrm{FS}^2}\}]=0$.

When first stages vary across sites, we conjecture that given the first and second moments of ITTs and first stages, without outcome-support restriction $\underline{\sigma}^2_{\mathrm{LATE}}$ is a sharp lower bound for
$\sigma^2_{\mathrm{LATE}}$. We leave the study of this conjecture for future work. If $\E_w[\mu_{\mathrm{FS}}(S)^2]=\mathrm{FS}$, strict positivity of first stages implies full compliance at every site; the complier-weighted variance then equals the ITT variance and is identified directly.

Turning to estimation, let
\[
 m_2=\E_w[\mu_{\mathrm{FS}}(S)^2],
 \qquad r=\E_w[\mu_{\mathrm{FS}}(S)\mu_{\mathrm{ITT}}(S)].
\]
Then
\begin{equation}\label{eq:late-fs2-moments}
 \mathrm{LATE}_{\mathrm{FS}^2}
 =\frac{r}{m_2}
 =\frac{\Cov_w(\mu_{\mathrm{FS}}(S),\mu_{\mathrm{ITT}}(S))
       +\mathrm{FS}\,\mathrm{ITT}}
 {\Var_w(\mu_{\mathrm{FS}}(S))+\mathrm{FS}^2}.
\end{equation}
For a scalar $l$, define
\begin{align*}
 \widehat\nu_i(l)
 &=\widehat\mu_{\mathrm{ITT}}(S_i)-l\widehat\mu_{\mathrm{FS}}(S_i),\\
 \widehat V_{\nu,i}(l)
 &=\widehat V_{\mathrm{ITT},i}
   +l^2\widehat V_{\mathrm{FS},i}
   -2l\widehat C_{\mathrm{FS},\mathrm{ITT},i},\\
 g_i(l)&=\widehat\nu_i(l)^2-\widehat V_{\nu,i}(l),
 \qquad
 \widehat Q(l)=\frac1n\sum_{i=1}^n\widehat w(S_i)g_i(l).
\end{align*}
The first two quantities are the estimated site effect on $O_{i,2}-lO_{i,1}$ and its estimated sampling variance. Define
\begin{align}
 \widehat r
 &=\frac1n\sum_{i=1}^n\widehat w(S_i)
 \{\widehat\mu_{\mathrm{FS}}(S_i)\widehat\mu_{\mathrm{ITT}}(S_i)
 -\widehat C_{\mathrm{FS},\mathrm{ITT},i}\},\notag\\
 \widehat m_2
 &=\frac1n\sum_{i=1}^n\widehat w(S_i)
 \{\widehat\mu_{\mathrm{FS}}(S_i)^2-\widehat V_{\mathrm{FS},i}\},\notag\\
 \widehat{\mathrm{LATE}}_{\mathrm{FS}^2}
 &=\widehat r/\widehat m_2,
 \qquad
 \widehat q=\widehat Q(\widehat{\mathrm{LATE}}_{\mathrm{FS}^2}).
 \label{eq:q-late-hat}
\end{align}
Our estimators are
\begin{equation}\label{eq:late-variance-hat}
 \widehat\sigma^2_{\mathrm{LATE},\mathrm{FS}^2}
 =\frac{\widehat q}{\widehat m_2}
\end{equation}
and
\begin{align}
 \widehat{\underline{\sigma}}^2_{\mathrm{LATE}}
 &=\frac{\widehat Q(\widehat{\mathrm{LATE}})}
 {\widehat{\mathrm{FS}}}
 +\frac{(\widehat r-\widehat m_2\widehat{\mathrm{LATE}})^2}
 {\widehat{\mathrm{FS}}(\widehat{\mathrm{FS}}-\widehat m_2)}
 \label{eq:late-variance-lower-bound-hat}\\
 &=\frac{\widehat m_2}{\widehat{\mathrm{FS}}}
 \left\{\widehat\sigma^2_{\mathrm{LATE},\mathrm{FS}^2}
 +\frac{\widehat{\mathrm{FS}}}{\widehat{\mathrm{FS}}-\widehat m_2}
 (\widehat{\mathrm{LATE}}_{\mathrm{FS}^2}
   -\widehat{\mathrm{LATE}})^2\right\}.
 \notag
\end{align}
These expressions are used whenever their denominators are nonzero.
Theorem~\ref{thm:late-variance} in Appendix~\ref{app:late-variance-inference} establishes asymptotic normality and gives the influence functions and standard-error estimators for both quantities. Estimating $\mathrm{LATE}_{\mathrm{FS}^2}$ has no first-order effect on $\widehat\sigma^2_{\mathrm{LATE},\mathrm{FS}^2}$, whereas estimating $\mathrm{LATE}$ generally contributes to the influence function of $\widehat{\underline{\sigma}}^2_{\mathrm{LATE}}$.

\paragraph{Comparison with existing approaches to LATE heterogeneity.}\label{subsubsec:late-comparison}
To estimate the variance of LATEs across sites, \citet{walters2015inputs} estimates a parametric selection model with jointly normal center-specific potential-outcome and compliance parameters.
\citet{adusumilli2024heterogeneity} use a grouped-random-effect model allowing compliance and treatment effects to be jointly heterogeneous.
Our approach identifies a squared-first-stage-weighted variance and a lower bound for the complier-weighted variance without parametric restrictions on their joint distribution.
\citet{angrist2023implementation} propose frequentist and Bayesian estimators of LATE heterogeneity with few sites and many units per site, a different setting from the many-sites, few-units-per-site setting considered here.
Similarly, \citet{walters2015inputs} uses an overidentification test of LATE homogeneity, which tests whether sites' ITTs and first stages lie on a common line through the origin, up to sampling error. The test's conventional asymptotic justification applies with few sites and many units per site.

\subsubsection{Estimating the covariance between sites' LATEs and their first-stage effects}\label{subsubsec:late-covariance}

In this subsection, our target parameter is
\begin{equation}\label{eq:late-fs-cov}
 \sigma_{\mathrm{LATE},\mathrm{FS}}
 =\E_w\!\left[
   \frac{\mu_{\mathrm{FS}}(S)}{\mathrm{FS}}
   \{\mu_{\mathrm{LATE}}(S)-\mathrm{LATE}\}
   \mu_{\mathrm{FS}}(S)
 \right].
\end{equation}
$\sigma_{\mathrm{LATE},\mathrm{FS}}$ is the covariance between site-specific LATEs and first stages under the same complier weights as those defining the overall LATE in \eqref{eq:overall-late}. It reveals whether sites with larger shares of compliers also tend to have compliers who benefit more from treatment. A positive covariance is consistent with Roy selection across sites: sites where compliers benefit more tend to have larger shares of compliers.

Let
\[
 \beta_{\mathrm{ITT},\mathrm{FS}}
 =\frac{\Cov_w(\mu_{\mathrm{FS}}(S),\mu_{\mathrm{ITT}}(S))}
        {\Var_w(\mu_{\mathrm{FS}}(S))}.
\]
\begin{proposition}\label{prop:late-fs-cov}
Suppose Assumption~\ref{ass:late} holds, the moments below exist, and
$\Var_w(\mu_{\mathrm{FS}}(S))>0$. Then,
\begin{equation}\label{eq:late-fs-identification}
 \sigma_{\mathrm{LATE},\mathrm{FS}}
 =\frac{\Var_w(\mu_{\mathrm{FS}}(S))}{\mathrm{FS}}
  \left(\beta_{\mathrm{ITT},\mathrm{FS}}-\mathrm{LATE}\right).
\end{equation}
\end{proposition}
Thus, $\sigma_{\mathrm{LATE},\mathrm{FS}}$ is positive if and only if the slope from the regression of sites’ ITTs on an intercept and their first stages exceeds the overall LATE. Equivalently, one can show that
\[
 \sigma_{\mathrm{LATE},\mathrm{FS}}
 =\frac{\E_w[\mu_{\mathrm{FS}}(S)^2]}{\mathrm{FS}}
 \bigl(\mathrm{LATE}_{\mathrm{FS}^2}-\mathrm{LATE}\bigr).
\]
Thus, the covariance is proportional to the change in the mean LATE when moving from first-stage to squared-first-stage weights.

We estimate \eqref{eq:late-fs-cov} by
\begin{equation}\label{eq:late-fs-cov-hat}
 \widehat\sigma_{\mathrm{LATE},\mathrm{FS}}
 =\frac{\widehat C_{\mathrm{FS},\mathrm{ITT}}
       -\widehat{\mathrm{LATE}}\,\widehat V_{\mathrm{FS}}}
       {\widehat{\mathrm{FS}}},
\end{equation}
where $\widehat V_{\mathrm{FS}}$ denotes the corrected variance estimator in
\eqref{eq:varx-hat} with $X=\mathrm{FS}$. Appendix~\ref{app:late-covariance-inference} gives a heuristic derivation of the influence function of $\widehat\sigma_{\mathrm{LATE},\mathrm{FS}}$.

\subsubsection{Relationships between LATEs across outcomes and treatment arms}\label{subsubsec:late-prediction}

In this section we study two targets: a weighted regression of one outcome's LATE on other outcomes' LATEs within a treatment arm, and a covariance--variance ratio relating LATEs across treatment arms. The latter uses different weights in its numerator and denominator and therefore is generally not a regression coefficient.

To define these targets, we consider a more general setting, namely an RCT with two treatment arms and $J+2$ outcomes. Outcomes $O_{i,1},\ldots,O_{i,J}$ are the $J\geq 2$ outcomes of interest, while $O_{i,J+1}$ and $O_{i,J+2}$ indicate take-up of treatments 1 and 2, respectively. For $d\in\{1,2\}$ and $k\in\{1,\ldots,J\}$, let
\[
 \mu_{\mathrm{FS}_d}(S)
 =\E[O_{i,J+d}(d)-O_{i,J+d}(0)\mid S_i=S],
 \qquad
 \mu_{\mathrm{ITT}_{d,k}}(S)
 =\E[O_{i,k}(d)-O_{i,k}(0)\mid S_i=S],
\]
respectively denote the site-specific first-stages of treatment $d$, and its ITT effects on outcome $k$. Then, let
\[
 \mu_{\mathrm{LATE}_{d,k}}(S)
 =\frac{\mu_{\mathrm{ITT}_{d,k}}(S)}{\mu_{\mathrm{FS}_d}(S)}
\]
denote the site-specific LATEs of treatment $d$ on outcome $k$.
We assume the analogues of Assumption~\ref{ass:late} for both treatment arms, and we assume that assignment to arm $d$ does not affect take-up of treatment $d'\ne d$.\footnote{For instance, this holds automatically with one-sided imperfect compliance, where one can take up treatment $d$ only if assigned to it.} For positive random variables $A(S)$, write
\[
 \E_A[Z(S)]=\frac{\E_w[A(S)Z(S)]}{\E_w[A(S)]},
\]
and use $\Var_A$ and $\Cov_A$ for the corresponding variance and covariance.

Our first target predicts the LATE of arm $d$ on outcome 1 using its LATEs on outcomes 2 through $J$. Let
\[
 \bm\mu_{\mathrm{LATE},d,-1}(S)
 =\big(\mu_{\mathrm{LATE}_{d,2}}(S),\ldots,
       \mu_{\mathrm{LATE}_{d,J}}(S)\big)\trans .
\]
For $d\in\{1,2\}$, define
\begin{equation}\label{eq:late-within-treatment-target}
 \beta^d
 =\Var_{\mathrm{FS}_d^2}(\bm\mu_{\mathrm{LATE},d,-1}(S))^{-1}
   \Cov_{\mathrm{FS}_d^2}
   (\bm\mu_{\mathrm{LATE},d,-1}(S),\mu_{\mathrm{LATE}_{d,1}}(S)),
\end{equation}
whenever the variance matrix is nonsingular. This is the coefficient on $\bm\mu_{\mathrm{LATE},d,-1}(S)$ from the regression of $\mu_{\mathrm{LATE}_{d,1}}(S)$ on an intercept and $\bm\mu_{\mathrm{LATE},d,-1}(S)$, under weights proportional to arm $d$'s squared first stage. $\beta^d$ can for instance be useful to estimate predictive mediation regressions as in Section~\ref{subsubsec:predictive-mediation} below, but on LATEs instead of ITT effects.

Our second target, for each outcome $k$, is the covariance--variance ratio
\begin{equation}\label{eq:late-cross-treatment-target}
 \beta^{12}_k
 =\frac{\Cov_{\mathrm{FS}_1\mathrm{FS}_2}
  (\mu_{\mathrm{LATE}_{1,k}}(S),\mu_{\mathrm{LATE}_{2,k}}(S))}
 {\Var_{\mathrm{FS}_1^2}(\mu_{\mathrm{LATE}_{1,k}}(S))}.
\end{equation}
$\beta^{12}_k$ resembles the coefficient from a regression of arm 2's outcome $k$ LATE on arm 1's outcome $k$ LATE. It is not literally a weighted least-squares coefficient, however, because its numerator and denominator use different weights. To make the comparison precise, when the corresponding weighted regression is well defined, let $\widetilde\beta^{12}_k$ denote the coefficient on $\mu_{\mathrm{LATE}_{1,k}}(S)$ from the regression of $\mu_{\mathrm{LATE}_{2,k}}(S)$ on an intercept and $\mu_{\mathrm{LATE}_{1,k}}(S)$, with weights proportional to $w(S)\mu_{\mathrm{FS}_1}(S)\mu_{\mathrm{FS}_2}(S)$:
\[
 \widetilde\beta^{12}_k
 =\frac{\Cov_{\mathrm{FS}_1\mathrm{FS}_2}
 (\mu_{\mathrm{LATE}_{1,k}}(S),\mu_{\mathrm{LATE}_{2,k}}(S))}
 {\Var_{\mathrm{FS}_1\mathrm{FS}_2}(\mu_{\mathrm{LATE}_{1,k}}(S))}
 =\frac{\Var_{\mathrm{FS}_1^2}(\mu_{\mathrm{LATE}_{1,k}}(S))}
 {\Var_{\mathrm{FS}_1\mathrm{FS}_2}(\mu_{\mathrm{LATE}_{1,k}}(S))}
 \beta^{12}_k.
\]
$\widetilde\beta^{12}_k$ and $\beta^{12}_k$ have the same sign. Moreover, $\widetilde\beta^{12}_k=\beta^{12}_k$ when the two arms have equal first stages within every site, or, more generally, when $\mu_{\mathrm{FS}_2}(S)=c\mu_{\mathrm{FS}_1}(S)$ almost surely for a constant $c>0$. We focus on \(\beta^{12}_k\), because it can be expressed using only second moments of site-specific ITTs and first stages, whereas \(\widetilde\beta^{12}_k\) generally cannot. Finally, note that the two arms may have different complier populations within each site. Consequently, the cross-arm association may reflect both differences in treatment effectiveness and differences in complier composition.

\begin{proposition}[Identification of cross-outcome and cross-treatment LATE relationships]\label{prop:late-prediction}
Suppose the analogues of Assumption~\ref{ass:late} hold for each treatment arm $d\in\{1,2\}$ and each outcome $k\in\{1,\ldots,J\}$, with binary take-up and $\mu_{\mathrm{FS}_d}(S)>0$ almost surely. Suppose also that assignment to one treatment arm does not change take-up of the other treatment:
\[
 O_{i,J+d'}(d)=O_{i,J+d'}(0)
 \quad\text{almost surely for }d,d'\in\{1,2\},\ d\ne d'.
\]
Assume $\E_w[\mu_{\mathrm{ITT}_{d,k}}(S)^2]<\infty$ for every $d\in\{1,2\}$ and $k\in\{1,\ldots,J\}$.
\begin{enumerate}
 \item For each $d\in\{1,2\}$ such that $\Var_{\mathrm{FS}_d^2}(\bm\mu_{\mathrm{LATE},d,-1}(S))$ is nonsingular, and for $j\in\{1,\ldots,J\}$, let
 \[
 \mathrm{LATE}_{d,j,\mathrm{FS}_d^2}
 =\E_{\mathrm{FS}_d^2}[\mu_{\mathrm{LATE}_{d,j}}(S)].
 \]
 Then $\beta^d$ is the coefficient from the $w$-weighted regression without an intercept of
 \[
 \mu_{\mathrm{ITT}_{d,1}}(S)
 -\mu_{\mathrm{FS}_d}(S)\mathrm{LATE}_{d,1,\mathrm{FS}_d^2}
 \]
 on
 \[
 \left(\mu_{\mathrm{ITT}_{d,2}}(S)
 -\mu_{\mathrm{FS}_d}(S)\mathrm{LATE}_{d,2,\mathrm{FS}_d^2},\ldots,
 \mu_{\mathrm{ITT}_{d,J}}(S)
 -\mu_{\mathrm{FS}_d}(S)\mathrm{LATE}_{d,J,\mathrm{FS}_d^2}\right)\trans.
 \]
 \item For each $k\in\{1,\ldots,J\}$ such that $\Var_{\mathrm{FS}_1^2}(\mu_{\mathrm{LATE}_{1,k}}(S))>0$,
 \begin{equation}\label{eq:late-cross-treatment-identification}
 \beta^{12}_k=
 \frac{
 \dfrac{\E_w[\mu_{\mathrm{ITT}_{1,k}}(S)\mu_{\mathrm{ITT}_{2,k}}(S)]}
 {\E_w[\mu_{\mathrm{FS}_1}(S)\mu_{\mathrm{FS}_2}(S)]}
 -\dfrac{\E_w[\mu_{\mathrm{FS}_2}(S)\mu_{\mathrm{ITT}_{1,k}}(S)]
 \E_w[\mu_{\mathrm{FS}_1}(S)\mu_{\mathrm{ITT}_{2,k}}(S)]}
 {\E_w[\mu_{\mathrm{FS}_1}(S)\mu_{\mathrm{FS}_2}(S)]^2}}
 {\dfrac{\E_w[\mu_{\mathrm{ITT}_{1,k}}(S)^2]}
 {\E_w[\mu_{\mathrm{FS}_1}(S)^2]}
 -\dfrac{\E_w[\mu_{\mathrm{FS}_1}(S)\mu_{\mathrm{ITT}_{1,k}}(S)]^2}
 {\E_w[\mu_{\mathrm{FS}_1}(S)^2]^2}}.
 \end{equation}
\end{enumerate}
\end{proposition}
Thus, both targets depend only on second moments of site-specific ITTs and first stages. Under the sampling and moment conditions ensuring consistency of the corrected second-moment estimators, both targets can therefore be consistently estimated. Appendix~\ref{app:late-regressions} gives estimators of $\beta^d$ and $\beta^{12}_k$ and heuristic derivations of their influence functions.

\subsection{Predictive and causal mediation analysis}\label{subsec:rct-mediation}

In this subsection, we return to assuming perfect compliance (or considering ITT effects), and we assume that $\mathcal D=1$ and
$L=2$: the RCT has one treatment arm, and we are interested in the treatment
effect on two outcomes. $O_{i,2}$ is the final outcome and $O_{i,1}$ is an intermediate outcome that might mediate the effect of the treatment on the final outcome.

\subsubsection{Predictive mediation}\label{subsubsec:predictive-mediation}

Let $\beta_{\mathrm{med}}$ denote the coefficient on
$\E[O_{i,1}(1)-O_{i,1}(0)\mid S_i]$ in a regression of
$\E[O_{i,2}(1)-O_{i,2}(0)\mid S_i]$ on an intercept and
$\E[O_{i,1}(1)-O_{i,1}(0)\mid S_i]$, weighted by $W(S_i)$. This yields a predictive mediation regression, which assesses whether sites in which
the treatment has larger effects on the mediator also tend to be
sites in which the treatment has larger effects on the final outcome. For example, in a
job-search counseling RCT, one may ask whether sites with larger effects on
job seekers' search effort also have larger effects on their job-finding rate. $\beta_{\mathrm{med}}$ can be helpful to generate hypotheses about the mechanisms
underlying the treatment effect. However, additional assumptions are needed for $\beta_{\mathrm{med}}$ to identify the mediator's effect.

To estimate $\beta_{\mathrm{med}}$, one lets
\[
 \begin{aligned}
 Y_i
 &=O_{i,2}(1)-O_{i,2}(0),\quad
 X_i=O_{i,1}(1)-O_{i,1}(0).
 \end{aligned}
\]
The corresponding estimators of $\mu_Y(S_i)$ and $\mu_X(S_i)$ and variance--covariance corrections are
\begin{align*}
 \widehat\mu_Y(S_i)&=\overline O_{1,i,2}-\overline O_{0,i,2},
 &\widehat\mu_X(S_i)&=\overline O_{1,i,1}-\overline O_{0,i,1},\\
 \widehat V_{Y,i}&=\widehat V_{1,i,2}+\widehat V_{0,i,2},
 &\widehat V_{X,i}&=\widehat V_{1,i,1}+\widehat V_{0,i,1},\\
 \widehat C_{XY,i}&=\widehat C_{1,i,12}+\widehat C_{0,i,12}.
\end{align*}
Plugging those into \eqref{eq:beta-hat} yields an estimator of $\beta_{\mathrm{med}}$. Note that one could also estimate a predictive mediation regression with several mediators. Then, $\mu_{X}$ is a vector collecting
site-specific effects on all mediators. Proposition~\ref{prop:rct-unbiased} supplies
the corresponding matrix of variance corrections and vector of covariance
corrections.

\subsubsection{Causal mediation}\label{subsubsec:causal-mediation}

We now give conditions under which the coefficients from the predictive-mediation regression have a causal interpretation. For a generic unit in site $i$, let
$O_{i,1}(d)$ denote the potential mediator under treatment $d$, and let
$O_{i,2}(d,o_1)$ denote the potential final outcome when treatment is set to
$d$ and the mediator is set to $o_1$. Thus, the potential final outcome under
treatment $d$ defined above satisfies
$O_{i,2}(d)=O_{i,2}(d,O_{i,1}(d))$, and
$\mu_{Y}(S_i)=\E[O_{i,2}(1,O_{i,1}(1))-O_{i,2}(0,O_{i,1}(0))\mid S_i].$
\begin{assumption}[Binary mediator and monotonicity]
\label{ass:mediator-monotonicity}
$O_{i,1}$ is binary and
satisfies the monotonicity
condition
 \[
  O_{i,1}(1)\geq O_{i,1}(0)\quad\text{almost surely}.
 \]
Moreover,
\[
 \Pp\{O_{i,1}(1)>O_{i,1}(0)\mid S_i\}>0
 \quad\text{almost surely}.
\]
\end{assumption}
For site $i$, define
\begin{align*}
 \mathrm{DE}(S_i)
 &=\E[O_{i,2}(1,O_{i,1}(0))-O_{i,2}(0,O_{i,1}(0))\mid S_i],\\
 \mathrm{IDE}(S_i)
 &=\E[O_{i,2}(1,1)-O_{i,2}(1,0)
       \mid O_{i,1}(1)>O_{i,1}(0),S_i].
\end{align*}
$\mathrm{DE}(S_i)$ is the site's average natural direct effect:
it measures the effect of switching treatment from zero to one while holding
the mediator at the value it would take under control. $\mathrm{IDE}(S_i)$ is
the average effect on the final outcome of switching the mediator
from zero to one while holding treatment fixed at one, among units in site $i$ whose mediator is switched from zero to one by the
treatment. By the usual natural
direct--indirect effect decomposition
\citep[see, e.g.,][]{vanderweele2013threeway},
\begin{align}
 \mu_{Y}(S_i)
 &=\mathrm{DE}(S_i)
   +\E[O_{i,2}(1,O_{i,1}(1))-O_{i,2}(1,O_{i,1}(0))
        \mid S_i]\nonumber\\
 &=\mathrm{DE}(S_i)
   +\E\!\left[
      \mathbf 1\{O_{i,1}(1)>O_{i,1}(0)\}
      \{O_{i,2}(1,1)-O_{i,2}(1,0)\}\mid S_i
     \right]\nonumber\\
 &=\mathrm{DE}(S_i)+\mu_{X}(S_i)\mathrm{IDE}(S_i).
 \label{eq:causal-mediation-decomposition}
\end{align}
The first equality follows by adding and
subtracting $O_{i,2}(1,O_{i,1}(0))$. The second equality follows from
Assumption~\ref{ass:mediator-monotonicity}, and the third from the law of iterated expectations and the fact that monotonicity implies $\Pp\{O_{i,1}(1)>O_{i,1}(0)\mid S_i\}=\E[O_{i,1}(1)-O_{i,1}(0)\mid S_i]=\mu_X(S_i)$.

\begin{assumption}[Cross-site independence for causal mediation]
\label{ass:causal-mediation}
Under the probability distribution that weights sites by $w(S)$,
$\mu_{X}(S)$ is independent of
$(\mathrm{IDE}(S),\mathrm{DE}(S))$.
\end{assumption}

Assumption~\ref{ass:causal-mediation} requires that sites' effects on the mediator be unrelated to both their direct effects and their compliers' gains from the mediator.

\begin{proposition}[Causal interpretation of the predictive-mediation regression]
\label{prop:causal-mediation}
Suppose that Assumptions~\ref{ass:mediator-monotonicity} and
\ref{ass:causal-mediation} hold, and that
$\Var_w(\mu_{X}(S))>0$. Then the slope $\beta_{\mathrm{med}}$ and intercept $\alpha_{\mathrm{med}}$ from the
weighted population regression of $\mu_{Y}(S)$ on an intercept and
$\mu_{X}(S)$ satisfy
\[
 \beta_{\mathrm{med}}=\E_w[\mathrm{IDE}(S)],
 \qquad
 \alpha_{\mathrm{med}}=\E_w[\mathrm{DE}(S)].
\]
\end{proposition}

Under Assumptions~\ref{ass:mediator-monotonicity} and
\ref{ass:causal-mediation}, $\beta_{\mathrm{med}}$ identifies
the $w$-weighted average across sites of the mediator's effect
among each site's mediator compliers. The corresponding average
indirect effect of treatment on the final outcome is therefore
also identified: it equals the average treatment effect on the
mediator multiplied by the mediation regression slope:
\[
 \E_w[\E[O_{i,2}(1,O_{i,1}(1))-O_{i,2}(1,O_{i,1}(0))
        \mid S_i]]=\E_w\!\left[\mu_X(S_i)\mathrm{IDE}(S_i)\right]
 =\E_w[\mu_X(S_i)]\,\beta_{\mathrm{med}}.
\]

\subsubsection{Testing homogeneity of direct and mediator effects}\label{subsubsec:mediation-test}

Consider the condition
\begin{equation}
 \mathrm{IDE}(S_i)=\mathrm{IDE},
 \qquad
 \mathrm{DE}(S_i)=\mathrm{DE}
 \quad\text{almost surely}.
 \label{eq:homogeneous-mediation}
\end{equation}
This condition requires both the direct effect and the mediator's effect among mediator compliers to be constant across sites. It is sufficient for the cross-site independence assumption in Assumption~\ref{ass:causal-mediation}. We derive a testable implication of this homogeneity condition, maintaining Assumption~\ref{ass:mediator-monotonicity}. Without that maintained assumption, rejection can instead reflect a failure of either condition.
\begin{proposition}[Testable implication of homogeneous direct and mediator effects]
	\label{prop:mediation-test-implication}
	Suppose that Assumption~\ref{ass:mediator-monotonicity} and \eqref{eq:homogeneous-mediation} hold, and that $\Var_w(\mu_X(S))>0$. Then $\mathrm{IDE}=\beta_{\mathrm{med}}$,
	\begin{equation}
		\mu_{Y}(S)-\mu_{X}(S)\beta_{\mathrm{med}}=\mathrm{DE}(S)+\mu_{X}(S)\{\mathrm{IDE}(S)-\beta_{\mathrm{med}}\}=\mathrm{DE}\quad\text{almost surely}, \label{eq:testimand}
	\end{equation}
	and therefore
	\begin{equation}
		\Var_w\!\left(
		\mu_{Y}(S)-\mu_{X}(S)\beta_{\mathrm{med}}
		\right)=0. \label{eq:testable_implication}
	\end{equation}
\end{proposition}
Under \eqref{eq:homogeneous-mediation} it follows from Proposition~\ref{prop:causal-mediation} that $\mathrm{IDE}=\beta_{\mathrm{med}}$. Then the first equality in \eqref{eq:testimand} follows from \eqref{eq:causal-mediation-decomposition}, while the second follows from the fact that $\mathrm{DE}(S)=\mathrm{DE}$ and $\mathrm{IDE}(S)=\mathrm{IDE}=\beta_{\mathrm{med}}$.
\begin{remark}[Power of a test based on \eqref{eq:testable_implication}]
	\label{rem:mediation-test-power}
Under the regularity conditions below, a test based on
\eqref{eq:testable_implication} has power tending to one against
fixed violations of \eqref{eq:homogeneous-mediation} that generate
a strictly positive residual variance. It cannot consistently
detect violations that preserve an exact linear relationship
between $\mu_Y(S)$ and $\mu_X(S)$.
Let us characterize such alternatives under the following models for $\mathrm{DE}(S)$ and $\mathrm{IDE}(S)$:
	\[
	\mathrm{DE}(S)=\gamma_0+\gamma_1\mu_{X}(S),
	\qquad
	\mathrm{IDE}(S)=\theta_0+\theta_1\mu_{X}(S).
	\]
	Suppose also that $\mu_{X}(S)$ has at least
	three support points. Then, \eqref{eq:testable_implication} holds if and only
	if $\theta_1=0$.\footnote{Indeed, if $\theta_1=0$ then
		$ \mu_{Y}(S)
		=\gamma_0+(\gamma_1+\theta_0)\mu_{X}(S),$
		so
		$\beta_{\mathrm{med}}=\gamma_1+\theta_0$ and \eqref{eq:testable_implication} holds. Conversely, if $\theta_1\neq0$,
		$\mu_{Y}(S)$ contains a nonzero quadratic term in
		$\mu_{X}(S)$ and therefore cannot be affine in
		$\mu_{X}(S)$ on three or more support points.} Therefore,
	\eqref{eq:homogeneous-mediation} fails while
	\eqref{eq:testable_implication} holds if and only if $\theta_1=0$ and
	$\gamma_1\neq0$. However, one may expect heterogeneity associated with the share of mediator compliers to arise more readily in the mediator's effect among compliers than in the treatment's direct effect. An additional assumption motivated by this view is
\[
 \theta_1=0\quad\Longrightarrow\quad\gamma_1=0.
\]
This restriction rules out variation in the direct effect with the share of mediator compliers when the mediator effect does not vary with that share. Under the linear model above, the requirement of at least three support points for $\mu_X(S)$, and this additional restriction, \eqref{eq:testable_implication} holds if and only if \eqref{eq:homogeneous-mediation} holds. Thus, within this restricted model, homogeneity is fully testable.
\end{remark}

Turning to the test's implementation, let
\begin{align*}
	\widehat\mu_{\varepsilon}(S_i)
	&=\widehat\mu_{Y}(S_i)
	-\widehat\beta_{\mathrm{med}}\,\widehat\mu_{X}(S_i),\\
	\widehat V_{\varepsilon,i}
	&=\widehat V_{Y,i}
	+\widehat\beta_{\mathrm{med}}^2\widehat V_{X,i}
	-2\widehat\beta_{\mathrm{med}}\,
	\widehat C_{XY,i}.
\end{align*}
To test \eqref{eq:homogeneous-mediation}, we use
\[
\widehat\sigma_{\varepsilon}^2
:=\frac1n\sum_{i=1}^n\widehat w(S_i)
\left[
\{\widehat\mu_{\varepsilon}(S_i)-\widehat\tau_{\varepsilon}\}^2
-\widehat V_{\varepsilon,i}
\right],
\qquad \text{ where }
\widehat\tau_{\varepsilon}:
=\frac1n\sum_{i=1}^n\widehat w(S_i)\widehat\mu_{\varepsilon}(S_i).
\]

Appendix~\ref{app:mediation-inference} derives the influence function and asymptotic variance of $\widehat\sigma_{\varepsilon}^2$ and describes how to use it to test homogeneity of direct and mediator effects. Under the maintained assumptions, rejection provides evidence against homogeneous direct and mediator effects, but does not necessarily reject Assumption~\ref{ass:causal-mediation}. Conversely, nonrejection does not establish homogeneity or validate the causal interpretation of the regression coefficients.

\subsubsection{Comparison to other approaches for mediation analysis in RCTs}\label{subsubsec:mediation-comparison}

Identifying the effect of a mediator on an outcome is notoriously
difficult. Standard approaches compare outcomes across units with the same
treatment assignment and covariates but different mediator values. Giving
those comparisons a causal interpretation requires sequential ignorability:
conditional on treatment and pretreatment covariates, the mediator must be as
good as randomly assigned \citep{imai2010identification}. This condition is often
hard to justify because the mediator is not randomized, even when the treatment
is.

Our approach instead regresses site-level treatment effects on the outcome
on site-level treatment effects on the mediator. Such regressions have a long history in meta-analyses \citep[see, e.g.,][]{daniels1997meta}. Prior work has shown that those regressions have a causal interpretation under the homogeneity condition in \eqref{eq:homogeneous-mediation} \citep[see, e.g.,][]{dunn2015evaluation}. Our Proposition \ref{prop:causal-mediation} slightly extends those results by showing that this condition can be relaxed, though our Assumption~\ref{ass:causal-mediation} remains a strong assumption. To our knowledge, the procedure in Section~\ref{subsubsec:mediation-test} provides the first sampling-error-corrected test of a necessary implication of \eqref{eq:homogeneous-mediation}. This provides a way to assess a sufficient condition for the causal interpretation of those regressions.

Another approach treats the randomized treatment as an instrument for the mediator. In this context, the IV exclusion restriction requires that the treatment affects the outcome
only through the mediator, referred to as the full mediation assumption. If i) the full mediation assumption holds, ii) the treatment is randomly assigned and iii) the treatment has a monotonic effect on the mediator, a 2SLS
regression of the outcome on the mediator using the treatment as the instrument  identifies the effect of the mediator for mediator compliers \citep{reardon2013assumptions}. Like Assumption \ref{ass:mediator-monotonicity} and \eqref{eq:homogeneous-mediation}, i), ii) and iii) are not fully testable but have a testable implication \citep{kwon2026testing}. The full-mediation
assumption and Assumption~\ref{ass:causal-mediation} are not nested. The former
does not restrict the mediator's effect on the outcome, but rules out
any direct effect of treatment. Assumption~\ref{ass:causal-mediation}, by
contrast, permits direct effects, but
requires that direct effects and mediator effects be independent of the treatment effect on
the mediator.


\subsection{Sites' responsiveness or providers' effectiveness?}\label{subsec:providers-sites}

In multi-site RCTs, treatment providers often vary across sites. Then, a natural question is whether heterogeneous effects across sites are driven by heterogeneity in providers' effectiveness, or by sites' heterogeneous responsiveness. In a multi-arm RCT, if in each site the treatments are delivered by different providers, as is the case in our empirical application, the variance and regression estimators proposed in this paper can help shed some light on this question.

In this subsection, we assume that
$\mathcal D=2$ and $L=1$: we consider an RCT with two treatment arms and one outcome. For $d\in\{1,2\}$, let $\Delta_d=O_{1}(d)-O_{1}(0)$. Then, let
\[
 \sigma^2_{\Delta}=\Var_w\!\left(\mu_{\Delta_2}(S_i)-\mu_{\Delta_1}(S_i)\right)
\]
denote the weighted variance across sites of the difference between the two treatments' ATEs. Let also $\beta_{21}$ denote the coefficient on
$\mu_{\Delta_1}(S_i)$ in a weighted regression of
$\mu_{\Delta_2}(S_i)$ on an intercept and
$\mu_{\Delta_1}(S_i)$. A small $\sigma^2_{\Delta}$ means that the difference between the two treatments' effects varies little across sites, while a positive $\beta_{21}$ means that the treatments tend to be more effective in the same sites. Both findings may indicate that treatment-effect heterogeneity is not entirely driven by heterogeneity in providers' effectiveness.

To formalize this idea, assume that for all $d\in \{1,2\}$,
\begin{equation}\label{eq:model_providers}
\mu_{\Delta_d}(S_i)=f_d(P_d(S_i),U_d(S_i)),
\end{equation}
where $P_d(S_i)$ is the effectiveness of the provider of treatment $d$ in site $S_i$, $U_d(S_i)$ is a vector of characteristics of site $S_i$, and $f_d(.)$ is a
possibly non-separable production function mapping the provider's effectiveness and the site's characteristics into the site's ATE. Let $R_d(S_i)=f_d(\E[P_d(S_i)],U_d(S_i))$, where $\E[P_d(S_i)]$ is providers' average effectiveness: $R_d(S_i)$, the responsiveness of $S_i$ to treatment $d$, is defined as $S_i$'s effect when $d$ is implemented with average effectiveness. Let $\overline{R}(S_i)=(R_1(S_i)+R_2(S_i))/2$ denote the average responsiveness of $S_i$ to the two treatments, let $\epsilon_d^S(S_i)=R_d(S_i)-\overline{R}(S_i)$ denote the difference between its responsiveness to $d$ and its average responsiveness, so that $\epsilon_1^S(S_i)=-\epsilon_2^S(S_i)=(R_1(S_i)-R_2(S_i))/2$. Finally, let $\epsilon_d^P(S_i,P_d(S_i))=f_d(P_d(S_i),U_d(S_i))-R_d(S_i)$ denote the difference between its treatment-$d$ ATE and what its treatment-$d$ ATE would be with an average-effectiveness provider.
It follows from these definitions that
\begin{equation}\label{eq:model_providers_decomposition}
\mu_{\Delta_d}(S_i)=\overline{R}(S_i)+\epsilon_d^S(S_i)+\epsilon_d^P(S_i,P_d(S_i)).
\end{equation}
\begin{propositionnopunct}[Does providers' effectiveness contribute to the variance of effects across sites?]\label{prop:providereffect}
\leavevmode\par
\begin{enumerate}
\item Assume that \eqref{eq:model_providers} holds.
\begin{enumerate}
\item Then $\sigma^2_{\Delta}=0$ if and only if
\[
\Var_w\!\left(\epsilon_2^S(S_i)-\epsilon_1^S(S_i)
+\epsilon_2^P(S_i,P_2(S_i))-\epsilon_1^P(S_i,P_1(S_i))\right)=0.
\]
\item If one further assumes that the $w$-weighted correlation between $\epsilon_2^S(S_i)-\epsilon_1^S(S_i)$ and
$\epsilon_2^P(S_i,P_2(S_i))-\epsilon_1^P(S_i,P_1(S_i))$ is strictly larger than $-1$,\footnote{With the convention that $0/0=0$, a convention we use here and in a few other places, when doing so avoids proving results separately in knife-edge cases where a ratio's numerator and denominator are both equal to zero.} and that the weighted correlation between $\epsilon_2^P(S_i,P_2(S_i))$ and $\epsilon_1^P(S_i,P_1(S_i))$ is strictly lower than $1$, then $\sigma^2_{\Delta}=0$ if and only if
\[
\Var_w(\epsilon_2^S(S_i)-\epsilon_1^S(S_i))=0,
\quad \Var_w(\epsilon_2^P(S_i,P_2(S_i)))=0,
\quad \Var_w(\epsilon_1^P(S_i,P_1(S_i)))=0.
\]
\end{enumerate}
\item If $\Var_w(\mu_{\Delta_1}(S_i))>0$ and
$\Var_w(\mu_{\Delta_2}(S_i))=\Var_w(\mu_{\Delta_1}(S_i))$, then $\sigma^2_{\Delta}=0$ if and only if $\beta_{21}=1$.
\end{enumerate}
\end{propositionnopunct}
Point 1.a) of Proposition~\ref{prop:providereffect} follows directly from \eqref{eq:model_providers_decomposition}: differencing the two treatments' effects cancels $\overline R(S_i)$, so $\sigma^2_{\Delta}$ only reflects sites' relative responsiveness to the two treatments and providers' contributions.
Point 1.b) shows that, except for knife-edge cases, $\sigma^2_{\Delta}=0$ rules out providers' effectiveness as a source of treatment-effect heterogeneity across sites. The first knife-edge case arises when sites' differences in responsiveness to the two treatments and differences in providers' contributions are perfectly negatively correlated, so that the two sources of heterogeneity exactly offset each other, a fairly implausible scenario. The second arises when providers' contributions to the effects of the two treatments are perfectly positively correlated, also a fairly implausible scenario if in each site the treatments are delivered by different providers. Outside these cases, $\sigma^2_{\Delta}=0$ implies that providers do not contribute to effect heterogeneity across sites, and that sites' relative responsiveness to the two treatments is constant. Hence, all treatment-effect heterogeneity comes from $\overline R(S_i)$, sites' average responsiveness to the two treatments. Point 2 shows that under the testable equal-variance restriction $\Var_w(\mu_{\Delta_2}(S_i))=\Var_w(\mu_{\Delta_1}(S_i))$, $\sigma^2_{\Delta}=0\Leftrightarrow\beta_{21}=1$.

To estimate $\sigma^2_{\Delta}$ from Proposition~\ref{prop:rct-unbiased} and Theorem~\ref{thm:sigma}, let
\[
 Y_i=O_{i,1}(2)-O_{i,1}(1).
\]
The corresponding estimator of $\mu_Y(S_i)$ and variance correction are
\begin{align*}
 \widehat\mu_Y(S_i)&=\overline O_{2,i,1}-\overline O_{1,i,1},
 &\widehat V_{Y,i}&=\widehat V_{2,i}+\widehat V_{1,i}.
\end{align*}
Plugging these quantities into \eqref{eq:sigma-hat} yields an estimator of $\sigma^2_{\Delta}$. The control-group mean cancels from the difference between the two estimated ATEs, so neither the estimator nor its variance correction uses observations from the control group.

To estimate $\beta_{21}$ from Proposition~\ref{prop:rct-unbiased} and Theorem \ref{thm:beta}, let
\[
 Y_i=O_{i,1}(2)-O_{i,1}(0),
 \qquad
 X_i=O_{i,1}(1)-O_{i,1}(0).
\]
The corresponding estimators of $\mu_Y(S_i)$ and $\mu_X(S_i)$ and
variance--covariance corrections are
\begin{align*}
 \widehat\mu_Y(S_i)&=\overline O_{2,i,1}-\overline O_{0,i,1},
 &\widehat\mu_X(S_i)&=\overline O_{1,i,1}-\overline O_{0,i,1},\\
 \widehat V_{Y,i}&=\widehat V_{2,i}+\widehat V_{0,i},
 &\widehat V_{X,i}&=\widehat V_{1,i}+\widehat V_{0,i},\\
 \widehat C_{XY,i}&=\widehat V_{0,i}.
\end{align*}
Plugging those into \eqref{eq:beta-hat} yields an estimator of $\beta_{21}$.

Results in this section relate to work that seeks to separate implementers' or
organizations' effectiveness from differences in the populations and contexts
in which they operate. \citet{bloom2003linking} pool multi-site welfare-to-work
experiments and relate local impacts to measured features of program
implementation. \citet{hong2024organizational} instead define an organization's
relative effectiveness by comparing its performance with that of organizations
facing similar ecological conditions, using control outcomes and observed
participant composition to adjust for those conditions. Relatedly,
\citet{angrist2023implementation} explain cross-study heterogeneity in an
educational intervention using take-up and whether teachers or volunteers
delivered the program. Our approach is complementary. It does not require the researcher to observe
and model all relevant implementation practices, or to assume that observed
variables suffice to adjust for differences in local conditions. Instead, it
uses a design in which two related treatments are delivered by different
providers within the same sites. Comparing their site-specific effects holds
the site fixed, and their covariance reveals whether the two treatments tend to
succeed in the same environments. Under the decomposition in
\eqref{eq:model_providers_decomposition}, the variance of the difference between
the treatments' effects can further indicate whether provider effectiveness
contributes to cross-site heterogeneity. This advantage comes with a narrower
scope: the comparison requires multiple treatments delivered in the same sites,
and does not identify which provider practices account for
differences in effectiveness.

\subsection{Extensions and avenues for future research}\label{subsec:extensions}

\subsubsection{Estimating effect heterogeneity in other RCTs}\label{subsubsec:other-rcts}

Replacing sites by strata, the estimators above can also be used to estimate
and predict treatment-effect heterogeneity across strata in any stratified RCT, irrespective of whether the strata correspond to geographic sites,
provided every treatment arm contains at least two units in each stratum (resp. three units if one wants to estimate centered third moments). The estimators above are therefore not
applicable to paired RCTs. If one is interested in estimating effect heterogeneity across strata, this may be an argument to use larger
strata, such as strata of four in a two-arm experiment.

We conjecture that our results are also applicable to nonstratified RCTs. Suppose that instead of separate randomizations within sites, a single completely randomized experiment is conducted, with fixed overall treatment-arm sizes. Let $C=((N_{d,i})_{d=0}^{\mathcal D})_{i=1}^n$ denote the realized site-by-arm count table. Conditional on $C$, the assignment vectors are independent across sites and, within each site $i$, assignment is uniform over all allocations having arm sizes $(N_{d,i})_{d=0}^{\mathcal D}$ \citep{miratrix2013adjusting}.\footnote{Their formal result considers binary treatment; the extension to multiple arms follows from the same multinomial counting argument.} Thus, conditional on $C$, the nonstratified experiment is equivalent to independent stratified experiments conducted within sites. This suggests modifying Assumption~\ref{hyp:iid} by excluding the arm sizes from the vectors drawn independently across sites, because the $N_{d,i}$ are now generated by the single randomization and are dependent across sites, and replacing Assumption~\ref{hyp:strat_completely_randomized} by its conditional-on-$C$ counterpart. For admissible sites, namely sites such that $N_{d,i}\geq2$ for every $d$, the site-specific estimators remain unchanged after inserting the realized arm sizes, and their conditional unbiasedness and variance-correction properties continue to hold. Assuming \(N_i\ge2(\mathcal D+1)\) almost surely, one possible
approach to retaining the original population target when aggregating
across admissible sites is to replace \(W(S_i)\) by
\[
\frac{A_i W(S_i)}{p_i},
\qquad
A_i:=\mathbf{1}\{N_{d,i}\geq2\ \text{for every }d\},
\]
where \(p_i\) is the probability that \(A_i=1\), conditional on
the realized site sizes and the overall treatment-arm totals.
This approach requires positive inclusion probabilities and suitable
moment conditions on the inverse-probability-weighted contributions.
Establishing consistency and asymptotic normality would require
accounting for both within-site estimation error and the randomness
of the count table \(C\). In particular, a conditional central limit
theorem would need to be complemented by an analysis of fluctuations
in the estimator's conditional mean around the population target.
Developing these results
is an interesting extension for future research.

\subsubsection{Application to other research designs}\label{subsubsec:other-designs}

Moment-correction arguments may also apply to other
research designs that provide unbiased site-specific effect estimators
and unbiased estimators of their variances. For instance, consider a two-period panel where no unit is treated at period one, and each site contains units that become treated at period two and units that remain untreated. If treated and
control units are independent i.i.d.\ samples within each site, under a no anticipation assumption and a within-site parallel trends assumption,
the difference between their average outcome changes is unbiased
for the site's average treatment effect on the treated.
With at least two observations in each arm, the estimator's variance can be unbiasedly estimated.
This provides a route to applying our methods to a canonical difference-in-differences design.
Teachers' value added offers another potential application, which would require specifying conditions
under which teacher-specific estimators and their variance and
covariance can be unbiasedly estimated.

\section{Application: the effects of publicly- and privately-provided counseling for job seekers}\label{sec:application}

\subsection{Study design and data}\label{subsec:application-design}

\citet{behaghel2014private} conduct an RCT in 216 local Public Employment Service (PES) offices in France to compare the public and private provision of counseling to job seekers. During their first interview at the local PES office, 43,977 job seekers are randomly assigned to one of three groups. The control group receives the standard services provided by the PES. The other two groups are assigned to intensive counseling programs delivered either by the PES or by a private provider. In this application, PES offices are the sites and job seekers are the randomization units. When analyzing the public program, we exclude 16 PES offices where less than two units were assigned to that arm or to the control group. For the private program, the corresponding sample restrictions lead us to exclude 12 PES offices.

Compliance with assignment is imperfect. Almost no job seeker assigned to the control group receives intensive counseling, but only 32\% of those assigned to public counseling and 43\% of those assigned to private counseling enroll. Our main outcome is an indicator for holding a job six months after randomization, one of the three main employment outcomes considered by \citet{behaghel2014private}. Results are similar for their other employment outcomes. Estimation is weighted using the weights in the original study.

The study has both limitations and strengths for our purposes. The data do not contain potential mediators such as measures of job-search effort, precluding the mediation analyses discussed above. Moreover, because randomization takes place within local labor markets, the programs may generate displacement effects. The estimated effects are therefore partial-equilibrium effects. On the other hand, this study has substantial imperfect compliance, thus allowing us to illustrate our results on LATEs' heterogeneity. Moreover, in every site two similar programs are delivered by different providers, thus allowing us to assess if effect heterogeneity is driven by providers' effectiveness or by sites' responsiveness.

\subsection{Estimating the variance of ITTs across sites}\label{subsec:application-itt-variance}

Table~\ref{tab:application-ate-variance} shows that the estimated average effects on employment after six months are 2.4 percentage points (pp) for public counseling and 1.9 pp for private counseling.
Those averages conceal substantial heterogeneity. The estimated variances are positive and statistically significant for both programs. The estimated standard deviations of site-specific effects are 9.2 pp for public counseling and 8.5 pp for private counseling, about four times their respective average effects.

The estimated centered third moments are positive but not statistically significant. Thus, there is no evidence that the distribution of site-specific effects is asymmetric. Dividing each estimator by the estimated variance on its estimation sample raised to the power $3/2$ yields implied skewnesses of 0.87 for public counseling and 0.13 for private counseling. Because the outcome is binary, site-specific ITTs are bounded between $-1$ and $1$, ruling out the unbounded tails of the heavy-tail, high-skewness simulation design in which confidence intervals based on analytic standard errors undercovered (Appendix~\ref{app:simulations}).

\begin{table}[htbp]
\centering
\caption{Variance and centered third moment across sites of counseling effects on employment}
\label{tab:application-ate-variance}
\begin{tabular}{lccccc}
\hline
& $\widehat{\mathrm{ITT}}$
& $\widehat\sigma^2_{\mathrm{ITT}}$
& $\sqrt{\widehat\sigma^2_{\mathrm{ITT}}}/\widehat{\mathrm{ITT}}$
& $\widehat\kappa_{\mathrm{ITT}}$
& Units\\
& (1)&(2)&(3)&(4)&(5)\\
\hline
Public counseling  & 0.024 & 0.0084 & 3.809 & 0.00063 & 7,198\\
                   & (0.011)&(0.0037)&&(0.00137)&\\
Private counseling & 0.019 & 0.0073 & 4.478 & 0.00008 & 34,768\\
                   & (0.008)&(0.0022)&&(0.00069)&\\
\hline
\end{tabular}
\begin{minipage}{0.95\linewidth}
\footnotesize
\textit{Notes:} The outcome is an indicator for holding a job six months after randomization. Column (1) reports the weighted average ITT effect. Column (2) reports the corrected variance estimator in \eqref{eq:sigma-hat}. Standard errors, based on the influence functions in Appendix~\ref{app:inference}, appear in parentheses. Column (3) divides the estimated standard deviation of site-specific effects by the corresponding average effect. Column (4) reports the corrected centered third-moment estimator in \eqref{eq:kappa-hat}, using sites with at least three treatment and three control observations: 194 sites and 7,154 units for public counseling, and 201 sites and 34,709 units for private counseling. Columns (1)--(3) and (5) use sites with at least two treatment and two control observations. Results use data from \citet{behaghel2014private}.
\end{minipage}
\end{table}

We compare the variance-based test of cross-site ITT heterogeneity with three alternative tests: the best linear predictor (BLP) test of \citet{cddf2025}, using the full set of site indicators as predictors and where the training is done by least squares (see Appendix~\ref{app:blp} for a formal definition); the BLP test implemented by default by \texttt{GenericML}, using several machine-learning learners; and the causal-forest calibration test of \cite{athey2019generalized}, implemented in \texttt{grf}. The variance-based test rejects homogeneous effects for both programs (one-sided $p$-values $0.011$ and $0.0004$). The OLS-based BLP test is less conclusive: over 1,000 random splits, its median one-sided $p$-values are $0.21$ and $0.058$. The BLP test computed by \texttt{GenericML} does not reject at the 5\% level for either program using its best learner (median one-sided $p$-values equal to $0.227$ and $0.062$). With default tuning, \texttt{grf} does not split on site indicators, making its calibration test uninformative; after relaxing the splitting restriction, it rejects homogeneity for both programs, albeit more marginally than the variance-based test (one-sided $p$-values $0.026$ and $0.049$). Note that the \texttt{grf} p-values are not clustered at the site level, and therefore do not account for the correlation of the treatments of units belonging to the same site. With this command, clustering at the site level would imply that the sample split is also clustered at that level. This would be problematic in our setting, as site indicators observed in the validation sample would not be represented in the training sample. Accordingly, these p-values should be interpreted cautiously.
\begin{table}[htbp]
\centering
\caption{One-sided p-values for tests of treatment-effect homogeneity across sites}
\label{tab:heterogeneity_tests_comparison}
\begin{tabular}{lcc}
\hline
& Public counseling & Private counseling \\
\hline
Corrected variance estimator & 0.011 & 0.0004 \\
BLP, using OLS for training & 0.210 & 0.058 \\
BLP, default implementation in GenericML & 0.227 & 0.062 \\
Generalized random forests (grf) & 0.026 & 0.049 \\
\hline
\end{tabular}

\begin{minipage}{0.95\textwidth}
\footnotesize
\emph{Notes:} The table reports one-sided p-values for tests of the null
hypothesis that treatment effects are constant across sites.
The first row reports p-values from tests based on the corrected
estimator of the variance of site-level ITT effects.
The second row implements the BLP test of \cite{cddf2025} using the full set of site indicators as predictors and where the training is done by least squares, as described in
Appendix \ref{app:blp}. Reported p-values are medians
across 1,000 random splits.
The third row uses the default implementation of the BLP test in the GenericML package, with site
indicators as covariates and the best learner selected by the package. Reported p-values are medians
across 1,000 random splits. The fourth row uses the calibration test from the
\texttt{grf} package \citep{athey2019generalized},
with 2,000 trees, site indicators as covariates,
and the differential-prediction coefficient.
We set \texttt{alpha = 0}, \texttt{min.node.size = 1},
and \texttt{imbalance.penalty = 0}, allowing the forest
to split on individual site indicators.
\end{minipage}
\end{table}

\subsection{Regressing sites' ITTs on their average untreated outcome}\label{subsec:application-itt-untreated}


Column (1) of Table~\ref{tab:application-predictors} reports measurement-error-corrected regressions of site-specific effects on sites' average untreated outcome. The coefficients are negative for both programs and especially large and precise for private counseling. For that program, the untreated outcome explains almost one half of the variance in site-specific effects. Intensive counseling thus tends to be more effective in sites where job seekers would otherwise have lower employment rates.
Column (2) reports regressions of site-specific effects on the local unemployment rate, which we are able to recover for all but one site. We find no statistically significant relationship between local unemployment and either program's effect. This does not contradict the results in Column (1). The local unemployment rate is an imperfect proxy for the labor-market conditions facing the RCT's population, which consists of job seekers at high risk of long-term unemployment. Regressing the control-group employment rate on local unemployment produces significantly negative coefficients, but measurement-error corrected $R^2$s are equal to only $0.134$ and $0.100$ for the public- and private-program samples, respectively. Finally, the naive regressions in Table~\ref{tab:application-predictors} illustrate the importance of correcting for estimation error when the predictor is estimated. In Column (1), the naive coefficients differ from the corrected coefficients, and their conventional standard errors are much smaller. On the other hand, when the predictor is measured without error, as is the case in Column (2), both approaches yield the same coefficient and very similar standard errors.

\begin{table}[htbp]
\centering
\caption{Predictors of site-specific counseling effects}
\label{tab:application-predictors}
\begin{tabular}{lcc}
\hline
& Untreated outcome & Unemployment rate\\
& (1)&(2)\\
\hline
\multicolumn{3}{l}{\textit{Panel A: Public counseling}}\\
$\widehat\beta$ & $-0.480$ & $-0.034$\\
                 & $(0.307)$ & $(0.265)$\\
$\widehat R^2$   & $0.124$ & $0.0002$\\
Naive coefficient& $-0.850$ & $-0.034$\\
                 & $(0.081)$ & $(0.266)$\\
Sites            & 200 & 199\\
\hline
\multicolumn{3}{l}{\textit{Panel B: Private counseling}}\\
$\widehat\beta$ & $-0.804$ & $0.095$\\
                 & $(0.126)$ & $(0.242)$\\
$\widehat R^2$   & $0.496$ & $0.002$\\
Naive coefficient& $-0.939$ & $0.095$\\
                 & $(0.036)$ & $(0.242)$\\
Sites            & 204 & 203\\
\hline
\end{tabular}
\begin{minipage}{0.95\linewidth}
\footnotesize
\textit{Notes:} Each column reports a separate weighted regression of site-specific ITT effects on the indicated site characteristic. The untreated outcome is the site's control-group employment rate after six months. The corrected coefficient and $R^2$ use \eqref{eq:beta-hat} and \eqref{eq:R2-hat}. Standard errors, based on the influence functions in Appendix~\ref{app:inference}, appear in parentheses. Naive regressions do not correct for measurement error. The corrected and naive coefficients coincide in Column (2) because the unemployment rate is measured without error. Results use data from \citet{behaghel2014private}.
\end{minipage}
\end{table}

The association between sites' ITT effects and untreated outcomes may just be due to heterogeneous job seeker characteristic across sites, which correlate with both the untreated outcome and the program's effect. To investigate this possibility, we follow \citet{behaghel2014private} and estimate, in the control group, a job-seeker-level logistic regression of employment on 43 characteristics measuring education, previous employment and unemployment histories, and reservation wages. We average the fitted values within each site to obtain its predicted employment rate without treatment. Finally we decompose the site's actual employment rate without treatment into this prediction and a residual. The predicted employment rate without treatment does not predict treatment effects. By contrast, the residual is strongly negatively related to effects, as shown in Table~\ref{tab:application-composition}. This suggests that the relationship between untreated outcomes and treatment effects is not driven by heterogeneous job seeker characteristic across sites.

\begin{table}[htbp]
\centering
\caption{Job-seeker composition and site-specific counseling effects}
\label{tab:application-composition}
\begin{tabular}{lcc}
\hline
& Predicted untreated outcome & Residual untreated outcome\\
& (1)&(2)\\
\hline
\multicolumn{3}{l}{\textit{Panel A: Public counseling}}\\
$\widehat\beta$ & $0.032$ & $-0.679$\\
                 & $(0.304)$ & $(0.356)$\\
$\widehat R^2$   & \multicolumn{2}{c}{$0.198$}\\
Sites            & 200 & 200\\
\hline
\multicolumn{3}{l}{\textit{Panel B: Private counseling}}\\
$\widehat\beta$ & $0.029$ & $-0.978$\\
                 & $(0.125)$ & $(0.113)$\\
$\widehat R^2$   & \multicolumn{2}{c}{$0.681$}\\
Sites            & 204 & 204\\
\hline
\end{tabular}
\begin{minipage}{0.95\linewidth}
\footnotesize
\textit{Notes:} Each panel reports one multivariate regression of site-specific ITT effects on the site's predicted untreated employment rate and the residual component of its control-group employment rate. Predictions are obtained from a job-seeker-level logistic regression using 43 baseline characteristics. Coefficients correct for measurement error in the site-level variables. Standard errors, based on the influence functions in Appendix~\ref{app:inference}, appear in parentheses. Results use data from \citet{behaghel2014private}.
\end{minipage}
\end{table}

The negative association between sites' ITT effects and untreated outcomes may be useful for targeting. Although a site's untreated outcome is unavailable before treatment, the employment rate of the site's prior cohort of eligible job seekers could serve as a proxy.

\subsection{Estimating and predicting LATEs' heterogeneity}\label{subsec:application-late}

We first document substantial heterogeneity in first-stage effects across sites. Table~\ref{tab:application-fs-variance} shows that assignment to public and private counseling increases enrollment in the corresponding program by 31.2 and 40.2 pp on average, respectively. The estimated standard deviations of first-stage effects across sites are approximately 10 and 11.8 pp, respectively, amounting to 32\% and 29\% of the corresponding average first stages. Then, we assess whether sites with more compliers also have compliers with systematically different gains. As shown in \eqref{eq:late-fs-identification}, the sign of the covariance between sites' first stage and LATE effects is the sign of
$\beta_{\mathrm{ITT},\mathrm{FS}}-\mathrm{LATE}$. Column (4) reports estimates of this difference. They are negative and insignificant for both programs, so we cannot reject the null that site-specific LATEs and first stages are uncorrelated.

\begin{table}[htbp]
\centering
\caption{First-stage effects: variance across sites, and association with LATEs}
\label{tab:application-fs-variance}
\begin{tabular}{lccccc}
\hline
& $\widehat{\mathrm{FS}}$
& $\widehat\sigma^2_{\mathrm{FS}}$
& $\sqrt{\widehat\sigma^2_{\mathrm{FS}}}/\widehat{\mathrm{FS}}$
& $\widehat\beta_{\mathrm{ITT},\mathrm{FS}}
   -\widehat{\mathrm{LATE}}$
& Units\\
& (1)&(2)&(3)&(4)&(5)\\
\hline
Public counseling  & 0.312 & 0.010 & 0.320 & -0.086 & 7,198\\
                   & (0.009)&(0.003)& & (0.224)&\\
Private counseling & 0.402 & 0.014 & 0.293 & -0.081 & 34,768\\
                   & (0.004)&(0.002)& & (0.095)&\\
\hline
\end{tabular}
\begin{minipage}{0.95\linewidth}
\footnotesize
\textit{Notes:} The outcome used to estimate first-stage effects
is enrollment in the assigned intensive counseling program.
Column (1) reports the weighted average first stage.
Column (2) reports its corrected variance across sites.
Column (3) reports the estimated standard deviation relative
to the average first stage.
Column (4) reports the statistic whose sign identifies the
sign of the complier-weighted covariance between site-specific
LATEs and first stages, as shown in
\eqref{eq:late-fs-identification}.
Standard errors, based on the influence functions in
Appendix~\ref{app:inference}, appear in parentheses
in columns (1), (2), and (4).
Results use data from \citet{behaghel2014private}.
\end{minipage}
\end{table}

Table~\ref{tab:application-late-variance} shows that the estimates of the average LATEs are equal to 7.7 pp for public counseling and 4.8 pp for private counseling. The lower bounds for the complier-weighted variances of LATEs are both positive and statistically significant. They imply estimated standard deviations of site-specific LATEs of at least 16.4 and 13.4 pp for public and private counseling, respectively, about 2.1 and 2.8 times the respective overall LATE. Under squared-first-stage weights, which give more importance to sites with stronger first stages, the estimated average LATEs are slightly lower, at 6.9 and 4.1 pp, respectively. Still, the similarity of LATEs under first-stage and squared-first-stage weights indicates that changing the weights does not materially shift the center of the LATE distribution in this application. With these weights, the estimated LATE variances are again positive and statistically significant, and the estimated standard deviations of site-specific LATEs are approximately 28.1 pp for public counseling and 20.5 pp for private counseling, about four and five times the respective weighted average effects.

\begin{table}[htbp]
 \centering
 \caption{Variance of LATEs across sites}
 \label{tab:application-late-variance}
 \begin{tabular}{lccccc}
 \hline
 & $\widehat{\mathrm{LATE}}$
 & $\widehat{\underline{\sigma}}^2_{\mathrm{LATE}}$
 & $\widehat{\mathrm{LATE}}_{\mathrm{FS}^2}$
 & $\widehat\sigma^2_{\mathrm{LATE},\mathrm{FS}^2}$
 & Units\\
 & (1)&(2)&(3)&(4)&(5)\\
 \hline
 Public counseling  & 0.077 & 0.027 & 0.069 & 0.079 & 7,198\\
                    & (0.044)&(0.012)&(0.048)&(0.034)&\\
 Private counseling & 0.048 & 0.018 & 0.041 & 0.042 & 34,768\\
                    & (0.025)&(0.005)&(0.024)&(0.013)&\\
 \hline
 \end{tabular}
 \begin{minipage}{0.95\linewidth}
 \footnotesize
 \textit{Notes:} Column (1) reports the overall LATE under the complier weights in \eqref{eq:overall-late}. Column (2) reports the estimated lower bound for the complier-weighted LATE variance in \eqref{eq:late-variance-lower-bound-hat}. Column (3) reports $\widehat{\mathrm{LATE}}_{\mathrm{FS}^2}=\widehat r/\widehat m_2$ in \eqref{eq:q-late-hat}. Column (4) reports the squared-first-stage-weighted variance estimator in \eqref{eq:late-variance-hat}. Standard errors appear in parentheses, using \eqref{eq:AL-lambda-late-proof} for Column (1), \eqref{eq:Vhat-late-variance-lower-bound} for Column (2), \eqref{eq:psi-late-fs2} for Column (3), and \eqref{eq:Vhat-late-variance} for Column (4). Results use data from \citet{behaghel2014private}.
 \end{minipage}
\end{table}

\subsection{Sites' responsiveness or providers' effectiveness?}\label{subsec:application-providers-sites}

%
%

If heterogeneity were driven entirely by differences in provider effectiveness, the effects of the two programs need not be related because public and private counseling are delivered by different providers within a site. Conversely, a positive association would indicate that the programs tend to work well in the same places and would therefore point toward a role for sites' responsiveness.

Following Proposition~\ref{prop:providereffect}, we first estimate $\sigma^2_{\Delta}$, the variance across sites of the difference between the private and public programs' effects, using the 200 sites in which all three arms have at least two units. The estimate is $0.0072$, with a standard error of $0.0025$, so we reject the null that $\sigma^2_{\Delta}=0$. Under the conditions of Point~1.b) of Proposition~\ref{prop:providereffect}, this implies that providers' effectiveness, and/or sites' relative responsiveness to the two programs, contribute to effect heterogeneity. Were the two programs' effects uncorrelated across sites, $\sigma^2_{\Delta}$ would be approximately equal to the sum of their variances in Table~\ref{tab:application-ate-variance}, about $0.016$. The estimate is less than half that, suggesting there is a positive covariance between the two programs' site-specific effects. Indeed, regressing the public program's site-specific effect on the private program's effect, correcting for estimation error in both variables, yields a coefficient of $0.563$ (standard error $0.215$) and an $R^2$ of $0.277$.

Overall, the difference between the programs' effects varies significantly across sites. This variation can reflect differences in provider effectiveness, differences in sites’ relative responsiveness to the two programs, or both. The positive association between the programs’ effects is consistent with a role for shared site responsiveness, although correlated provider effectiveness could also generate this association. These results do not separately identify the contributions of providers and sites.

Finally, following Section~\ref{subsubsec:late-prediction} we relate the two programs' site-specific LATEs. Letting the private program be arm 1 and the public program be arm 2, $\beta^{12}$ in \eqref{eq:late-cross-treatment-target} is the analogue, for LATEs, of the regression of the public program's ITT effect on the private program's ITT effect we reported above. Using the 200 sites where all three arms have at least two units, the estimator in \eqref{eq:late-cross-treatment-estimator-app} is equal to $0.728$, with a standard error based on \eqref{eq:late-cross-beta-if-app} equal to $0.274$. The positive correlation between the programs' effects therefore persists when considering LATEs instead of ITTs.

\section{Conclusion}\label{sec:conclusion}

We develop methods to investigate treatment-effect heterogeneity across sites in multi-site RCTs, even when each site contains too few observations to estimate its effect precisely. Combining established and new measurement-error corrections, we estimate the variance and centered third moment of site effects, as well as coefficients from regressions relating site effects to their untreated outcomes or to their effects on other outcomes, including potential mediators. We construct confidence intervals that are asymptotically valid as the number of sites grows, even when the number of units per site remains bounded. With imperfect compliance, we also identify a squared-first-stage-weighted variance of site-specific LATEs and a lower bound for their complier-weighted variance, allowing researchers to assess whether differences in take-up alone explain ITT heterogeneity.

Our application to public and private counseling programs for French job seekers illustrates why looking beyond average effects matters. Average employment gains of approximately 2 percentage points conceal cross-site standard deviations of ITT effects of 8–9 percentage points. We also find substantial LATE heterogeneity, so differences in take-up cannot fully explain the variation in ITT effects. Estimated effects are larger in sites with lower untreated employment rates. Finally, the public and private programs tend to be more effective in the same sites despite being implemented by different providers. This pattern is consistent with a role for site responsiveness, although correlated provider effectiveness could also explain it.

\bibliographystyle{apalike}
\bibliography{biblio,blp_appendix_references}

\appendix

\section{Estimators' influence function and inference}\label{app:inference}

\subsection{Variance, regression, and centered-third-moment estimators}
\label{app:general-inference}

Let
\begin{align}
\psi_{\sigma^2,i}
 &=w(S_i)\left[
   \{\widehat\mu_Y(S_i)-\E_w[\mu_Y(S)]\}^2
   -\widehat V_{Y,i}-\sigma^2\right],
 \label{eq:psi-sigma}\\
 \psi_{\beta,i}
 &=\Var_w(\mu_X(S))^{-1}w(S_i)
 \Big[
   \{\widehat\mu_X(S_i)-\E_w[\mu_X(S)]\}
   \big\{\widehat\mu_Y(S_i)-\E_w[\mu_Y(S)]
        -\{\widehat\mu_X(S_i)-\E_w[\mu_X(S)]\}\trans\beta\big\}
 \notag\\[-2mm]
 &\hspace{30mm}
   -\widehat C_{XY,i}+\widehat V_{X,i}\beta
 \Big]\label{eq:psi-beta}\\
 \psi_{\kappa_Y,i}
 &=w(S_i)\Big[
 \{\widehat\mu_Y(S_i)-\E_w[\mu_Y(S)]\}^3
 -3\{\widehat\mu_Y(S_i)-\E_w[\mu_Y(S)]\}\widehat V_{Y,i}
 +3\widehat C_{YV,i}-\widehat K_{Y,i}-\kappa_Y\notag\\[-2mm]
 &\hspace{30mm}-3\sigma^2
 \{\widehat\mu_Y(S_i)-\E_w[\mu_Y(S)]\}\Big].
 \label{eq:psi-kappa}
\end{align}
Lemma~\ref{lem:corrected} and the definition of $\beta$ imply
$\E[\psi_{\sigma^2,i}]=0$, $\E[\psi_{\beta,i}]=0$, and $\E[\psi_{\kappa_Y,i}]=0$.
\begin{assumption}[Conditions for estimating $\sigma^2$]
\label{ass:sigma}
\begin{enumerate}[label=(\roman*)]
 \item\label{ass:sigma-mom1}
 $\E[w(S_i)^2\{1+\widehat\mu_Y(S_i)^2\}]<\infty$.
 \item\label{ass:sigma-mom2}
 $\E[\psi_{\sigma^2,i}^2]<\infty$.
\end{enumerate}
\end{assumption}
\begin{assumption}[Conditions for estimating $\beta$]\label{ass:beta}
\begin{enumerate}[label=(\roman*)]
 \item\label{ass:beta-second}
 $\E[w(S_i)^2\{1+\norm{\widehat\mu_X(S_i)}^2+
 \widehat\mu_Y(S_i)^2\}]<\infty$ and
 $\E[w(S_i)\norm{\widehat\mu_X(S_i)}^2]<\infty$.
\item\label{ass:beta-vx-first}
 For every $(j,k)\in\{1,\ldots,K\}^2$,
 \[
 \E\!\left[\,|w(S_i)\widehat V_{X,i,jk}|\,\right]<\infty,
 \]
 where $\widehat V_{X,i,jk}$ denotes the $(j,k)$ entry of the matrix $\widehat V_{X,i}$.
 \item\label{ass:beta-ident}
 $\Var_w(\mu_X(S))$ is positive definite.
 \item\label{ass:beta-ifmoment}
 $\E[\norm{\psi_{\beta,i}}^2]<\infty$.
\end{enumerate}
\end{assumption}
\begin{assumption}[Conditions for estimating $\kappa_Y$]\label{ass:kappa}
\begin{enumerate}[label=(\roman*)]
 \item\label{ass:kappa-mean}
 $\E[w(S_i)^2\{1+\widehat\mu_Y(S_i)^2\}]<\infty$.
 \item\label{ass:kappa-second}
 $\E[|w(S_i)\{\widehat\mu_Y(S_i)^2-\widehat V_{Y,i}\}|]<\infty$.
 \item\label{ass:kappa-ifmoment}
 $\E[\psi_{\kappa_Y,i}^2]<\infty$.
\end{enumerate}
\end{assumption}
Assumptions~\ref{ass:sigma}\ref{ass:sigma-mom2},
\ref{ass:beta}\ref{ass:beta-ifmoment}, and Assumption~\ref{ass:kappa}\ref{ass:kappa-ifmoment} are imposed directly on the influence functions. Stronger primitive sufficient conditions could instead be imposed on the estimated conditional means and on the estimated variances and covariance.

Let
\begin{align}
\widehat\psi_{\sigma^2,i}
 &=\widehat w(S_i)\left[
   \{\widehat\mu_Y(S_i)-\overline{\widehat\mu}_Y\}^2
   -\widehat V_{Y,i}-\widehat\sigma^2\right],
 \label{eq:psihat-sigma}\\
 \widehat\psi_{\beta,i}
 &=\widehat{\Var}(\mu_X(S))^{-1}\widehat w(S_i)
 \Big[
   \{\widehat\mu_X(S_i)-\overline{\widehat\mu}_X\}
   \big\{\widehat\mu_Y(S_i)-\overline{\widehat\mu}_Y
       -\{\widehat\mu_X(S_i)-\overline{\widehat\mu}_X\}\trans
        \widehat\beta\big\}
 \notag\\[-2mm]
 &\hspace{30mm}
   -\widehat C_{XY,i}+\widehat V_{X,i}\widehat\beta
 \Big]\label{eq:psihat-beta}\\
\widehat\psi_{\kappa_Y,i}
 &=\widehat w(S_i)\Big[
 \{\widehat\mu_Y(S_i)-\overline{\widehat\mu}_Y\}^3
 -3\{\widehat\mu_Y(S_i)-\overline{\widehat\mu}_Y\}\widehat V_{Y,i}
 +3\widehat C_{YV,i}-\widehat K_{Y,i}-\widehat\kappa_Y\notag\\[-2mm]
 &\hspace{30mm}-3\widehat\sigma^2
 \{\widehat\mu_Y(S_i)-\overline{\widehat\mu}_Y\}\Big].
 \label{eq:psihat-kappa}
\end{align}
Let
\begin{align}
 \widehat V_{\sigma^2}
 &=\frac1n\sum_{i=1}^n\widehat\psi_{\sigma^2,i}^2,
 \label{eq:Vhat-sigma}\\
 \widehat V_\beta
 &=\frac1n\sum_{i=1}^n
 \widehat\psi_{\beta,i}\widehat\psi_{\beta,i}\trans
 \label{eq:Vhat-beta}\\
\widehat V_{\kappa_Y}&=\frac1n\sum_{i=1}^n\widehat\psi_{\kappa_Y,i}^2.
\end{align}
\begin{assumption}[Additional conditions for estimating $V_\beta$]\label{ass:variance}
\begin{equation*}\label{ass:variance-beta}
 \E[w(S_i)^2\norm{\widehat\mu_X(S_i)}^4]<\infty,
 \qquad
 \E[w(S_i)^2\widehat V_{X,i,jk}^2]<\infty
 \quad\text{for every }(j,k)\in\{1,\ldots,K\}^2.
\end{equation*}
\end{assumption}
The following theorems establish conditions under which $\widehat{\sigma}^2$, $\widehat{\beta}$, and $\widehat\kappa_Y$ are $\sqrt{n}$-consistent and asymptotically normal. They derive the estimators' influence functions and show that the asymptotic variances of $\widehat\sigma^2$ and $\widehat\beta$ can be consistently estimated. To preserve space, we do not prove that $\widehat V_{\kappa_Y}$ is consistent for the asymptotic variance of $\widehat\kappa_Y$, but this result holds as well.

\begin{theorem}\label{thm:sigma}
Suppose Assumptions~\ref{ass:sampling} and
\ref{ass:sigma} hold. Then
\begin{equation}\label{eq:AL-sigma}
 \sqrt n(\widehat\sigma^2-\sigma^2)
 =\frac1{\sqrt n}\sum_{i=1}^n\psi_{\sigma^2,i}+\op(1).
\end{equation}
Consequently,
\[
 \sqrt n(\widehat\sigma^2-\sigma^2)
 \overset{d}{\longrightarrow}N(0,V_{\sigma^2}),\qquad
 V_{\sigma^2}:=\E[\psi_{\sigma^2,i}^2].
\]
Moreover,
\[
 \widehat V_{\sigma^2}
 \overset{\Pp}{\longrightarrow}V_{\sigma^2}.
\]
\end{theorem}
\begin{remark}[Possibly negative variance estimator]
\eqref{eq:sigma-hat} shows that $\widehat\sigma^2$ can be
negative in finite samples, even though $\sigma^2\geq0$. Truncating the
estimator at zero changes its asymptotic distribution if
$\sigma^2=0$ and is therefore not imposed here.
\end{remark}
\begin{theorem}\label{thm:beta}
Suppose Assumptions~\ref{ass:sampling} and \ref{ass:beta} hold. Then
$\widehat{\Var}(\mu_X(S))$ is nonsingular with probability approaching one and
\begin{equation}\label{eq:AL-beta}
 \sqrt n(\widehat\beta-\beta)
 =\frac1{\sqrt n}\sum_{i=1}^n\psi_{\beta,i}+\op(1).
\end{equation}
Consequently,
\[
 \sqrt n(\widehat\beta-\beta)
 \overset{d}{\longrightarrow}N(0,V_\beta),
 \qquad
 V_\beta:=\E[\psi_{\beta,i}\psi_{\beta,i}\trans].
\]
Moreover, if Assumption
\ref{ass:variance} further holds,
\[
 \widehat V_\beta\overset{\Pp}{\longrightarrow}V_\beta.
\]
\end{theorem}
If $\sigma^2>0$, the $R^2$ of the weighted population regression of $\mu_Y(S)$ on a constant and $\mu_X(S)$ is
\[
 R^2
 =\frac{\Var_w\!\left(\mu_X(S)\trans\beta\right)}
        {\Var_w\!\left(\mu_Y(S)\right)}
 =\frac{\beta\trans\Var_w(\mu_X(S))\beta}{\sigma^2}.
\]
Whenever $\widehat\beta$ and the ratio below are well defined, we estimate
$R^2$ by the plug-in estimator
\begin{equation}\label{eq:R2-hat}
 \widehat R^2
 =\frac{\widehat\beta\trans
 \widehat{\Var}(\mu_X(S))\widehat\beta}
 {\widehat\sigma^2}.
\end{equation}
Because the corrected variance estimators need not be positive semidefinite
in finite samples, $\widehat R^2$ need not belong to $[0,1]$.
\begin{theorem}\label{thm:kappa}
Suppose Assumptions~\ref{ass:sampling} and \ref{ass:kappa} hold. Then
\begin{equation}\label{eq:AL-kappa}
 \sqrt n(\widehat\kappa_Y-\kappa_Y)
 =\frac1{\sqrt n}\sum_{i=1}^n\psi_{\kappa_Y,i}+\op(1).
\end{equation}
Consequently,
\[
 \sqrt n(\widehat\kappa_Y-\kappa_Y)
 \overset{d}{\longrightarrow}N(0,V_{\kappa_Y}),\qquad
 V_{\kappa_Y}:=\E[\psi_{\kappa_Y,i}^2].
\]
\end{theorem}

\subsection{Inference on $\underline{\sigma}^2_{\mathrm{LATE}}$ and $\sigma^2_{\mathrm{LATE},\mathrm{FS}^2}$}
\label{app:late-variance-inference}

Use the estimators and notation from Section~\ref{subsubsec:late-variance}, and let
\[
 Q(l)=\E_w[\{\mu_{\mathrm{ITT}}(S)-l\mu_{\mathrm{FS}}(S)\}^2],
 \qquad q=Q(\mathrm{LATE}_{\mathrm{FS}^2}),
 \qquad h=Q(\mathrm{LATE}).
\]
For the lower bound, write $\delta=\mathrm{FS}-m_2>0$ and
$d=r-m_2\mathrm{LATE}$, so that
$\underline{\sigma}^2_{\mathrm{LATE}}=(h+d^2/\delta)/\mathrm{FS}$.
Define
\begin{align}
 \psi_{q,i}
 &=w(S_i)\{g_i(\mathrm{LATE}_{\mathrm{FS}^2})-q\},
 \label{eq:psi-q-late}\\
 \psi_{\mathrm{FS}^2,i}
 &=w(S_i)\{\widehat\mu_{\mathrm{FS}}(S_i)^2
 -\widehat V_{\mathrm{FS},i}-m_2\},
 \label{eq:psi-fs-second-moment}\\
 \psi_{\sigma^2_{\mathrm{LATE},\mathrm{FS}^2},i}
 &=\frac{\psi_{q,i}
 -\sigma^2_{\mathrm{LATE},\mathrm{FS}^2}\psi_{\mathrm{FS}^2,i}}{m_2}.
 \label{eq:psi-late-variance}
\end{align}
As $\mathrm{LATE}_{\mathrm{FS}^2}$ minimizes $Q(l)$, its estimation has no first-order effect on $\widehat q$. With
\[
 \psi_{r,i}=w(S_i)\{
 \widehat\mu_{\mathrm{FS}}(S_i)\widehat\mu_{\mathrm{ITT}}(S_i)
 -\widehat C_{\mathrm{FS},\mathrm{ITT},i}-r\},
\]
the influence function of its estimator is
\begin{equation}\label{eq:psi-late-fs2}
 \psi_{\mathrm{LATE}_{\mathrm{FS}^2},i}
 =\frac{\psi_{r,i}-\mathrm{LATE}_{\mathrm{FS}^2}\psi_{\mathrm{FS}^2,i}}{m_2}.
\end{equation}

For the lower bound, define the influence functions of the average first stage and overall LATE:
\begin{equation}\label{eq:psi-fs-and-late}
 \psi_{\mathrm{FS},i}
 =w(S_i)\{\widehat\mu_{\mathrm{FS}}(S_i)-\mathrm{FS}\},
 \qquad
 \psi_{\mathrm{LATE},i}
 =\frac{w(S_i)\{\widehat\mu_{\mathrm{ITT}}(S_i)
 -\mathrm{LATE}\widehat\mu_{\mathrm{FS}}(S_i)\}}{\mathrm{FS}}.
\end{equation}
The derivative
$Q'(\mathrm{LATE})=2(m_2\mathrm{LATE}-r)$ need not vanish. Consequently, define
\begin{align}
 \psi_{h,i}
 &=w(S_i)\{g_i(\mathrm{LATE})-h\}
 +2(m_2\mathrm{LATE}-r)\psi_{\mathrm{LATE},i},
 \label{eq:psi-late-bound-numerator}\\
 \psi_{d,i}
 &=\psi_{r,i}-\mathrm{LATE}\psi_{\mathrm{FS}^2,i}
 -m_2\psi_{\mathrm{LATE},i},\notag\\
 \psi_{\delta,i}
 &=\psi_{\mathrm{FS},i}-\psi_{\mathrm{FS}^2,i},\notag\\
 \psi_{\underline{\sigma}^2_{\mathrm{LATE}},i}
 &=\frac{\psi_{h,i}
 +(2d/\delta)\psi_{d,i}-(d^2/\delta^2)\psi_{\delta,i}
 -\underline{\sigma}^2_{\mathrm{LATE}}\psi_{\mathrm{FS},i}}
 {\mathrm{FS}}.
 \label{eq:psi-late-variance-lower-bound}
\end{align}
The extra term in \eqref{eq:psi-late-bound-numerator} accounts for estimating the centering constant $\mathrm{LATE}$.

\begin{assumption}[Conditions for estimating $\sigma^2_{\mathrm{LATE},\mathrm{FS}^2}$]
\label{ass:late-variance-asymptotics}
The random variables $\psi_{r,i}$, $\psi_{q,i}$, and
$\psi_{\mathrm{FS}^2,i}$ have finite second moments.
\end{assumption}

\begin{assumption}[Additional conditions for estimating the lower bound]
\label{ass:late-bound-asymptotics}
The first-stage moments satisfy $\delta=\mathrm{FS}-m_2>0$, and
\[
 \E[w(S_i)^2\{1+\widehat\mu_{\mathrm{ITT}}(S_i)^2\}]<\infty.
\]
\end{assumption}

\begin{theorem}\label{thm:late-variance}
Suppose Assumptions~\ref{hyp:iid},
\ref{hyp:strat_completely_randomized}, \ref{ass:late}, and
\ref{ass:late-variance-asymptotics} hold.
\begin{enumerate}
 \item Then
 \begin{equation}\label{eq:AL-late-variance}
 \sqrt n(\widehat\sigma^2_{\mathrm{LATE},\mathrm{FS}^2}
 -\sigma^2_{\mathrm{LATE},\mathrm{FS}^2})
 =\frac1{\sqrt n}\sum_{i=1}^n
 \psi_{\sigma^2_{\mathrm{LATE},\mathrm{FS}^2},i}+\op(1).
 \end{equation}
 Consequently,
 \[
 \sqrt n(\widehat\sigma^2_{\mathrm{LATE},\mathrm{FS}^2}
 -\sigma^2_{\mathrm{LATE},\mathrm{FS}^2})
 \overset{d}{\longrightarrow}N(0,V_{\sigma^2_{\mathrm{LATE},\mathrm{FS}^2}}),\qquad
 V_{\sigma^2_{\mathrm{LATE},\mathrm{FS}^2}}
 :=\E[\psi_{\sigma^2_{\mathrm{LATE},\mathrm{FS}^2},i}^2].
 \]
 \item Suppose Assumption~\ref{ass:late-bound-asymptotics} also holds. Then
 \begin{equation}\label{eq:AL-late-variance-lower-bound}
 \sqrt n(\widehat{\underline{\sigma}}^2_{\mathrm{LATE}}
 -\underline{\sigma}^2_{\mathrm{LATE}})
 =\frac1{\sqrt n}\sum_{i=1}^n
 \psi_{\underline{\sigma}^2_{\mathrm{LATE}},i}+\op(1).
 \end{equation}
 Consequently,
 \[
 \sqrt n(\widehat{\underline{\sigma}}^2_{\mathrm{LATE}}
 -\underline{\sigma}^2_{\mathrm{LATE}})
 \overset{d}{\longrightarrow}N(0,V_{\underline{\sigma}^2_{\mathrm{LATE}}}),\qquad
 V_{\underline{\sigma}^2_{\mathrm{LATE}}}
 :=\E[\psi_{\underline{\sigma}^2_{\mathrm{LATE}},i}^2].
 \]
\end{enumerate}
\end{theorem}

For implementation, set
\begin{align*}
 \widehat\psi_{q,i}
 &=\widehat w(S_i)\{g_i(\widehat{\mathrm{LATE}}_{\mathrm{FS}^2})
 -\widehat q\},\\
 \widehat\psi_{\mathrm{FS}^2,i}
 &=\widehat w(S_i)\{\widehat\mu_{\mathrm{FS}}(S_i)^2
 -\widehat V_{\mathrm{FS},i}-\widehat m_2\},\\
 \widehat\psi_{\sigma^2_{\mathrm{LATE},\mathrm{FS}^2},i}
 &=\frac{\widehat\psi_{q,i}
 -\widehat\sigma^2_{\mathrm{LATE},\mathrm{FS}^2}
 \widehat\psi_{\mathrm{FS}^2,i}}{\widehat m_2},\\
 \widehat\psi_{\mathrm{FS},i}
 &=\widehat w(S_i)\{\widehat\mu_{\mathrm{FS}}(S_i)
 -\widehat{\mathrm{FS}}\},\\
 \widehat\psi_{\mathrm{LATE},i}
 &=\frac{\widehat w(S_i)\{\widehat\mu_{\mathrm{ITT}}(S_i)
 -\widehat{\mathrm{LATE}}\widehat\mu_{\mathrm{FS}}(S_i)\}}
 {\widehat{\mathrm{FS}}},\\
 \widehat h&=\widehat Q(\widehat{\mathrm{LATE}}),\\
 \widehat\psi_{h,i}
 &=\widehat w(S_i)\{g_i(\widehat{\mathrm{LATE}})-\widehat h\}
 +2(\widehat m_2\widehat{\mathrm{LATE}}-\widehat r)
 \widehat\psi_{\mathrm{LATE},i},\\
 \widehat d&=\widehat r-\widehat m_2\widehat{\mathrm{LATE}},
 \qquad \widehat\delta=\widehat{\mathrm{FS}}-\widehat m_2,\\
 \widehat\psi_{r,i}
 &=\widehat w(S_i)\{\widehat\mu_{\mathrm{FS}}(S_i)
 \widehat\mu_{\mathrm{ITT}}(S_i)-\widehat C_{\mathrm{FS},\mathrm{ITT},i}
 -\widehat r\},\\
 \widehat\psi_{d,i}
 &=\widehat\psi_{r,i}-\widehat{\mathrm{LATE}}\widehat\psi_{\mathrm{FS}^2,i}
 -\widehat m_2\widehat\psi_{\mathrm{LATE},i},\\
 \widehat\psi_{\delta,i}
 &=\widehat\psi_{\mathrm{FS},i}-\widehat\psi_{\mathrm{FS}^2,i},\\
 \widehat\psi_{\underline{\sigma}^2_{\mathrm{LATE}},i}
 &=\frac{\widehat\psi_{h,i}
 +(2\widehat d/\widehat\delta)\widehat\psi_{d,i}
 -(\widehat d^2/\widehat\delta^2)\widehat\psi_{\delta,i}
 -\widehat{\underline{\sigma}}^2_{\mathrm{LATE}}
 \widehat\psi_{\mathrm{FS},i}}{\widehat{\mathrm{FS}}}.
\end{align*}
The asymptotic-variance estimators are
\begin{align}
 \widehat V_{\sigma^2_{\mathrm{LATE},\mathrm{FS}^2}}
 &=\frac1n\sum_{i=1}^n
 \widehat\psi_{\sigma^2_{\mathrm{LATE},\mathrm{FS}^2},i}^{\,2},
 \label{eq:Vhat-late-variance}\\
 \widehat V_{\underline{\sigma}^2_{\mathrm{LATE}}}
 &=\frac1n\sum_{i=1}^n
 \widehat\psi_{\underline{\sigma}^2_{\mathrm{LATE}},i}^{\,2}.
 \label{eq:Vhat-late-variance-lower-bound}
\end{align}

\subsection{Inference on the covariance between sites' LATEs and first-stage effects}\label{app:late-covariance-inference}

We give a heuristic calculation of
$\widehat\sigma_{\mathrm{LATE},\mathrm{FS}}$'s influence function, using
$\psi_{\mathrm{FS},i}$ and $\psi_{\mathrm{LATE},i}$ defined in
\eqref{eq:psi-fs-and-late}. The influence functions of
$\widehat C_{\mathrm{FS},\mathrm{ITT}}$ and $\widehat V_{\mathrm{FS}}$ are, respectively,
\begin{align*}
 \psi_{C_{\mathrm{FS},\mathrm{ITT}},i}
 &=w(S_i)\Big[
  \{\widehat\mu_{\mathrm{FS}}(S_i)-\mathrm{FS}\}
  \{\widehat\mu_{\mathrm{ITT}}(S_i)-\mathrm{ITT}\}
  -\widehat C_{\mathrm{FS},\mathrm{ITT},i}
  -\Cov_w(\mu_{\mathrm{FS}}(S),\mu_{\mathrm{ITT}}(S))\Big],\\
 \psi_{V_{\mathrm{FS}},i}
 &=w(S_i)\Big[
  \{\widehat\mu_{\mathrm{FS}}(S_i)-\mathrm{FS}\}^2
  -\widehat V_{\mathrm{FS},i}-\Var_w(\mu_{\mathrm{FS}}(S))\Big].
\end{align*}
A ratio expansion gives
\begin{equation}\label{eq:AL-lambda-late-proof}
 \sqrt n(\widehat{\mathrm{LATE}}-\mathrm{LATE})
 =\frac1{\sqrt n}\sum_{i=1}^n
 \psi_{\mathrm{LATE},i}+\op(1).
\end{equation}
Applying the delta method to
\eqref{eq:late-fs-cov-hat} gives
\begin{equation}\label{eq:psi-late-fs-cov}
 \psi_{\mathrm{LATE},\mathrm{FS},i}
 =\frac{1}{\mathrm{FS}}\left(
   \psi_{C_{\mathrm{FS},\mathrm{ITT}},i}
   -\mathrm{LATE}\,\psi_{V_{\mathrm{FS}},i}
   -\Var_w(\mu_{\mathrm{FS}}(S))\psi_{\mathrm{LATE},i}
   -\sigma_{\mathrm{LATE},\mathrm{FS}}\psi_{\mathrm{FS},i}
 \right).
\end{equation}
This suggests estimating the asymptotic variance by replacing all population quantities in \eqref{eq:psi-late-fs-cov} with their sample analogues and averaging the squared estimated influence functions across sites.

\subsection{Estimation of, and inference on, $\beta^d$ and $\beta^{12}_k$}\label{app:late-regressions}

For any two site effects $A$ and $B$ considered below, let
\begin{align}
 \mathcal M_{A,B}
 &=\E_w[\mu_A(S)\mu_B(S)],\notag\\
 \widehat{\mathcal M}_{A,B}
 &=\frac1n\sum_{i=1}^n\widehat w(S_i)
 \{\widehat\mu_A(S_i)\widehat\mu_B(S_i)-\widehat C_{A,B,i}\},
 \label{eq:late-reg-corrected-moment}
\end{align}
where $\widehat C_{A,B,i}$ estimates the conditional covariance of the two site-effect estimators. The influence function of $\widehat{\mathcal M}_{A,B}$ is
\begin{equation}\label{eq:late-reg-moment-if}
 \psi_{A,B,i}=w(S_i)
 \{\widehat\mu_A(S_i)\widehat\mu_B(S_i)
 -\widehat C_{A,B,i}-\mathcal M_{A,B}\}.
\end{equation}
The formulas below are heuristic and require the same types of moment and nonsingularity conditions as Theorem~\ref{thm:beta}.

\subsubsection{LATEs for several outcomes under the same treatment arm}

For $j\in\{1,\ldots,J\}$, estimate the squared-first-stage-weighted mean LATE by
\begin{equation}\label{eq:late-dj-fs2-estimator-app}
 \widehat{\mathrm{LATE}}_{d,j,\mathrm{FS}_d^2}
 =\frac{\widehat{\mathcal M}_{\mathrm{FS}_d,\mathrm{ITT}_{d,j}}}
 {\widehat{\mathcal M}_{\mathrm{FS}_d,\mathrm{FS}_d}}.
\end{equation}
For each site, form
\[
 \widehat U_{d,j}(S_i)
 =\widehat\mu_{\mathrm{ITT}_{d,j}}(S_i)
 -\widehat\mu_{\mathrm{FS}_d}(S_i)
  \widehat{\mathrm{LATE}}_{d,j,\mathrm{FS}_d^2}.
\]
The conditional covariance correction for $\widehat U_{d,j}(S_i)$ and $\widehat U_{d,\ell}(S_i)$ is
\begin{align}
 \widehat C_{U_{d,j},U_{d,\ell},i}
 &=\widehat C_{\mathrm{ITT}_{d,j},\mathrm{ITT}_{d,\ell},i}
 -\widehat{\mathrm{LATE}}_{d,\ell,\mathrm{FS}_d^2}
  \widehat C_{\mathrm{ITT}_{d,j},\mathrm{FS}_d,i}\notag\\
 &\quad-\widehat{\mathrm{LATE}}_{d,j,\mathrm{FS}_d^2}
  \widehat C_{\mathrm{FS}_d,\mathrm{ITT}_{d,\ell},i}
 +\widehat{\mathrm{LATE}}_{d,j,\mathrm{FS}_d^2}
  \widehat{\mathrm{LATE}}_{d,\ell,\mathrm{FS}_d^2}
  \widehat V_{\mathrm{FS}_d,i}.
 \label{eq:late-transformed-correction-app}
\end{align}
Stacking outcomes $2,\ldots,J$ in $\widehat U_{d,-1}(S_i)$, the estimator is
\begin{equation}\label{eq:late-within-beta-estimator-app}
 \widehat\beta^d=
 \left[\frac1n\sum_{i=1}^n\widehat w(S_i)
 \{\widehat U_{d,-1}(S_i)\widehat U_{d,-1}(S_i)\trans
 -\widehat V_{U_{d,-1},i}\}\right]^{-1}
 \left[\frac1n\sum_{i=1}^n\widehat w(S_i)
 \{\widehat U_{d,-1}(S_i)\widehat U_{d,1}(S_i)
 -\widehat C_{U_{d,-1},U_{d,1},i}\}\right].
\end{equation}

Let $U_{d,j}(S)=\mu_{\mathrm{ITT}_{d,j}}(S)-\mu_{\mathrm{FS}_d}(S)
\mathrm{LATE}_{d,j,\mathrm{FS}_d^2}$, and
\[
 \widetilde U_{d,j,i}
 =\widehat\mu_{\mathrm{ITT}_{d,j}}(S_i)
 -\widehat\mu_{\mathrm{FS}_d}(S_i)
  \mathrm{LATE}_{d,j,\mathrm{FS}_d^2}.
\]
For $j,\ell\in\{1,\ldots,J\}$, let
$\widetilde C_{U_{d,j},U_{d,\ell},i}$ denote
\eqref{eq:late-transformed-correction-app} with the estimated weighted
LATEs replaced by their population counterparts.
Stack $\widetilde U_{d,2,i},\ldots,\widetilde U_{d,J,i}$ in
$\widetilde U_{d,-1,i}$. Let the $(j,\ell)$ entry of
$\widetilde V_{U_{d,-1},i}$ be
$\widetilde C_{U_{d,j},U_{d,\ell},i}$ and the $j$th entry of
$\widetilde C_{U_{d,-1},U_{d,1},i}$ be
$\widetilde C_{U_{d,j},U_{d,1},i}$, for $j,\ell\in\{2,\ldots,J\}$.
The influence function of \eqref{eq:late-within-beta-estimator-app} is
\begin{align}
 \psi_{\beta^d,i}
 &=\E_w[U_{d,-1}(S)U_{d,-1}(S)\trans]^{-1}w(S_i)
 \Big[\widetilde U_{d,-1,i}
 \{\widetilde U_{d,1,i}-\widetilde U_{d,-1,i}\trans\beta^d\}\notag\\
 &\qquad-\widetilde C_{U_{d,-1},U_{d,1},i}
 +\widetilde V_{U_{d,-1},i}\beta^d\Big].
 \label{eq:late-within-beta-if-app}
\end{align}
Estimation of the weighted mean LATEs in \eqref{eq:late-dj-fs2-estimator-app} has no first-order effect. Indeed, their definition implies the orthogonality condition
\[
 \E_w[\mu_{\mathrm{FS}_d}(S)U_{d,j}(S)]=0.
\]
Then, one can show that the derivatives of
$\E_w[U_{d,j}(S)U_{d,\ell}(S)]$ with respect to
$\mathrm{LATE}_{d,j,\mathrm{FS}_d^2}$ and
$\mathrm{LATE}_{d,\ell,\mathrm{FS}_d^2}$ are zero.

\subsubsection{LATEs for the same outcome under two treatment arms}

The plug-in estimator of \eqref{eq:late-cross-treatment-identification} is
\begin{equation}\label{eq:late-cross-treatment-estimator-app}
 \widehat\beta^{12}_k=\frac{\widehat C^{12}_k}{\widehat V^1_k},
\end{equation}
where
\begin{align*}
 \widehat C^{12}_k
 &=\frac{\widehat{\mathcal M}_{\mathrm{ITT}_{1,k},\mathrm{ITT}_{2,k}}}
 {\widehat{\mathcal M}_{\mathrm{FS}_1,\mathrm{FS}_2}}
 -\frac{\widehat{\mathcal M}_{\mathrm{FS}_2,\mathrm{ITT}_{1,k}}
 \widehat{\mathcal M}_{\mathrm{FS}_1,\mathrm{ITT}_{2,k}}}
 {\widehat{\mathcal M}_{\mathrm{FS}_1,\mathrm{FS}_2}^2},\\
 \widehat V^1_k
 &=\frac{\widehat{\mathcal M}_{\mathrm{ITT}_{1,k},\mathrm{ITT}_{1,k}}}
 {\widehat{\mathcal M}_{\mathrm{FS}_1,\mathrm{FS}_1}}
 -\frac{\widehat{\mathcal M}_{\mathrm{FS}_1,\mathrm{ITT}_{1,k}}^2}
 {\widehat{\mathcal M}_{\mathrm{FS}_1,\mathrm{FS}_1}^2}.
\end{align*}
Let $C^{12}_k$ and $V^1_k$ denote the probability limits of the preceding two displays. Applying the quotient and product rules to their corrected moments gives
\begin{align}
 \psi_{C^{12}_k,i}
 &=\frac{\psi_{\mathrm{ITT}_{1,k},\mathrm{ITT}_{2,k},i}}
 {\mathcal M_{\mathrm{FS}_1,\mathrm{FS}_2}}
 -\frac{\mathcal M_{\mathrm{ITT}_{1,k},\mathrm{ITT}_{2,k}}
 \psi_{\mathrm{FS}_1,\mathrm{FS}_2,i}}
 {\mathcal M_{\mathrm{FS}_1,\mathrm{FS}_2}^2}\notag\\
 &\quad-\frac{\mathcal M_{\mathrm{FS}_1,\mathrm{ITT}_{2,k}}
 \psi_{\mathrm{FS}_2,\mathrm{ITT}_{1,k},i}
 +\mathcal M_{\mathrm{FS}_2,\mathrm{ITT}_{1,k}}
 \psi_{\mathrm{FS}_1,\mathrm{ITT}_{2,k},i}}
 {\mathcal M_{\mathrm{FS}_1,\mathrm{FS}_2}^2}\notag\\
 &\quad+\frac{2\mathcal M_{\mathrm{FS}_2,\mathrm{ITT}_{1,k}}
 \mathcal M_{\mathrm{FS}_1,\mathrm{ITT}_{2,k}}
 \psi_{\mathrm{FS}_1,\mathrm{FS}_2,i}}
 {\mathcal M_{\mathrm{FS}_1,\mathrm{FS}_2}^3},
 \label{eq:late-cross-cov-if-app}\\
 \psi_{V^1_k,i}
 &=\frac{\psi_{\mathrm{ITT}_{1,k},\mathrm{ITT}_{1,k},i}}
 {\mathcal M_{\mathrm{FS}_1,\mathrm{FS}_1}}
 -\frac{\mathcal M_{\mathrm{ITT}_{1,k},\mathrm{ITT}_{1,k}}
 \psi_{\mathrm{FS}_1,\mathrm{FS}_1,i}}
 {\mathcal M_{\mathrm{FS}_1,\mathrm{FS}_1}^2}\notag\\
 &\quad-\frac{2\mathcal M_{\mathrm{FS}_1,\mathrm{ITT}_{1,k}}
 \psi_{\mathrm{FS}_1,\mathrm{ITT}_{1,k},i}}
 {\mathcal M_{\mathrm{FS}_1,\mathrm{FS}_1}^2}
 +\frac{2\mathcal M_{\mathrm{FS}_1,\mathrm{ITT}_{1,k}}^2
 \psi_{\mathrm{FS}_1,\mathrm{FS}_1,i}}
 {\mathcal M_{\mathrm{FS}_1,\mathrm{FS}_1}^3}.
 \label{eq:late-cross-var-if-app}
\end{align}
Therefore,
\begin{equation}\label{eq:late-cross-beta-if-app}
 \psi_{\beta^{12}_k,i}
 =\frac{\psi_{C^{12}_k,i}-\beta^{12}_k\psi_{V^1_k,i}}{V^1_k}.
\end{equation}

\subsection{Inference on the mediation residual variance}
\label{app:mediation-inference}

The first-step estimation of $\beta_{\mathrm{med}}$ does not contribute to the
asymptotic variance of $\widehat\sigma_{\varepsilon}^2$. To see this, write
\[
\sigma_{\varepsilon}^2
=\Var_w\!\left(
\mu_{Y}(S)-\beta_{\mathrm{med}}\mu_{X}(S)
\right),
\]
and, for any scalar $b$, let $\widehat\sigma_{\varepsilon}^2(b)$ denote the estimator in Section~\ref{subsubsec:mediation-test} with $\widehat\beta_{\mathrm{med}}$ replaced by $b$. By construction,
\begin{align*}
	\widehat\sigma_{\varepsilon}^2(b)
	&=\widehat{\Var}(\mu_{Y}(S))
	-2b\widehat{\Cov}(\mu_{X}(S),
	\mu_{Y}(S))\\
	&\quad+b^2\widehat{\Var}(\mu_{X}(S)).
\end{align*}
For compactness, let
\[
\widehat V_2=\widehat{\Var}(\mu_{Y}(S)),\qquad
\widehat V_1=\widehat{\Var}(\mu_{X}(S)),\qquad
\widehat C_{12}=\widehat{\Cov}(\mu_{X}(S),
\mu_{Y}(S)).
\]
With this notation,
$\widehat\beta_{\mathrm{med}}=\widehat C_{12}/\widehat V_1$. Therefore, for any scalar $b$,
\begin{align*}
	\widehat\sigma_{\varepsilon}^2(b)
	&=\widehat V_2-2b\widehat C_{12}+b^2\widehat V_1\\
	&=\widehat V_2+\widehat V_1
	\left\{b^2-2b\frac{\widehat C_{12}}{\widehat V_1}\right\}\\
	&=\widehat V_2+\widehat V_1
	\left\{\left(b-\frac{\widehat C_{12}}{\widehat V_1}\right)^2
	-\left(\frac{\widehat C_{12}}{\widehat V_1}\right)^2\right\}\\
	&=\widehat V_2-\frac{\widehat C_{12}^2}{\widehat V_1}
	+\widehat V_1(b-\widehat\beta_{\mathrm{med}})^2.
\end{align*}
Since $\widehat\sigma_{\varepsilon}^2=\widehat\sigma_{\varepsilon}^2(\widehat\beta_{\mathrm{med}})$, evaluating
the last display first at $b=\widehat\beta_{\mathrm{med}}$ and then at $b=\beta_{\mathrm{med}}$ yields
\begin{align*}
	\widehat\sigma_{\varepsilon}^2-\widehat\sigma_{\varepsilon}^2(\beta_{\mathrm{med}})
	&=-\widehat V_1(\widehat\beta_{\mathrm{med}}-\beta_{\mathrm{med}})^2
	=\op(n^{-1/2}).
\end{align*}
Here, the final equality follows because $\widehat\beta_{\mathrm{med}}$ is $\sqrt n$-consistent and the corrected regressor variance converges.
Therefore,
the influence function of $\widehat\sigma_{\varepsilon}^2$ is
\begin{align*}
	\psi_{\sigma_{\varepsilon}^2,i}
	=w(S_i)\Big[&
	\big\{\widehat\mu_{Y}(S_i)
	-\beta_{\mathrm{med}}\widehat\mu_{X}(S_i)
	-\E_w[\mu_{Y}(S)-\beta_{\mathrm{med}}\mu_{X}(S)]\big\}^2\\
	&-\widehat V_{Y,i}
	-\beta_{\mathrm{med}}^2\widehat V_{X,i}
	+2\beta_{\mathrm{med}}\widehat C_{XY,i}
	-\sigma_{\varepsilon}^2\Big],
\end{align*}
which can be estimated by
\[
\widehat\psi_{\sigma_{\varepsilon}^2,i}
=\widehat w(S_i)\left[
\{\widehat\mu_{\varepsilon}(S_i)-\widehat\tau_{\varepsilon}\}^2
-\widehat V_{\varepsilon,i}-\widehat\sigma_{\varepsilon}^2
\right].
\]
Under the conditions of Theorems \ref{thm:sigma} and \ref{thm:beta} and the corresponding moment conditions for the transformed outcome
\[
 Y_i-\beta_{\mathrm{med}}X_i
 =O_{i,2}(1)-O_{i,2}(0)
 -\beta_{\mathrm{med}}\{O_{i,1}(1)-O_{i,1}(0)\},
\]
one can show that
\[
\sqrt n\big(\widehat\sigma_{\varepsilon}^2-\sigma_{\varepsilon}^2\big)
=\frac1{\sqrt n}\sum_{i=1}^n\psi_{\sigma_{\varepsilon}^2,i}+\op(1),
\]
and
\[
\widehat V_{\sigma_{\varepsilon}^2}
=\frac1n\sum_{i=1}^n\widehat\psi_{\sigma_{\varepsilon}^2,i}^{\,2}
\]
is a consistent estimator of the asymptotic variance of
$\widehat\sigma_{\varepsilon}^2$.

One can therefore test the implication of \eqref{eq:homogeneous-mediation} in
\eqref{eq:testable_implication} by comparing
$\sqrt n\,\widehat\sigma_{\varepsilon}^2/
\widehat V_{\sigma_{\varepsilon}^2}^{1/2}$
to standard normal critical values.

\section{Simulations}\label{app:simulations}

We simulate multi-site RCTs with $n$ sites. Site $i$ has an untreated mean $\alpha_i\sim N(0,0.2^2)$ and an average treatment effect $\tau_i=0.1-0.5\alpha_i+e_i$, where $e_i$ is independent of $\alpha_i$, has mean zero and standard deviation $0.3$, and is drawn from one of four distributions: normal; Laplace, which is symmetric with heavier tails than the normal; a Beta(2,5) distribution, which is right-skewed with bounded support; and an exponential distribution, which is right-skewed with a long tail. Their standardized skewnesses are $0$, $0$, $0.6$, and $2$. Site $i$ has $N_{1,i}$ treated and $N_{0,i}$ control units, drawn independently and uniformly from $\{8,\ldots,25\}$, and receives weight $W(S_i)=N_{1,i}+N_{0,i}$. Unit outcomes are $O_{u,i}(0)=\alpha_i+0.5(E_{u,i}-1)$ and $O_{u,i}(1)=O_{u,i}(0)+\tau_i+0.3(E'_{u,i}-1)$, where $E_{u,i}$ and $E'_{u,i}$ are independent standard exponential variables, so that unit-level errors are skewed. The targets are $\sigma^2=\Var_w(\tau_i)=0.1$, $\beta=-0.5$, the coefficient from the regression of $\tau_i$ on $\alpha_i$ as in Section~\ref{subsec:rct-ate-untreated}, and $\kappa=\E_w[(\tau_i-\E_w[\tau_i])^3]$, which equals $0.3^3$ times the standardized skewness of $e_i$.

For each design we draw 1,000 samples and compute $\widehat\sigma^2$, $\widehat\beta$, and $\widehat\kappa_Y$ with the estimators of Section~\ref{sec:framework}, applied to the site-level estimators and corrections of Proposition~\ref{prop:rct-unbiased}. Table~\ref{tab:simulations} reports the coverage rates of nominal 95\% confidence intervals based on the influence functions (IFs) in Theorems~\ref{thm:sigma} to \ref{thm:kappa}. For $\sigma^2$ and $\kappa_Y$, it also reports the coverage of intervals from a studentized site-level bootstrap with 999 replications: sites are resampled with replacement, the estimator and its influence-function standard error are recomputed in each bootstrap sample, and the interval is formed with critical values equal to the 2.5th and 97.5th percentiles of the bootstrapped $t$-statistics. Panels A to C use 50, 100, and 200 sites and all four distributions. Panel D uses larger numbers of sites for the exponential distribution.

\begin{table}[H]
\centering
\caption{Coverage rates of nominal 95\% confidence intervals}
\label{tab:simulations}
\begin{tabular}{lccccc}
\hline
& \multicolumn{2}{c}{$\sigma^2$ (Theorem~\ref{thm:sigma})} & $\beta$ (Theorem~\ref{thm:beta}) & \multicolumn{2}{c}{$\kappa_Y$ (Theorem~\ref{thm:kappa})}\\
Distribution of site effects & IF & Bootstrap & IF & IF & Bootstrap\\
& (1) & (2) & (3) & (4) & (5)\\
\hline
\multicolumn{6}{l}{\textit{Panel A: 50 sites}}\\
Normal       & 0.904 & 0.947 & 0.919 & 0.966 & 0.904\\
Laplace      & 0.840 & 0.928 & 0.920 & 0.954 & 0.896\\
Beta(2,5)    & 0.895 & 0.935 & 0.916 & 0.879 & 0.857\\
Exponential  & 0.839 & 0.908 & 0.929 & 0.621 & 0.748\\
\hline
\multicolumn{6}{l}{\textit{Panel B: 100 sites}}\\
Normal       & 0.909 & 0.930 & 0.943 & 0.952 & 0.906\\
Laplace      & 0.896 & 0.944 & 0.954 & 0.966 & 0.909\\
Beta(2,5)    & 0.917 & 0.929 & 0.935 & 0.914 & 0.896\\
Exponential  & 0.887 & 0.925 & 0.949 & 0.727 & 0.831\\
\hline
\multicolumn{6}{l}{\textit{Panel C: 200 sites}}\\
Normal       & 0.944 & 0.953 & 0.941 & 0.954 & 0.911\\
Laplace      & 0.908 & 0.935 & 0.938 & 0.965 & 0.898\\
Beta(2,5)    & 0.938 & 0.954 & 0.946 & 0.923 & 0.904\\
Exponential  & 0.892 & 0.929 & 0.938 & 0.769 & 0.862\\
\hline
\multicolumn{6}{l}{\textit{Panel D: more sites, exponential distribution}}\\
Exponential, 500 sites    & 0.926 & 0.949 & 0.939 & 0.839 & 0.908\\
Exponential, 1,000 sites  & 0.943 & 0.950 & 0.948 & 0.889 & 0.926\\
Exponential, 5,000 sites  & 0.944 & 0.953 & 0.943 & 0.928 & 0.947\\
\hline
\end{tabular}
\begin{minipage}{0.95\linewidth}
\footnotesize
\textit{Notes:} IF denotes influence function. Each cell reports the share of 1,000 simulated samples in which a nominal 95\% confidence interval contains the true parameter. Columns (1), (3), and (4) use intervals of the form estimate $\pm1.96$ standard error, with standard errors based on the influence functions in Theorems~\ref{thm:sigma} to \ref{thm:kappa}. Columns (2) and (5) use a studentized site-level bootstrap with 999 replications. The data-generating process is described in the text; the standardized skewness of $e_i$ is $0$ for the normal and Laplace distributions, $0.6$ for the Beta(2,5), and $2$ for the exponential.
\end{minipage}
\end{table}

\section{Comparison between $\widehat \sigma^2_{\mathrm{ATE}}$ and the BLP estimator of \cite{cddf2025}}\label{app:blp}

Throughout, we maintain the sampling and complete-randomization assumptions of Section~\ref{sec:multi-site-rcts}, with \(\mathcal D=L=1\). Moreover, to simplify we let \(W(S_i)=N_i\).

\subsection{Best linear predictor with site fixed effects}

We consider the best linear predictor (BLP) of conditional treatment effects in \citet{cddf2025}, specialized to the case where
the only heterogeneity predictors are the full set of site
indicators, and where the training is done by saturated least squares, without
regularization.

To simplify, suppose $N_{0,i}$ and
$N_{1,i}$ are even. Uniformly split each arm into two equally sized
subsamples, independently of outcomes and independently across sites.
Let $\mathcal I_i^k$ denote subsample $k\in\{1,2\}$ and
$N_{d,i}^k=N_{d,i}/2$, $N_i^k=N_i/2$. Then
\[
 p_i:=N_{1,i}/N_i=N_{1,i}^k/N_i^k,
 \qquad
 \sum_{u\in\mathcal I_i^k}(D_{u,i}-p_i)=0.
\]
Define the subsample estimators
\[
 \widehat\mu_Y^k(S_i)=\overline O_{1,i,1}^k-\overline O_{0,i,1}^k,
 \qquad
 \overline O_{d,i,1}^k=
 \frac{1}{N_{d,i}^k}\sum_{u\in\mathcal I_i^k:D_{u,i}=d}O_{u,i,1}.
\]
Conditional on subsample membership, each subsample has
complete randomization with its prescribed arm counts.
Thus, under within-site i.i.d. sampling, these two effect estimates are
independent conditional on $S_i$, unbiased for $\mu_Y(S_i)$, and
identically distributed conditional on $S_i$.

When subsample $k$ is the validation
sample and subsample $3-k$ is the training sample, site $i$'s predicted effect in the validation sample is
\[
 A_i^k=\widehat\mu_Y^{3-k}(S_i).
\]
Let $B_i^k=\overline O_{0,i,1}^{3-k}$ be the average untreated outcome in site $i$ in the training sample, and let $C_i^k=\widehat\mu_Y^k(S_i)$ be its estimated effect in the validation sample.

Specializing the weighted residual BLP regression in Section~3.1 of
\citet{cddf2025}, estimate on subsample $k$
\begin{equation}\label{eq:individual-blp}
 O_{u,i,1}=\gamma^{k\trans}X_{1,i}^k
 +\widehat\beta_1^k(D_{u,i}-p_i)
 +\widehat\beta_2^k(D_{u,i}-p_i)(A_i^k-\overline A_N^k)
 +\widehat\varepsilon_{u,i}^k,
\end{equation}
by weighted least squares with individual weight $[p_i(1-p_i)]^{-1}$,
where
\[
 X_{1,i}^k=(1,B_i^k,p_i,p_iA_i^k)\trans,
 \qquad
 \overline A_N^k=\frac{\sum_iN_i A_i^k}{\sum_iN_i}.
\]
\begin{lemma}[Exact site-level representation]\label{lem:representation}
Under the split described above, if
$\sum_iN_i(A_i^k-\overline A_N^k)^2>0$, the interaction coefficient
in \eqref{eq:individual-blp} is
\begin{equation}\label{eq:site-blp}
 \widehat\beta_2^k
 =\frac{\sum_{i=1}^nN_i(A_i^k-\overline A_N^k)
                       (C_i^k-\overline C_N^k)}
        {\sum_{i=1}^nN_i(A_i^k-\overline A_N^k)^2},
 \qquad
 \overline C_N^k=\frac{\sum_iN_iC_i^k}{\sum_iN_i}.
\end{equation}
\end{lemma}

Thus the BLP regresses the estimated effect in the validation sample
on the estimated effect in the training sample, with an intercept.
The individual-level implementation naturally weights sites by their
size. Write $A,C$ for the two half-sample estimates and
$r_w=\E_w[\Var(A\mid S)]$. Under laws of large numbers for the relevant
weighted moments, Lemma~\ref{lem:representation} implies
\begin{equation}\label{eq:plim}
 \widehat\beta_2^k\ \longrightarrow_p\
 b_w:=\frac{\sigma^2_{\mathrm{ATE}}}
 {\sigma^2_{\mathrm{ATE}}+r_w}.
\end{equation}
The BLP slope thus targets a different parameter than the variance of treatment effects: in our setting, it converges to the ratio of that variance to the sum of that variance and the estimation noise in the predicted effects.
By contrast, its numerator, normalized by the sum of site weights,
is consistent for $\sigma^2_{\mathrm{ATE}}$ because independent
half-sample estimation errors do not contribute to the cross-sample
covariance. Finally, we consider
\begin{equation}\label{eq:average}
 \widehat b=\frac12(\widehat\beta_2^1+\widehat\beta_2^2),
\end{equation}
the average of the two BLP estimators.

\subsection{Power comparison}
Our goal is to compare the power of tests of homogeneous effects based on $\widehat b$ and on our variance estimator $ \widehat\sigma^2_{\mathrm{ATE}}$.
To be able to make this comparison, we need to assume that all sites have the same number of units, that potential outcomes are normally distributed within sites, and that sites' effects are normally distributed across sites:
\begin{assumption}[Gaussian comparison experiment]\label{ass:gaussian}
For each $n$:
\begin{enumerate}[label=(\roman*)]
\item $N_{d,i}=N_d$ for $d=0,1$, where $N_d\geq4$ is a fixed even integer.
\item Conditional on $S_i$, potential outcomes have Gaussian marginals
with means $\mu_d(S_i)$ and variances $\omega_d^2>0$, constant across
sites.
\item The site ATE is
$\mu_Y(S_i)=\mu+\sqrt{\theta}\,Z_i$, where $Z_i\sim N(0,1)$ and
$\theta=\sigma^2_{\mathrm{ATE}}$.
\end{enumerate}
\end{assumption}

Define the full-sample effect-estimation variance $v$, half-sample
variance $r$, and variance $c$ of the estimated correction by
\begin{align}
 v&=\frac{\omega_1^2}{N_1}+\frac{\omega_0^2}{N_0},\qquad r=2v,
 \label{eq:v}\\
 c&=\Var(\widehat V_{Y,i})
 =\frac{2\omega_1^4}{N_1^2(N_1-1)}
  +\frac{2\omega_0^4}{N_0^2(N_0-1)}.
 \label{eq:c}
\end{align}
Write $A_i=\widehat\mu_Y^1(S_i)$, $C_i=\widehat\mu_Y^2(S_i)$,
$\widehat X_i=A_i-\bar A$, $\widehat Z_i=C_i-\bar C$, and
\[
 \widehat a_A=\frac1n\sum_i\widehat X_i^2,\quad
 \widehat a_C=\frac1n\sum_i\widehat Z_i^2,\quad
 \widehat q=\frac1n\sum_i\widehat X_i\widehat Z_i.
\]
A feasible influence-function estimate for \eqref{eq:average} is
\begin{align}
 \widehat\psi_{b,i}
 &=\frac12\left(\widehat a_A^{-1}+\widehat a_C^{-1}\right)
       (\widehat X_i\widehat Z_i-\widehat q)\notag\\
 &\quad-\frac{\widehat q}{2}
 \left\{\frac{\widehat X_i^2-\widehat a_A}{\widehat a_A^2}
       +\frac{\widehat Z_i^2-\widehat a_C}{\widehat a_C^2}\right\}.
 \label{eq:if-b-hat}
\end{align}
For our estimator, use
\begin{equation}\label{eq:if-v-hat}
 \widehat\psi_{\sigma^2,i}
 =(\widehat\mu_Y(S_i)-\overline{\widehat\mu}_Y)^2
 -\widehat V_{Y,i}-\widehat\sigma^2_{\mathrm{ATE}}.
\end{equation}
Let $\widehat V_b=n^{-1}\sum_i\widehat\psi_{b,i}^2$ and
$\widehat V_{\sigma^2}=n^{-1}\sum_i\widehat\psi_{\sigma^2,i}^2$.
These account for the covariance between the reversed-split slopes;
the two slope estimates must not be treated as independent.

\begin{proposition}[Local-power comparison]\label{prop:power}
Consider a sequence of models satisfying Assumption~\ref{ass:gaussian}
with $\theta=\theta_n=\delta/\sqrt n$ for a fixed $\delta\geq0$. Then
\begin{align}
 T_b:=\frac{\sqrt n\,\widehat b}{\sqrt{\widehat V_b}}
 &\ \Rightarrow\ N\!\left(\frac{\delta}{2v},1\right),
 \label{eq:limit-b}\\
 T_{\sigma^2}:=\frac{\sqrt n\,\widehat\sigma^2_{\mathrm{ATE}}}
 {\sqrt{\widehat V_{\sigma^2}}}
 &\ \Rightarrow\ N\!\left(\frac{\delta}{\sqrt{2v^2+c}},1\right).
 \label{eq:limit-v}
\end{align}
Both tests that reject homogeneity when their statistic exceeds
$z_{1-\alpha}=\Phi^{-1}(1-\alpha)$ have asymptotic size $\alpha$.
Their limiting powers are
\begin{align*}
 \pi_b(\delta)&=1-\Phi\left(z_{1-\alpha}-\frac{\delta}{2v}\right),\\
 \pi_{\sigma^2}(\delta)&=1-\Phi\left(z_{1-\alpha}
                                  -\frac{\delta}{\sqrt{2v^2+c}}\right).
\end{align*}
For every $\delta>0$, $\pi_{\sigma^2}(\delta)>\pi_b(\delta)$.
\end{proposition}
Proposition~\ref{prop:power} shows that, under Assumption \ref{ass:gaussian}, a test of homogeneous effects based on our corrected variance estimator has strictly greater local
power than the averaged-BLP test for every $\delta>0$. Under Assumption~\ref{ass:gaussian} $c<2/3v^2$, so $\sqrt{2v^2+c}<2v$, and therefore $\pi_{\sigma^2}(\delta)>\pi_b(\delta)$.

\citet{cddf2025} generally aggregate p-values across splits, rather than
test using \eqref{eq:average}.
Let $T_1,T_2$ be the two BLP t-statistics, and
$p_k=1-\Phi(T_k)$. Under the local alternatives in
Proposition~\ref{prop:power},
\begin{equation}\label{eq:directional}
 T_1-T_b=\op(1),\qquad T_2-T_b=\op(1).
\end{equation}
Thus a test based on the unadjusted median of these two p-values, as recommended by \citet{cddf2025}, has
the same limiting power as the averaged-BLP t-test.\footnote{A test based on the adjusted p-value
$\min\{1,2\operatorname{median}(p_1,p_2)\}$ instead is more conservative.}
The equivalence in \eqref{eq:directional} concerns two reversed
directions of one partition. It may not apply to the authors' general procedure using many fresh partitions. It may also not apply
to implementations of their method where the training is done with some regularization.

\section{Proofs}\label{sec:proofs}

\subsection{Proof of Lemma~\ref{lem:corrected}}\label{subsec:proof-lemma-corrected}

Let
\[
 e_{X,i}=\widehat\mu_X(S_i)-\mu_X(S_i),\qquad
 e_{Y,i}=\widehat\mu_Y(S_i)-\mu_Y(S_i).
\]
Because $w(S_i)$ is measurable with respect to $S_i$, weighted expectations
obey $\E_w[R_i]=\E_w\{\E[R_i\mid S_i]\}$ whenever the relevant
expectations exist.  Assumption~\ref{ass:sampling}\ref{ass:sampling-means}
implies that $\E[e_{X,i}\mid S_i]=0$ and
$\E[e_{Y,i}\mid S_i]=0$.  Therefore,
\begin{align}
&\E_w\!\left[
 \{\widehat\mu_Y(S_i)-\E_w[\mu_Y(S)]\}^2-\widehat V_{Y,i}
\right]\nonumber\\
&\quad=\E_w\!\left[
 \{\mu_Y(S_i)-\E_w[\mu_Y(S)]\}^2
 +\Var(\widehat\mu_Y(S_i)\mid S_i)-\widehat V_{Y,i}
\right]\nonumber\\
&\quad=\Var_w(\mu_Y(S)),
\label{eq:corr-delta-proof}
\end{align}
where the last equality uses the conditional unbiasedness of
$\widehat V_{Y,i}$.  The matrix-valued analogue proves \eqref{eq:corr-x}.
Similarly,
\begin{align}
&\E_w\!\left[
 \{\widehat\mu_X(S_i)-\E_w[\mu_X(S)]\}
 \{\widehat\mu_Y(S_i)-\E_w[\mu_Y(S)]\}
 -\widehat C_{XY,i}
\right]\nonumber\\
&\quad=\E_w\!\left[
 \{\mu_X(S_i)-\E_w[\mu_X(S)]\}
 \{\mu_Y(S_i)-\E_w[\mu_Y(S)]\}
 +\Cov(\widehat\mu_X(S_i),\widehat\mu_Y(S_i)\mid S_i)
 -\widehat C_{XY,i}
\right]\nonumber\\
&\quad=\Cov_w(\mu_X(S),\mu_Y(S)),
\label{eq:corr-xdelta-proof}
\end{align}
where the last equality uses the conditional unbiasedness of
$\widehat C_{XY,i}$.  This proves \eqref{eq:corr-delta}--\eqref{eq:corr-xdelta}.

To prove \eqref{eq:corr-third}, let
\[
 e_i=\widehat\mu_Y(S_i)-\mu_Y(S_i),\qquad
 d_i=\mu_Y(S_i)-\E_w[\mu_Y(S)].
\]
Conditional on $S_i$, $\E[e_i\mid S_i]=0$, so
\[
 \E[(e_i+d_i)^3\mid S_i]
 =d_i^3+3d_i\Var(\widehat\mu_Y(S_i)\mid S_i)+\E[e_i^3\mid S_i].
\]
Moreover,
\[
 \E[(e_i+d_i)\widehat V_{Y,i}\mid S_i]
 =d_i\Var(\widehat\mu_Y(S_i)\mid S_i)
 +\Cov(\widehat\mu_Y(S_i),\widehat V_{Y,i}\mid S_i).
\]
For brevity, write
\[
 v_i=\Var(\widehat\mu_Y(S_i)\mid S_i),\qquad
 c_i=\Cov(\widehat\mu_Y(S_i),\widehat V_{Y,i}\mid S_i),\qquad
 k_i=\E[e_i^3\mid S_i].
\]
The conditional unbiasedness restrictions then imply
\begin{align*}
&\E\!\left[
 (e_i+d_i)^3-3(e_i+d_i)\widehat V_{Y,i}
 +3\widehat C_{YV,i}-\widehat K_{Y,i}
 \mathrel{\Big|}S_i\right]\\
&\qquad=d_i^3+3d_iv_i+k_i-3(d_iv_i+c_i)+3c_i-k_i\\
&\qquad=d_i^3.
\end{align*}
Taking a weighted expectation of the previous display, the result follows.

\begin{lemma}[Conditional product representation]\label{lem:rct-product-law}
For each $d$, order the indices assigned to arm $d$ as
$U_{d,1}<\cdots<U_{d,N_{d,i}}$. Conditional on $S_i$, the collections
\[
 \big(\bm O_{U_{d,1},i}(d),\ldots,
       \bm O_{U_{d,N_{d,i}},i}(d)\big),\qquad d=0,\ldots,\mathcal D,
\]
are mutually independent, and the collection for arm $d$ consists of
$N_{d,i}$ i.i.d. draws from the conditional marginal distribution of
$\bm O_i(d)$ given $S_i$.
\end{lemma}
\begin{proof}
Write $\bm D_i=(D_{u,i})_{u=1}^{N_i}$ and let
$\bm P_i=((\bm O_{u,i}(d))_{d=0}^{\mathcal D})_{u=1}^{N_i}$
collect the sampled potential outcomes. Fix a feasible assignment vector
$\bm d$, and let $u_{d,1}<\cdots<u_{d,N_{d,i}}$ be the indices it assigns
to arm $d$. For arbitrary Borel sets $A_{d,j}\subseteq\R^L$, define
\[
 E_{\bm d}:=\bigcap_{d=0}^{\mathcal D}\bigcap_{j=1}^{N_{d,i}}
 \{\bm O_{u_{d,j},i}(d)\in A_{d,j}\},
 \qquad
 q_i:=\binom{N_i}{N_{0,i},\ldots,N_{\mathcal D,i}}^{-1}.
\]
Since the arm sizes are measurable with respect to $S_i$, we have
\begin{align*}
 &\Pp\!\left(
 \bigcap_{d=0}^{\mathcal D}\bigcap_{j=1}^{N_{d,i}}
 \{\bm O_{U_{d,j},i}(d)\in A_{d,j}\}
 \mathrel{\Big|}\bm D_i=\bm d,S_i\right)\\
 &\quad=\Pp\!\left(
 \bigcap_{d=0}^{\mathcal D}\bigcap_{j=1}^{N_{d,i}}
 \{\bm O_{u_{d,j},i}(d)\in A_{d,j}\}
 \mathrel{\Big|}\bm D_i=\bm d,S_i\right)\\
 &\quad=\frac{\Pp(E_{\bm d},\bm D_i=\bm d\mid S_i)}
 {\Pp(\bm D_i=\bm d\mid S_i)}\\
 &\quad=\frac{\E[\mathbf 1\{E_{\bm d}\}
 \Pp(\bm D_i=\bm d\mid\bm P_i,S_i)\mid S_i]}
 {\E[\Pp(\bm D_i=\bm d\mid\bm P_i,S_i)\mid S_i]}\\
 &\quad=\frac{q_i\Pp(E_{\bm d}\mid S_i)}{q_i}
 =\prod_{d=0}^{\mathcal D}\prod_{j=1}^{N_{d,i}}
 \Pp(\bm O_i(d)\in A_{d,j}\mid S_i).
\end{align*}
The first equality uses $U_{d,j}=u_{d,j}$ on the event
$\bm D_i=\bm d$. The second is the definition of conditional probability,
and the third follows by iterated expectations, since $E_{\bm d}$ is
measurable with respect to $(\bm P_i,S_i)$. The fourth uses
Assumption~\ref{hyp:strat_completely_randomized}, which makes the
conditional assignment probability equal to $q_i$. Finally, all indices
$u_{d,j}$ are distinct, so Point~2 of Assumption~\ref{hyp:iid} gives the
product in the last equality. This product does not depend on $\bm d$.
Averaging over the assignment conditional on $S_i$ therefore gives the
same factorization conditional on $S_i$ alone, proving both mutual
independence across arms and i.i.d. sampling within each arm.
\end{proof}

\begin{lemma}[Normalized-weight expansion]
\label{lem:normalized-weight-expansion}
Let $(u_i,Z_i)_{i=1}^n$ be i.i.d., where $u_i>0$, $Z_i\in\R^d$ for
some fixed $d$, and $\E[u_i]=1$. Define
\[
 \overline u=\frac1n\sum_{i=1}^n u_i,
 \qquad
 \widehat\theta=\frac1n\sum_{i=1}^n\frac{u_i}{\overline u}Z_i,
 \qquad
 \theta=\E[u_iZ_i].
\]
If $\E[\norm{u_i(Z_i-\theta)}^2]<\infty$, then
\begin{equation}\label{eq:normalized-weight-expansion}
 \sqrt n(\widehat\theta-\theta)
 =\frac1{\sqrt n}\sum_{i=1}^n u_i(Z_i-\theta)+\op(1).
\end{equation}
\end{lemma}

\begin{proof}
Because $n^{-1}\sum_i(u_i/\overline u)=1$, the following identity is exact:
\[
 \sqrt n(\widehat\theta-\theta)
 =\frac1{\overline u}\frac1{\sqrt n}
  \sum_{i=1}^n u_i(Z_i-\theta).
\]
The weak law of large numbers gives $\overline u\overset{\Pp}{\longrightarrow}1$,
while the finite-second-moment condition and the multivariate central limit
theorem give
$n^{-1/2}\sum_i u_i(Z_i-\theta)=\Op(1)$. Hence replacing
$1/\overline u$ by one changes the right-hand side by $\op(1)$, proving
\eqref{eq:normalized-weight-expansion}.
\end{proof}

\subsection{Proof of Theorem~\ref{thm:sigma}}\label{subsec:proof-theorem-sigma}

For brevity, let
\[
 m_Y=\E_w[\mu_Y(S)],\qquad
 u_i=w(S_i),\qquad \overline u=\frac1n\sum_{i=1}^nu_i.
\]
Then, $\widehat w(S_i)=u_i/\overline u$.
By Assumption~\ref{ass:sampling}\ref{ass:sampling-means} and the law of
iterated expectations,
\[
 \E[u_i\widehat\mu_Y(S_i)]
 =\E\!\left[u_i\E\{\widehat\mu_Y(S_i)\mid S_i\}\right]
 =\E[u_i\mu_Y(S_i)]
 =m_Y.
\]
For every $i$,
\[
 \widehat\mu_Y(S_i)-\overline{\widehat\mu}_Y
 =\{\widehat\mu_Y(S_i)-m_Y\}
  -\{\overline{\widehat\mu}_Y-m_Y\}.
\]
Therefore,
\begin{align*}
&\frac1n\sum_{i=1}^n\widehat w(S_i)
 \{\widehat\mu_Y(S_i)-\overline{\widehat\mu}_Y\}^2\\
=&\frac1n\sum_{i=1}^n\widehat w(S_i)
 \Big[
  \{\widehat\mu_Y(S_i)-m_Y\}^2
  -2\{\widehat\mu_Y(S_i)-m_Y\}
    \{\overline{\widehat\mu}_Y-m_Y\}
  +\{\overline{\widehat\mu}_Y-m_Y\}^2
  \Big]\\
=&\frac1n\sum_{i=1}^n\widehat w(S_i)
  \{\widehat\mu_Y(S_i)-m_Y\}^2
 -2\{\overline{\widehat\mu}_Y-m_Y\}
   \frac1n\sum_{i=1}^n\widehat w(S_i)
   \{\widehat\mu_Y(S_i)-m_Y\}
 +\{\overline{\widehat\mu}_Y-m_Y\}^2\\
=&\frac1n\sum_{i=1}^n\widehat w(S_i)
  \{\widehat\mu_Y(S_i)-m_Y\}^2
 -d_Y^2,
\end{align*}
where $d_Y=\overline{\widehat\mu}_Y-m_Y.$
Subtracting $n^{-1}\sum_{i=1}^n\widehat w(S_i)\widehat V_{Y,i}$ from both sides and
using the definition of $\widehat\sigma^2$ gives
\begin{align}
 \widehat\sigma^2
 &=\frac1n\sum_{i=1}^n\widehat w(S_i)
 \left[
  \{\widehat\mu_Y(S_i)-m_Y\}^2
  -\widehat V_{Y,i}
 \right]
 -d_Y^2.
 \label{eq:sigma-exact}
\end{align}
Apply Lemma~\ref{lem:normalized-weight-expansion} first with
$Z_i=\widehat\mu_Y(S_i)$. Because
$\E[u_i\widehat\mu_Y(S_i)]=m_Y$, its moment condition is
$\E[u_i^2\{\widehat\mu_Y(S_i)-m_Y\}^2]<\infty$, which follows from
Assumption~\ref{ass:sigma}\ref{ass:sigma-mom1}. Therefore,
$d_Y=\overline{\widehat\mu}_Y-m_Y=\Op(n^{-1/2})$.
Next apply the lemma with
\[
 Z_i=\{\widehat\mu_Y(S_i)-m_Y\}^2-\widehat V_{Y,i}.
\]
Lemma~\ref{lem:corrected} gives $\E[u_iZ_i]=\sigma^2$, and
Assumption~\ref{ass:sigma}\ref{ass:sigma-mom2} is the required
second-moment condition. Hence
\[
 \sqrt n\left\{
 \frac1n\sum_{i=1}^n\widehat w(S_i)Z_i-\sigma^2
 \right\}
 =\frac1{\sqrt n}\sum_{i=1}^n\psi_{\sigma^2,i}+\op(1).
\]
Combining this display with \eqref{eq:sigma-exact} and
$\sqrt n\,d_Y^2=\op(1)$ proves \eqref{eq:AL-sigma}. The stated normal limit then follows from
\eqref{eq:AL-sigma}, Assumption~\ref{ass:sampling}\ref{ass:sampling-iid},
Assumption~\ref{ass:sigma}\ref{ass:sigma-mom2}, and the i.i.d. central limit
theorem and Slutsky's theorem.

To prove consistency of $\widehat V_{\sigma^2}$, write
\[
 d_\sigma=\widehat\sigma^2-\sigma^2,
 \qquad
 A_i=\widehat\mu_Y(S_i)-m_Y.
\]
Using \eqref{eq:psihat-sigma} and \eqref{eq:psi-sigma},
\begin{align}
 \widehat\psi_{\sigma^2,i}-\psi_{\sigma^2,i}
 &=\frac{u_i}{\overline u}\left[
   \{\widehat\mu_Y(S_i)-\overline{\widehat\mu}_Y\}^2
   -\widehat V_{Y,i}-\widehat\sigma^2\right]\nonumber\\
 &\quad-\left[
   u_i\left[
   \{\widehat\mu_Y(S_i)-m_Y\}^2
   -\widehat V_{Y,i}-\sigma^2\right]
  \right]\nonumber\\
 &=\frac{u_i}{\overline u}
   \left[(A_i-d_Y)^2-\widehat V_{Y,i}
   -(\sigma^2+d_\sigma)\right]
   -u_i\left[A_i^2-\widehat V_{Y,i}
   -\sigma^2\right]\nonumber\\
 &=\frac{u_i}{\overline u}
   \left[A_i^2-2d_Y A_i+d_Y^2
   -\widehat V_{Y,i}-\sigma^2-d_\sigma\right]
   -u_i\left[A_i^2-\widehat V_{Y,i}
   -\sigma^2\right]\nonumber\\
 &=\left(\frac{u_i}{\overline u}-u_i\right)
   \left[A_i^2-\widehat V_{Y,i}-\sigma^2\right]
   +\frac{u_i}{\overline u}
   \left[-2d_Y A_i+d_Y^2-d_\sigma\right]\nonumber\\
 &=\left(\frac1{\overline u}-1\right)
   \psi_{\sigma^2,i}
   +\frac{u_i}{\overline u}
   \left[-2d_Y A_i+d_Y^2-d_\sigma\right].\label{eq:psi-sigma-difference}
\end{align}
Let
$r_i=\widehat\psi_{\sigma^2,i}-\psi_{\sigma^2,i}$. From
$(a_1+a_2+a_3+a_4)^2\leq4\sum_{j=1}^4a_j^2$ and
\eqref{eq:psi-sigma-difference},
\begin{align*}
 \frac1n\sum_{i=1}^n r_i^2
 &\leq4\left(\frac1{\overline u}-1\right)^2
      \frac1n\sum_{i=1}^n\psi_{\sigma^2,i}^2
      +\frac{16d_Y^2}{\overline u^2}
       \frac1n\sum_{i=1}^nu_i^2A_i^2\\
 &\qquad
      +\frac{4d_Y^4}{\overline u^2}
       \frac1n\sum_{i=1}^nu_i^2
      +\frac{4d_\sigma^2}{\overline u^2}
       \frac1n\sum_{i=1}^nu_i^2.
\end{align*}
Assumption~\ref{ass:sigma}\ref{ass:sigma-mom1} and the weak law of large
numbers imply
\[
 \frac1n\sum_{i=1}^n u_i^2A_i^2
 \overset{\Pp}{\longrightarrow}\E[u_i^2A_i^2]<\infty,\qquad
 \frac1n\sum_{i=1}^nu_i^2
 \overset{\Pp}{\longrightarrow}\E[u_i^2]<\infty.
\]
Moreover, $d_Y=\op(1)$ by Assumption~\ref{ass:sigma}\ref{ass:sigma-mom1} and the weak law of large numbers and
$d_\sigma=\op(1)$ by \eqref{eq:AL-sigma}. Also,
$\overline u\overset{\Pp}{\longrightarrow}1$ and
$n^{-1}\sum_{i=1}^n\psi_{\sigma^2,i}^2=\Op(1)$ by the weak law of large
numbers and
Assumption~\ref{ass:sigma}\ref{ass:sigma-mom1} and Assumption~\ref{ass:sigma}\ref{ass:sigma-mom2}, respectively. Consequently, every term on the
right-hand side of the preceding bound is $\op(1)$, and hence
\begin{equation}\label{eq:l2-sigma}
 \frac1n\sum_{i=1}^n
 \{\widehat\psi_{\sigma^2,i}
   -\psi_{\sigma^2,i}\}^2=\op(1).
\end{equation}
Next,
\begin{align*}
&\left|
 \frac1n\sum_{i=1}^n\widehat\psi_{\sigma^2,i}^2
 -\frac1n\sum_{i=1}^n\psi_{\sigma^2,i}^2
 \right|\\
&=\left|
 \frac1n\sum_{i=1}^n
 r_i\{2\psi_{\sigma^2,i}+r_i\}
 \right|\\
&\leq
 2\left(\frac1n\sum_{i=1}^n r_i^2\right)^{1/2}
  \left(\frac1n\sum_{i=1}^n\psi_{\sigma^2,i}^2\right)^{1/2}
 +\frac1n\sum_{i=1}^n r_i^2,
\end{align*}
where the inequality follows from the triangle inequality and Cauchy--Schwarz. By
Assumption~\ref{ass:sigma}\ref{ass:sigma-mom2} and the weak law of large
numbers,
\[
 \frac1n\sum_{i=1}^n\psi_{\sigma^2,i}^2
 \overset{\Pp}{\longrightarrow}
 \E[\psi_{\sigma^2,i}^2]<\infty.
\]
Together with \eqref{eq:l2-sigma}, this proves that
\[
 \frac1n\sum_{i=1}^n\widehat\psi_{\sigma^2,i}^2
 -\frac1n\sum_{i=1}^n\psi_{\sigma^2,i}^2=\op(1).
\]
Therefore,
$\widehat V_{\sigma^2}\overset{\Pp}{\longrightarrow}
V_{\sigma^2}$.

\subsection{Proof of Theorem~\ref{thm:beta}}\label{subsec:proof-theorem-beta}

For brevity, let
$u_i=w(S_i)$ and $\overline u=n^{-1}\sum_{i=1}^nu_i$. Then,
$\widehat w(S_i)=u_i/\overline u$. Let
\begin{align*}
 \widetilde{\Var}(\mu_X(S))
 &=\frac1n\sum_{i=1}^nu_i
 \left[
  \{\widehat\mu_X(S_i)-\E_w[\mu_X(S)]\}
  \{\widehat\mu_X(S_i)-\E_w[\mu_X(S)]\}\trans-\widehat V_{X,i}
 \right],\\
 \widetilde{\Cov}(\mu_X(S),\mu_Y(S))
 &=\frac1n\sum_{i=1}^nu_i
 \left[
  \{\widehat\mu_X(S_i)-\E_w[\mu_X(S)]\}
  \{\widehat\mu_Y(S_i)-\E_w[\mu_Y(S)]\}
  -\widehat C_{XY,i}
 \right].
\end{align*}
Write
\[
d_X=\overline{\widehat\mu}_X-\E_w[\mu_X(S)],\quad
d_Y=\overline{\widehat\mu}_Y-\E_w[\mu_Y(S)],
\]
\[
R_X=\frac1n\sum_{i=1}^nu_i
\{\widehat\mu_X(S_i)-\E_w[\mu_X(S)]\},\quad
R_Y=\frac1n\sum_{i=1}^nu_i
\{\widehat\mu_Y(S_i)-\E_w[\mu_Y(S)]\}.
\]
Then $d_X=R_X/\overline u$ and $d_Y=R_Y/\overline u$.
Expanding the centered products, one has
\begin{align}
 \widehat{\Var}(\mu_X(S))
 &=\frac{1}{\overline u}\widetilde{\Var}(\mu_X(S))
 -d_Xd_X\trans,
 \label{eq:varx-exact}\\
 \widehat{\Cov}(\mu_X(S),\mu_Y(S))
 &=\frac{1}{\overline u}
 \widetilde{\Cov}(\mu_X(S),\mu_Y(S))-d_Xd_Y.
 \label{eq:cov-exact}
\end{align}
Applying Lemma~\ref{lem:normalized-weight-expansion} with the stacked vector
$Z_i=(\widehat\mu_X(S_i)\trans,\widehat\mu_Y(S_i))\trans$,
Assumption~\ref{ass:sampling}\ref{ass:sampling-means} gives
\[
 \E[u_iZ_i]=m_Z,
 \qquad
 m_Z:=\big(\E_w[\mu_X(S)]\trans,\E_w[\mu_Y(S)]\big)\trans,
\]
while Assumption~\ref{ass:beta}\ref{ass:beta-second} implies the lemma's
moment condition
\[
 \E[\norm{u_i(Z_i-m_Z)}^2]<\infty.
\]
Therefore, the lemma gives
\[
d_X=\Op(n^{-1/2}),\qquad d_Y=\Op(n^{-1/2}).
\]

We first establish consistency of $\widehat{\Var}(\mu_X(S))$. Let
\[
 Z_i^{XX}=\operatorname{vec}\!\left(
 \{\widehat\mu_X(S_i)-\E_w[\mu_X(S)]\}
 \{\widehat\mu_X(S_i)-\E_w[\mu_X(S)]\}\trans-\widehat V_{X,i}
 \right),
\]
where $\operatorname{vec}(A)$ stacks the columns of $A$ into a column vector.
Fix $(j,k)\in\{1,\ldots,K\}^2$ and write
$m_{X,j}=\E_w[\mu_{X,j}(S)]$. The $(j,k)$ entry of
$\widetilde{\Var}(\mu_X(S))$ is the sample average of
\[
 Z_{i,jk}
 =u_i\left[\{\widehat\mu_{X,j}(S_i)-m_{X,j}\}
  \{\widehat\mu_{X,k}(S_i)-m_{X,k}\}
  -\widehat V_{X,i,jk}\right].
\]
This random variable is integrable. Indeed,
\begin{align*}
 |Z_{i,jk}|
 &\leq u_i\left[
 \left|\widehat\mu_{X,j}(S_i)-m_{X,j}\right|
 \left|\widehat\mu_{X,k}(S_i)-m_{X,k}\right|
 +|\widehat V_{X,i,jk}|\right]\\
 &\leq \frac{u_i}{2}
 \left\{\widehat\mu_{X,j}(S_i)-m_{X,j}\right\}^2
 +\frac{u_i}{2}
 \left\{\widehat\mu_{X,k}(S_i)-m_{X,k}\right\}^2
 +u_i|\widehat V_{X,i,jk}|,
\end{align*}
where the second inequality uses $|ab|\leq(a^2+b^2)/2$. The expectation of
the right-hand side is finite by Assumption~\ref{ass:beta}\ref{ass:beta-second}
and Assumption~\ref{ass:beta}\ref{ass:beta-vx-first}. Therefore, the weak law
of large numbers and Lemma~\ref{lem:corrected} imply, for every $(j,k)$,
\[
 \widetilde{\Var}(\mu_X(S))_{jk}
 \overset{\Pp}{\longrightarrow}
 \Var_w(\mu_X(S))_{jk}.
\]
Equivalently,
\[
 \operatorname{vec}\{\widetilde{\Var}(\mu_X(S))\}
 \overset{\Pp}{\longrightarrow}
 \operatorname{vec}\{\Var_w(\mu_X(S))\}.
\]
Because $K$ is fixed, this implies convergence of the whole matrix. Moreover,
$\overline u^{-1}=1+\op(1)$ and each entry of
$d_Xd_X\trans$ is $\Op(n^{-1})$. Hence
\begin{align}
 \widehat{\Var}(\mu_X(S))
 &\overset{\Pp}{\longrightarrow}\Var_w(\mu_X(S)).
 \label{eq:varx-cons}
\end{align}
Assumption~\ref{ass:beta}\ref{ass:beta-ident}, continuity of the determinant,
and \eqref{eq:varx-cons} imply that
$\widehat{\Var}(\mu_X(S))$ is nonsingular with probability approaching one.
Continuity of matrix inversion at nonsingular matrices then gives
\[
 \widehat{\Var}(\mu_X(S))^{-1}
 \overset{\Pp}{\longrightarrow}\Var_w(\mu_X(S))^{-1}.
\]

Using the definition of $\beta$, we have the exact identity
\begin{align}
 \widehat\beta-\beta
 &=\widehat{\Var}(\mu_X(S))^{-1}
 \Big[
  \widehat{\Cov}(\mu_X(S),\mu_Y(S))
  -\widehat{\Var}(\mu_X(S))\beta
 \Big].
 \label{eq:beta-exact}
\end{align}
Combining \eqref{eq:varx-exact}, \eqref{eq:cov-exact}, and
\eqref{eq:beta-exact}, and using the $\Op(n^{-1/2})$ rates for
$d_X$ and $d_Y$, gives
\begin{align}
 \sqrt n(\widehat\beta-\beta)
 &=\widehat{\Var}(\mu_X(S))^{-1}\frac1{\overline u}
 \frac1{\sqrt n}\sum_{i=1}^n
 u_i\Big[
  \{\widehat\mu_X(S_i)-\E_w[\mu_X(S)]\}
  \notag\\[-2mm]
 &\hspace{35mm}\times
  \big\{\widehat\mu_Y(S_i)-\E_w[\mu_Y(S)]
    -\{\widehat\mu_X(S_i)-\E_w[\mu_X(S)]\}\trans\beta\big\}\notag\\
 &\hspace{35mm}
  -\widehat C_{XY,i}+\widehat V_{X,i}\beta
   \Big]+\op(1).\label{eq:beta-linalmost}
\end{align}
Let $H_i$ denote the bracketed vector in \eqref{eq:beta-linalmost}.
Lemma~\ref{lem:corrected} and the definition of $\beta$ give
$\E[u_iH_i]=0$. Moreover,
Assumption~\ref{ass:beta}\ref{ass:beta-ifmoment} ensures the moment condition
of Lemma~\ref{lem:normalized-weight-expansion}. Applying that lemma with
$Z_i=H_i$ gives
\[
 \frac1{\overline u}\frac1{\sqrt n}\sum_{i=1}^nu_iH_i
 =\frac1{\sqrt n}\sum_{i=1}^nu_iH_i+\op(1).
\]
We also have
\[
\widehat{\Var}(\mu_X(S))^{-1}
=
 \Var_w(\mu_X(S))^{-1}
+
o_{\Pp}(1),
\]
and
\[
\frac{1}{\sqrt n}\sum_{i=1}^n\psi_{\beta,i}
=
O_{\Pp}(1),
\]
where the last display follows from the multivariate central limit theorem and
Assumption~\ref{ass:beta}\ref{ass:beta-ifmoment}. It follows from
\eqref{eq:beta-linalmost}, the preceding displays, and Slutsky's theorem that
\[
\sqrt n(\widehat\beta-\beta)
=
 \frac{1}{\sqrt n}\sum_{i=1}^n\psi_{\beta,i}
+
o_{\Pp}(1),
\]
establishing \eqref{eq:AL-beta}.
The stated normal limit then follows from
\eqref{eq:AL-beta}, Assumption~\ref{ass:sampling}\ref{ass:sampling-iid},
Assumption~\ref{ass:beta}\ref{ass:beta-ifmoment}, the multivariate i.i.d. central limit theorem and Slutsky's theorem.

Now suppose that Assumption~\ref{ass:variance} also holds. We prove consistency of $\widehat V_\beta$. For brevity, let
\[
 m_X=\E_w[\mu_X(S)],\qquad m_Y=\E_w[\mu_Y(S)],\qquad
 M=\Var_w(\mu_X(S)),
\]
and define
\[
 A_i=\widehat\mu_X(S_i)-m_X,\qquad
 B_i=\widehat\mu_Y(S_i)-m_Y,\qquad
 d_\beta=\widehat\beta-\beta, \qquad\widehat M=\widehat{\Var}(\mu_X(S)).
\]
Write
\begin{align*}
 q_i
 &=u_i\left\{A_i(B_i-A_i\trans\beta)
   -\widehat C_{XY,i}+\widehat V_{X,i}\beta\right\},\\
 \widehat q_i
 &=\frac{u_i}{\overline u}\left[
   (A_i-d_X)
   \{B_i-d_Y-(A_i-d_X)\trans\widehat\beta\}
   -\widehat C_{XY,i}+\widehat V_{X,i}\widehat\beta
   \right].
\end{align*}
Then $\psi_{\beta,i}=M^{-1}q_i$ and
$\widehat\psi_{\beta,i}=\widehat M^{-1}\widehat q_i$. Expanding and
collecting terms gives
\begin{align}
 \widehat q_i-q_i
 &=\left(\frac1{\overline u}-1\right)q_i
   +\frac{u_i}{\overline u}\big[
   -d_XB_i-d_Y A_i+d_Xd_Y
   +A_i d_X\trans\beta+d_XA_i\trans\beta-d_Xd_X\trans\beta
 \notag\\
 &\quad+
 \{\widehat V_{X,i}-(A_i-d_X)(A_i-d_X)\trans\}d_\beta
 \big].
 \label{eq:qbeta-difference}
\end{align}
We show componentwise that the empirical mean squared difference between
$\widehat q_i$ and $q_i$ converges to zero. Fix
$j\in\{1,\ldots,K\}$. The $j$th component of
\eqref{eq:qbeta-difference} equals
\begin{align*}
 \widehat q_{i,j}-q_{i,j}
 &=\left(\frac1{\overline u}-1\right)q_{i,j}
   +\frac{u_i}{\overline u}\Big[
   -d_{X,j}B_i-d_Y A_{i,j}+d_{X,j}d_Y
   +A_{i,j}d_X\trans\beta+d_{X,j}A_i\trans\beta
   -d_{X,j}d_X\trans\beta\\
 &\quad+
 \sum_{k=1}^K
 \left\{\widehat V_{X,i,jk}
 -(A_{i,j}-d_{X,j})(A_{i,k}-d_{X,k})\right\}d_{\beta,k}
 \Big].
\end{align*}
Since $K$ is fixed, repeated use of
$(a_1+\cdots+a_J)^2\leq J\sum_{\ell=1}^J a_\ell^2$ gives a finite constant
$C$, independent of $n$, such that
\begin{align*}
\frac1n\sum_{i=1}^n(\widehat q_{i,j}-q_{i,j})^2
 &\leq C\Bigg[
 \left(\frac1{\overline u}-1\right)^2
 \frac1n\sum_{i=1}^nq_{i,j}^2\notag\\
&\qquad+\frac1{\overline u^2}\left[
 d_{X,j}^2\frac1n\sum_{i=1}^n u_i^2B_i^2
 +d_Y^2\frac1n\sum_{i=1}^n u_i^2A_{i,j}^2
 +d_{X,j}^2d_Y^2\frac1n\sum_{i=1}^nu_i^2\right]\notag\\
&\qquad
 +\frac1{\overline u^2}\left[
 (d_X\trans\beta)^2\frac1n\sum_{i=1}^n u_i^2A_{i,j}^2
 +d_{X,j}^2\frac1n\sum_{i=1}^nu_i^2(A_i\trans\beta)^2
 +d_{X,j}^2(d_X\trans\beta)^2\frac1n\sum_{i=1}^nu_i^2\right]\notag\\
&\qquad
 +\frac1{\overline u^2}\sum_{k=1}^K d_{\beta,k}^2
 \frac1n\sum_{i=1}^nu_i^2
 \left\{\widehat V_{X,i,jk}
 -(A_{i,j}-d_{X,j})(A_{i,k}-d_{X,k})\right\}^2
 \Bigg].
\end{align*}
Assumption~\ref{ass:beta}\ref{ass:beta-second} and the weak law of large
numbers imply
\[
 \frac1n\sum_{i=1}^n u_i^2B_i^2=\Op(1),\qquad
 \frac1n\sum_{i=1}^n u_i^2A_{i,j}^2=\Op(1),\qquad\frac1n\sum_{i=1}^n u_i^2=\Op(1)\qquad
 \frac1n\sum_{i=1}^nu_i^2(A_i\trans\beta)^2=\Op(1).
\]
For every $j,k$, the inequality $(a-b)^2\leq2a^2+2b^2$ yields
\begin{align*}
&\frac1n\sum_{i=1}^nu_i^2
 \left\{\widehat V_{X,i,jk}
 -(A_{i,j}-d_{X,j})(A_{i,k}-d_{X,k})\right\}^2\\
&\qquad\leq
 2\frac1n\sum_{i=1}^nu_i^2\widehat V_{X,i,jk}^2
 +2\frac1n\sum_{i=1}^nu_i^2
 (A_{i,j}-d_{X,j})^2(A_{i,k}-d_{X,k})^2.
\end{align*}
The first term is $\Op(1)$ by Assumption~\ref{ass:variance} and the weak law
of large numbers. For the second term, using
$(x-y)^2\leq2x^2+2y^2$ twice and $2uv\leq u^2+v^2$ gives
\begin{align*}
 &(A_{i,j}-d_{X,j})^2(A_{i,k}-d_{X,k})^2\\
 &\qquad\leq
 4A_{i,j}^2A_{i,k}^2
 +4A_{i,j}^2d_{X,k}^2
 +4d_{X,j}^2A_{i,k}^2
 +4d_{X,j}^2d_{X,k}^2\\
 &\qquad\leq
 2A_{i,j}^4+2A_{i,k}^4
 +4A_{i,j}^2d_{X,k}^2
 +4d_{X,j}^2A_{i,k}^2
 +4d_{X,j}^2d_{X,k}^2.
\end{align*}
Multiplying this inequality by $u_i^2$,
Assumption~\ref{ass:variance}, the weak law of large numbers, and
$d_X=\Op(n^{-1/2})$ therefore imply that the second term
is also $\Op(1)$.
Consequently, for every $j,k$,
\[
 \frac1n\sum_{i=1}^nu_i^2
 \left\{\widehat V_{X,i,jk}
 -(A_{i,j}-d_{X,j})(A_{i,k}-d_{X,k})\right\}^2
 =\Op(1).
\]
Since $q_i=M\psi_{\beta,i}$,
Assumption~\ref{ass:beta}\ref{ass:beta-ifmoment} and the weak law of large
numbers imply $n^{-1}\sum_{i=1}^nq_{i,j}^2=\Op(1)$. Moreover,
$\overline u\overset{\Pp}{\longrightarrow}1$.
Also, $d_X=\Op(n^{-1/2})$, $d_Y=\Op(n^{-1/2})$, and
\eqref{eq:AL-beta} implies $d_\beta=\Op(n^{-1/2})$. Hence, for every $j$,
\begin{equation}
 \frac1n\sum_{i=1}^n(\widehat q_{i,j}-q_{i,j})^2=\op(1).
 \label{eq:l2-qbeta}
\end{equation}

Next fix $j\in\{1,\ldots,K\}$. Since
\[
 \widehat\psi_{\beta,i}-\psi_{\beta,i}
 =\widehat M^{-1}(\widehat q_i-q_i)
  +(\widehat M^{-1}-M^{-1})q_i,
\]
the $j$th component is a finite sum of scalar products. Entrywise convergence
of $\widehat M^{-1}$ to $M^{-1}$, \eqref{eq:l2-qbeta}, and repeated use of
$(a_1+\cdots+a_K)^2\leq K\sum_{k=1}^K a_k^2$ imply
\begin{equation}
 \frac1n\sum_{i=1}^n
 (\widehat\psi_{\beta,i,j}-\psi_{\beta,i,j})^2=\op(1).
 \label{eq:l2-beta}
\end{equation}
Finally, fix $(j,k)\in\{1,\ldots,K\}^2$ and let
$r_{i,j}=\widehat\psi_{\beta,i,j}-\psi_{\beta,i,j}$. Then
\begin{align}
&\frac1n\sum_{i=1}^n
 \widehat\psi_{\beta,i,j}\widehat\psi_{\beta,i,k}
 -\frac1n\sum_{i=1}^n
 \psi_{\beta,i,j}\psi_{\beta,i,k}\nonumber\\
=&
 \frac1n\sum_{i=1}^n r_{i,j}\psi_{\beta,i,k}
 +\frac1n\sum_{i=1}^n \psi_{\beta,i,j}r_{i,k}
 +\frac1n\sum_{i=1}^n r_{i,j}r_{i,k}\label{eq:l3-beta}\\
=&\op(1)\label{eq:l4-beta}.
\end{align}
The second equality follows from the fact that by Cauchy--Schwarz, the absolute value of the first term is bounded by
\[
 \left\{\frac1n\sum_{i=1}^n r_{i,j}^2\right\}^{1/2}
 \left\{\frac1n\sum_{i=1}^n\psi_{\beta,i,k}^2\right\}^{1/2}.
\]
By
\eqref{eq:l2-beta}, $\frac1n\sum_{i=1}^n r_{i,j}^2=\op(1)$, while
Assumption~\ref{ass:beta}\ref{ass:beta-ifmoment} and the weak law of large
numbers imply
\[
 \frac1n\sum_{i=1}^n\psi_{\beta,i,j}^2=\Op(1)
 \qquad\text{for every }j.
\]
Thus, the first term in \eqref{eq:l3-beta} is an $\op(1)$, and one can use similar steps to show that the two remaining terms are also $\op(1)$.
Moreover, for each
$j,k$, the weak law of large numbers gives
\[
 \frac1n\sum_{i=1}^n\psi_{\beta,i,j}\psi_{\beta,i,k}
 \overset{\Pp}{\longrightarrow}
 \E[\psi_{\beta,i,j}\psi_{\beta,i,k}],
\]
because
$\E[|\psi_{\beta,i,j}\psi_{\beta,i,k}|]<\infty$ by Cauchy--Schwarz and
Assumption~\ref{ass:beta}\ref{ass:beta-ifmoment}. Combined with \eqref{eq:l4-beta}, this
proves that
\[
 \widehat V_\beta\overset{\Pp}{\longrightarrow}V_\beta.
\]

\subsection{Proof of Theorem~\ref{thm:kappa}}\label{subsec:proof-theorem-kappa}
Let
\begin{align*}
 m_Y&=\E_w[\mu_Y(S)],&A_i&=\widehat\mu_Y(S_i)-m_Y,&
 d_Y&=\overline{\widehat\mu}_Y-m_Y,\\
 u_i&=w(S_i),&\overline u&=n^{-1}\sum_{i=1}^nu_i,&
 H_i&=A_i^3-3A_i\widehat V_{Y,i}+3\widehat C_{YV,i}-\widehat K_{Y,i}.
\end{align*}
To derive the exact representation, first expand each centered term in
\eqref{eq:kappa-hat}:
\begin{align*}
\widehat\kappa_Y
&=\frac1n\sum_{i=1}^n\widehat w(S_i)
 \left[H_i-3d_Y\{A_i^2-\widehat V_{Y,i}\}
 +3d_Y^2A_i-d_Y^3\right].
\end{align*}
By the definition of $d_Y$,
\[
 \frac1n\sum_{i=1}^n\widehat w(S_i)A_i=d_Y,
\]
while \eqref{eq:sigma-exact} implies
\[
 \frac1n\sum_{i=1}^n\widehat w(S_i)
 (A_i^2-\widehat V_{Y,i})=\widehat\sigma^2+d_Y^2.
\]
Also, $n^{-1}\sum_i\widehat w(S_i)=1$. Consequently,
\begin{align*}
 \widehat\kappa_Y
 &=\frac1n\sum_{i=1}^n\widehat w(S_i)H_i
 -3d_Y(\widehat\sigma^2+d_Y^2)+3d_Y^3-d_Y^3,
\end{align*}
which simplifies to
\begin{equation}\label{eq:kappa-exact}
 \widehat\kappa_Y=\frac1n\sum_{i=1}^n\widehat w(S_i)H_i
 -3d_Y\widehat\sigma^2-d_Y^3.
\end{equation}
We next derive the linearization. We start by rewriting
\eqref{eq:psi-kappa} with the notation used in the proof:
\begin{align*}
 \psi_{\kappa_Y,i}
 &=u_i\left[
 A_i^3-3A_i\widehat V_{Y,i}
 +3\widehat C_{YV,i}-\widehat K_{Y,i}-\kappa_Y
 -3\sigma^2A_i\right]\\
 &=u_i\{H_i-\kappa_Y-3\sigma^2A_i\}.
\end{align*}
Since $d_Y=\frac{n^{-1}\sum_i u_iA_i}{\overline u}$,
\begin{align*}
 \frac1{\overline u}\frac1n\sum_{i=1}^n\psi_{\kappa_Y,i}
 &=\frac1{\overline u}\frac1n\sum_{i=1}^n
 u_i(H_i-\kappa_Y)-3\sigma^2d_Y.
\end{align*}
Subtracting $\kappa_Y$ from \eqref{eq:kappa-exact} and using this equality,
\begin{align*}
 \widehat\kappa_Y-\kappa_Y
 &=\frac1{\overline u}\frac1n\sum_{i=1}^n\psi_{\kappa_Y,i}
 +3\sigma^2d_Y-3d_Y\widehat\sigma^2-d_Y^3\\
 &=\frac1{\overline u}\frac1n\sum_{i=1}^n\psi_{\kappa_Y,i}
 -3d_Y(\widehat\sigma^2-\sigma^2)-d_Y^3.
\end{align*}
Multiplication by $\sqrt n$ yields
\begin{align}
 \sqrt n(\widehat\kappa_Y-\kappa_Y)
 &=\frac1{\overline u}\frac1{\sqrt n}\sum_{i=1}^n\psi_{\kappa_Y,i}
 -3\sqrt n\,d_Y(\widehat\sigma^2-\sigma^2)
 -\sqrt n\,d_Y^3.\label{eq:kappa-linearization}
\end{align}
Assumption~\ref{ass:kappa}\ref{ass:kappa-mean} implies
$\overline u-1=\Op(n^{-1/2})$ and $d_Y=\Op(n^{-1/2})$.
Lemma~\ref{lem:corrected}, Assumptions~\ref{ass:kappa}\ref{ass:kappa-mean}--
\ref{ass:kappa-second}, the weak law of large numbers, and
\eqref{eq:sigma-exact} imply $\widehat\sigma^2-\sigma^2=\op(1)$.
Assumption~\ref{ass:kappa}\ref{ass:kappa-ifmoment} and the central limit
theorem give $n^{-1/2}\sum_i\psi_{\kappa_Y,i}=\Op(1)$. Thus
\eqref{eq:kappa-linearization} proves \eqref{eq:AL-kappa}; its stated normal
limit follows from the i.i.d. central limit theorem and Slutsky's theorem.

\subsection{Proof of Proposition~\ref{prop:rct-unbiased}}\label{subsec:proof-rct-unbiased}

By Lemma~\ref{lem:rct-product-law}, conditional on $S_i$, each arm contains
an i.i.d. sample from its arm-specific conditional marginal distribution, and
the samples are independent across arms. The usual properties of an i.i.d.
sample therefore apply separately within each arm. In particular, its sample
mean is unbiased and its sample covariance matrix is unbiased for the
population covariance matrix. Dividing the latter by $N_{d,i}$ and stacking
the arms gives
\begin{align*}
 \E[\widehat\mu_O(S_i)\mid S_i]&=\mu_O(S_i),\\
 \E[\widehat V_{O,i}\mid S_i]
 &=\Var(\widehat\mu_O(S_i)\mid S_i),
\end{align*}
where the off-diagonal arm blocks of $\widehat V_{O,i}$ and $\Var(\widehat\mu_O(S_i)\mid S_i)$ are zero by cross-arm
independence. This proves \eqref{eq:unbiased_estimator} and
\eqref{eq:unbiased_variance}.

Then, consider an i.i.d. scalar sample $Z_1,\ldots,Z_m$, let
$\mu=\E[Z_j]$, and define $\kappa_3=\E[(Z_j-\mu)^3]$. One has
\begin{align}
 \E[(\overline Z-\mu)^3]
 &=\frac{\kappa_3}{m^2},\label{eq:mean-third-comoment}\\
 \E\!\left[\frac1m\sum_{j=1}^m(Z_j-\overline Z)^3\right]
 &=\frac{(m-1)(m-2)}{m^2}\kappa_3.
 \label{eq:sample-third-comoment}
\end{align}
The first identity follows by expanding the cube of the sample mean: all
summands with unequal observation indices have expectation zero, leaving
$m\kappa_3/m^3$. The second follows by expanding the cube and collecting the
terms with coincident indices. Now apply \eqref{eq:sample-third-comoment}
within arm $d$, conditionally on $S_i$, to the scalar variable
$Z_j=c_d\trans\bm O_{j,i}(d)$, with $m=N_{d,i}$. By
Lemma~\ref{lem:rct-product-law}, these variables are i.i.d. within arms and
the samples are independent across arms. Therefore,
\[
 \E[\widehat M_{3,O,i}(c)\mid S_i]
 =\sum_{d=0}^{\mathcal D}\frac{1}{N_{d,i}^2}
 \E\!\left[\{c_d\trans(\bm O_i(d)-\E[\bm O_i(d)\mid S_i])\}^3
 \mid S_i\right].
\]
Applying \eqref{eq:mean-third-comoment}, and using conditional independence
across arms from Lemma~\ref{lem:rct-product-law}, it follows that the right-hand side
of the last display equals
\[
 \E\!\left[
 \{c\trans(\widehat\mu_O(S_i)-\mu_O(S_i))\}^3\mid S_i
 \right],
\]
which proves \eqref{eq:unbiased_third_moment} for every non-random $c$.

It remains to verify the statements for the linear transformations. The first
two identities of the proposition, premultiplied and postmultiplied by $B$
and/or $c$, give the conditional-unbiasedness restrictions for
$\widehat\mu_X(S_i)$, $\widehat\mu_Y(S_i)$, $\widehat V_{X,i}$,
$\widehat V_{Y,i}$, and $\widehat C_{XY,i}$. Moreover,
\eqref{eq:unbiased_third_moment} and the definition of $\widehat K_{Y,i}$ yield
\begin{align*}
 \E[\widehat K_{Y,i}\mid S_i]
 =\E[\{\widehat\mu_Y(S_i)-\mu_Y(S_i)\}^3\mid S_i].
\end{align*}
Finally, for an i.i.d. scalar sample $Z_1,\ldots,Z_m$,
\begin{align*}
 &\Cov\!\left(\overline Z,
 \frac{1}{m(m-1)}\sum_{j=1}^m
 (Z_j-\overline Z)^2\right)
 =\frac{\kappa_3}{m^2}.
\end{align*}
Applying the previous display to
$Z_j=c_d\trans\bm O_{j,i}(d)$
shows that
\[
 \Cov(\widehat\mu_Y(S_i),\widehat V_{Y,i}\mid S_i)
 =\E[\widehat M_{3,O,i}(c)\mid S_i]
 =\E[\widehat C_{YV,i}\mid S_i].
\]
Thus all the conditional-unbiasedness restrictions in
Assumption~\ref{ass:sampling} hold. The i.i.d. requirement follows because
all the displayed statistics are fixed functions of the i.i.d. site vectors
and of randomizations conducted independently across sites.

\subsection{Proof of Proposition~\ref{prop:late-variance-identification}}\label{subsec:proof-late-variance-identification}

For brevity, let $F=\mu_{\mathrm{FS}}(S)$,
$T=\mu_{\mathrm{LATE}}(S)$, $l_1=\mathrm{LATE}$, and
$l_2=\mathrm{LATE}_{\mathrm{FS}^2}$.
Then $\mu_{\mathrm{ITT}}(S)=FT$,
$\mathrm{FS}=\E_w[F]$, and $m_2=\E_w[F^2]$.

For Point~1,
\[
 \E_w[\{\mu_{\mathrm{ITT}}(S)-l_2F\}^2]
 =\E_w[F^2(T-l_2)^2]
 =m_2\sigma^2_{\mathrm{LATE},\mathrm{FS}^2}.
\]
Dividing by $m_2>0$ proves the equality.

For Point~2, Assumption~\ref{ass:late} implies $0<F\leq1$ almost surely.
Let $X=F(T-l_1)$, $h=\E_w[X^2]$, $d=\E_w[FX]=r-m_2l_1$,
and $\delta=\mathrm{FS}-m_2>0$. Since $\E_w[X]=0$,
$\E_w[(1-F)X]=-d$. Also,
\[
 \mathrm{FS}\,\sigma^2_{\mathrm{LATE}}
 =\E_w[X^2/F]
 =h+\E_w[(1-F)X^2/F].
\]
If the target variance is infinite, the bound is immediate. Otherwise,
Cauchy--Schwarz gives
\[
 d^2
 =\left\{\E_w\!\left[
 \sqrt{\frac{1-F}{F}}X\sqrt{F(1-F)}
 \right]\right\}^2
 \leq \E_w[(1-F)X^2/F]\,\delta.
\]
Consequently,
$\sigma^2_{\mathrm{LATE}}\geq(h+d^2/\delta)/\mathrm{FS}$.
Moreover, the definition of $l_2$ gives
$\E_w[F^2(T-l_2)]=0$, so
\[
 \E_w[F^2(T-l_1)^2]
 =\E_w[F^2(T-l_2)^2]+m_2(l_2-l_1)^2.
\]
Combining this identity with $d=m_2(l_2-l_1)$ proves the final equality in
\eqref{eq:late-variance-lower-bound}.

\subsection{Proof of Theorem~\ref{thm:late-variance}}\label{subsec:proof-theorem-late-variance}

\paragraph{Point 1.}
We first establish consistency and root-$n$ convergence of
$\widehat{\mathrm{LATE}}_{\mathrm{FS}^2}$. Conditional unbiasedness gives
\[
 \E_w\!\left[\widehat\mu_{\mathrm{FS}}(S_i)
 \widehat\mu_{\mathrm{ITT}}(S_i)
 -\widehat C_{\mathrm{FS},\mathrm{ITT},i}\right]
 =\E_w[\mu_{\mathrm{FS}}(S)\mu_{\mathrm{ITT}}(S)]
\]
and
\[
 \E_w[\widehat\mu_{\mathrm{FS}}(S_i)^2-\widehat V_{\mathrm{FS},i}]
 =\E_w[\mu_{\mathrm{FS}}(S)^2].
\]
The normalized-weight expansion and the second-moment conditions therefore imply
that $\widehat r$ and $\widehat m_2$ are root-$n$ consistent for these two
expectations. Since $\E_w[\mu_{\mathrm{FS}}(S)^2]>0$, a ratio expansion yields
\begin{equation}\label{eq:rate-late-fs2-proof}
 \widehat{\mathrm{LATE}}_{\mathrm{FS}^2}
 -\mathrm{LATE}_{\mathrm{FS}^2}=\Op(n^{-1/2}).
\end{equation}

We next expand the numerator of
$\widehat\sigma^2_{\mathrm{LATE},\mathrm{FS}^2}$.
By the unbiasedness of the variance and covariance corrections,
\[
 \E[g_i(l)\mid S_i]
 =\{\mu_{\mathrm{ITT}}(S_i)
      -l\mu_{\mathrm{FS}}(S_i)\}^2.
\]
At $l=\mathrm{LATE}_{\mathrm{FS}^2}$, this implies
$\E_w[g_i(\mathrm{LATE}_{\mathrm{FS}^2})]=q$.
Assumption~\ref{ass:late-variance-asymptotics}
allows us to apply Lemma~\ref{lem:normalized-weight-expansion} with
$Z_i=g_i(\mathrm{LATE}_{\mathrm{FS}^2})$:
\begin{equation}\label{eq:AL-Q-lambda-proof}
 \sqrt n\{\widehat Q(\mathrm{LATE}_{\mathrm{FS}^2})-q\}
 =\frac1{\sqrt n}\sum_{i=1}^n
 w(S_i)\{g_i(\mathrm{LATE}_{\mathrm{FS}^2})-q\}+\op(1).
\end{equation}

Differentiating $g_i(l)$ gives
\begin{align*}
 \frac{\partial g_i(l)}{\partial l}
 &=-2\widehat\mu_{\mathrm{FS}}(S_i)
   \{\widehat\mu_{\mathrm{ITT}}(S_i)
      -l\widehat\mu_{\mathrm{FS}}(S_i)\}-2l\widehat V_{\mathrm{FS},i}
       +2\widehat C_{\mathrm{FS},\mathrm{ITT},i}.
\end{align*}
The first and third finite-second-moment conditions in
Assumption~\ref{ass:late-variance-asymptotics} and the preceding display imply
\[
 \E\!\left[\left|w(S_i)
 g_i'(\mathrm{LATE}_{\mathrm{FS}^2})\right|\right]<\infty.
\]
The conditional unbiasedness identities in Proposition~\ref{prop:rct-unbiased} therefore imply
\begin{align}
 \E_w\!\left[\frac{\partial g_i(\mathrm{LATE}_{\mathrm{FS}^2})}{\partial l}\right]
 &=-2\E_w\!\left[
  \mu_{\mathrm{FS}}(S)
  \{\mu_{\mathrm{ITT}}(S)
    -\mathrm{LATE}_{\mathrm{FS}^2}\mu_{\mathrm{FS}}(S)\}
  \right]\notag\\
 &=0,
 \label{eq:derivative-Q-late-proof}
\end{align}
Therefore, the weak law of large numbers and
\eqref{eq:derivative-Q-late-proof} give
\begin{equation}\label{eq:LLN-Q-prime-late-proof}
 \widehat Q'(\mathrm{LATE}_{\mathrm{FS}^2})\overset{\Pp}{\longrightarrow}0.
\end{equation}
Also, direct differentiation shows that
\[
 \widehat Q''(l)
 =2\frac1n\sum_{i=1}^n\widehat w(S_i)
 \left\{\widehat\mu_{\mathrm{FS}}(S_i)^2
 -\widehat V_{\mathrm{FS},i}\right\},
\]
which does not depend on $l$. The weak law of large numbers,
Assumption~\ref{ass:late-variance-asymptotics},
and unbiasedness of $\widehat V_{\mathrm{FS},i}$ imply that this expression converges in
probability to $2\E_w[\mu_{\mathrm{FS}}(S)^2]$ and hence is $\Op(1)$. The exact second-order Taylor identity for
the quadratic function $\widehat Q(l)$ is
\begin{align*}
 \widehat q
 &=\widehat Q(\mathrm{LATE}_{\mathrm{FS}^2})
  +\widehat Q'(\mathrm{LATE}_{\mathrm{FS}^2})
   (\widehat{\mathrm{LATE}}_{\mathrm{FS}^2}-\mathrm{LATE}_{\mathrm{FS}^2})\\
 &\quad+\frac12\widehat Q''(\mathrm{LATE}_{\mathrm{FS}^2})
   (\widehat{\mathrm{LATE}}_{\mathrm{FS}^2}-\mathrm{LATE}_{\mathrm{FS}^2})^2.
\end{align*}
By \eqref{eq:LLN-Q-prime-late-proof} and
\eqref{eq:rate-late-fs2-proof}, $\widehat Q'(\mathrm{LATE}_{\mathrm{FS}^2})
   \sqrt n(\widehat{\mathrm{LATE}}_{\mathrm{FS}^2}-\mathrm{LATE}_{\mathrm{FS}^2})=\op(1)$.
   Similarly, $$\frac12\widehat Q''(\mathrm{LATE}_{\mathrm{FS}^2})
   \sqrt n(\widehat{\mathrm{LATE}}_{\mathrm{FS}^2}-\mathrm{LATE}_{\mathrm{FS}^2})^2=\frac12\widehat Q''(\mathrm{LATE}_{\mathrm{FS}^2})
   n^{-1/2}(\sqrt n(\widehat{\mathrm{LATE}}_{\mathrm{FS}^2}-\mathrm{LATE}_{\mathrm{FS}^2}))^2=\op(1),$$ because $\widehat Q''(\mathrm{LATE}_{\mathrm{FS}^2})=\Op(1)$ and in view of \eqref{eq:rate-late-fs2-proof}. Therefore,
\begin{align}
 \sqrt n(\widehat q-q)
 &=\sqrt n\{\widehat Q(\mathrm{LATE}_{\mathrm{FS}^2})-q\}+\op(1)\notag\\
 &=\frac1{\sqrt n}\sum_{i=1}^n\psi_{q,i}+\op(1),
 \label{eq:AL-q-late-proof}
\end{align}
where the last equality follows from \eqref{eq:AL-Q-lambda-proof}
and the definition in
\eqref{eq:psi-q-late}.

For the denominator, unbiasedness of $\widehat V_{\mathrm{FS},i}$ implies
\[
 \E_w\!\left[
  \widehat\mu_{\mathrm{FS}}(S_i)^2
  -\widehat V_{\mathrm{FS},i}
 \right]
 =\E_w[\mu_{\mathrm{FS}}(S)^2].
\]
Assumption~\ref{ass:late-variance-asymptotics}
allows us to apply Lemma~\ref{lem:normalized-weight-expansion} with
$Z_i=\widehat\mu_{\mathrm{FS}}(S_i)^2-\widehat V_{\mathrm{FS},i}$.
Therefore,
\begin{align}
 &\sqrt n\left\{
  \frac1n\sum_{i=1}^n\widehat w(S_i)
  \left[
   \widehat\mu_{\mathrm{FS}}(S_i)^2
   -\widehat V_{\mathrm{FS},i}
  \right]
  -\E_w[\mu_{\mathrm{FS}}(S)^2]
 \right\}\notag\\
 &\qquad=\frac1{\sqrt n}\sum_{i=1}^n
  \psi_{\mathrm{FS}^2,i}+\op(1).
 \label{eq:AL-fs-second-moment-proof}
\end{align}

By Proposition~\ref{prop:late-variance-identification},
$q=\E_w[\mu_{\mathrm{FS}}(S)^2]\sigma^2_{\mathrm{LATE},\mathrm{FS}^2}$.
Assumption~\ref{ass:late} implies
$\E_w[\mu_{\mathrm{FS}}(S)^2]>0$, while
\eqref{eq:AL-fs-second-moment-proof} implies that the denominator of
\eqref{eq:late-variance-hat} converges in probability to this strictly
positive quantity. It is therefore bounded away from zero with probability
approaching one. Expanding the ratio in \eqref{eq:late-variance-hat} and using
\eqref{eq:AL-q-late-proof} and \eqref{eq:AL-fs-second-moment-proof}, we obtain
\begin{align*}
 \sqrt n\left(
  \widehat\sigma^2_{\mathrm{LATE},\mathrm{FS}^2}-\sigma^2_{\mathrm{LATE},\mathrm{FS}^2}
 \right)
  &=\frac1{\sqrt n}\sum_{i=1}^n
   \frac{\psi_{q,i}
    -\sigma^2_{\mathrm{LATE},\mathrm{FS}^2}\psi_{\mathrm{FS}^2,i}}
    {\E_w[\mu_{\mathrm{FS}}(S)^2]}+\op(1)\\
 &=\frac1{\sqrt n}\sum_{i=1}^n
  \psi_{\sigma^2_{\mathrm{LATE},\mathrm{FS}^2},i}+\op(1),
\end{align*}
which proves \eqref{eq:AL-late-variance}.

Each influence function in the last display has mean zero. The observations are i.i.d. across sites by Assumptions~\ref{hyp:iid} and
\ref{hyp:strat_completely_randomized}, and
$\E[\psi_{\sigma^2_{\mathrm{LATE},\mathrm{FS}^2},i}^2]<\infty$ by
Assumption~\ref{ass:late-variance-asymptotics}. The central limit theorem and Slutsky's lemma therefore imply
\[
 \sqrt n\left(
  \widehat\sigma^2_{\mathrm{LATE},\mathrm{FS}^2}-\sigma^2_{\mathrm{LATE},\mathrm{FS}^2}
 \right)
 \overset d\longrightarrow
 N(0,V_{\sigma^2_{\mathrm{LATE},\mathrm{FS}^2}}),
\]
which proves Point~1.

\paragraph{Point 2.}
For this point impose Assumption~\ref{ass:late-bound-asymptotics} as well.
Write $F_i^\ast=\widehat\mu_{\mathrm{FS}}(S_i)$,
$I_i^\ast=\widehat\mu_{\mathrm{ITT}}(S_i)$, and
\[
 a_i=(I_i^\ast)^2-\widehat V_{\mathrm{ITT},i},\qquad
 b_i=F_i^\ast I_i^\ast-\widehat C_{\mathrm{FS},\mathrm{ITT},i},\qquad
 c_i=(F_i^\ast)^2-\widehat V_{\mathrm{FS},i}.
\]
Thus $g_i(l)=a_i-2lb_i+l^2c_i$.
The additional assumption implies $\E[w(S_i)^2]<\infty$ and
$\E[w(S_i)^2(I_i^\ast)^2]<\infty$.
Binary take-up gives $|F_i^\ast|\leq1$.
Assumption~\ref{ass:late-variance-asymptotics}, together with
$\E[w(S_i)^2]<\infty$, implies that
$w(S_i)b_i$, $w(S_i)c_i$, and
$w(S_i)g_i(\mathrm{LATE}_{\mathrm{FS}^2})$ have finite second moments.
Since
\[
 a_i=g_i(\mathrm{LATE}_{\mathrm{FS}^2})
 +2\mathrm{LATE}_{\mathrm{FS}^2}b_i
 -\mathrm{LATE}_{\mathrm{FS}^2}^{\,2}c_i,
\]
$w(S_i)a_i$ also has a finite second moment.
Consequently, for
$Z_i=(1,F_i^\ast,I_i^\ast,a_i,b_i,c_i)\trans$,
\begin{equation}\label{eq:late-joint-moment-condition}
 \E[\|w(S_i)Z_i\|^2]<\infty.
\end{equation}

The normalized-weight expansion applied jointly to these variables gives
root-$n$ consistency and joint asymptotic linearity of their weighted averages.
In particular,
\[
 \sqrt n(\widehat{\mathrm{LATE}}-\mathrm{LATE})
 =\frac1{\sqrt n}\sum_i\psi_{\mathrm{LATE},i}+\op(1).
\]
For a fixed $l=\mathrm{LATE}$ it also gives
\[
 \sqrt n\{\widehat Q(l)-Q(l)\}
 =\frac1{\sqrt n}\sum_iw(S_i)\{g_i(l)-h\}+\op(1).
\]
Because $\widehat Q$ is quadratic, its exact Taylor expansion yields
\[
 \widehat h
 =\widehat Q(\mathrm{LATE})
 +\widehat Q'(\mathrm{LATE})
   (\widehat{\mathrm{LATE}}-\mathrm{LATE})
 +\widehat m_2(\widehat{\mathrm{LATE}}-\mathrm{LATE})^2.
\]
Here $\widehat Q'(\mathrm{LATE})
=2(\widehat m_2\mathrm{LATE}-\widehat r)
\overset{\Pp}{\longrightarrow}2(m_2\mathrm{LATE}-r)$,
and the last term is $\Op(n^{-1})$.
It follows that
\[
 \sqrt n(\widehat h-h)
 =\frac1{\sqrt n}\sum_i\psi_{h,i}+\op(1).
\]
The same joint expansion and the product rule give
\[
 \sqrt n(\widehat d-d)=\frac1{\sqrt n}\sum_i\psi_{d,i}+\op(1),
 \qquad
 \sqrt n(\widehat\delta-\delta)
 =\frac1{\sqrt n}\sum_i\psi_{\delta,i}+\op(1).
\]
Applying the delta method to
$(h,d,\delta,\mathrm{FS})\mapsto(h+d^2/\delta)/\mathrm{FS}$,
whose denominators are strictly positive, yields
\[
 \psi_{\underline{\sigma}^2_{\mathrm{LATE}},i}
 =\frac1{\mathrm{FS}}\left\{
 \psi_{h,i}+\frac{2d}{\delta}\psi_{d,i}
 -\frac{d^2}{\delta^2}\psi_{\delta,i}
 -\underline{\sigma}^2_{\mathrm{LATE}}\psi_{\mathrm{FS},i}
 \right\}.
\]
This proves \eqref{eq:AL-late-variance-lower-bound}.
The joint central limit theorem implied by
\eqref{eq:late-joint-moment-condition} gives the stated normal limit.

\subsection{Proof of Proposition~\ref{prop:late-fs-cov}}\label{subsec:proof-late-fs-cov}

First note that from \eqref{eq:overall-late},
\[
 \E_w\!\left[
  \mu_{\mathrm{FS}}(S)
  \{\mu_{\mathrm{LATE}}(S)-\mathrm{LATE}\}
 \right]=0.
\]
Therefore,
\begin{equation*}
 \sigma_{\mathrm{LATE},\mathrm{FS}}
 =\E_w\!\left[
   \frac{\mu_{\mathrm{FS}}(S)}{\mathrm{FS}}
   \{\mu_{\mathrm{LATE}}(S)-\mathrm{LATE}\}
   \{\mu_{\mathrm{FS}}(S)-\mathrm{FS}\}
 \right].
\end{equation*}
Then, using the fact that
$\mu_{\mathrm{ITT}}(S)=\mu_{\mathrm{FS}}(S)
\mu_{\mathrm{LATE}}(S)$,
\begin{align*}
 \mathrm{FS}\,\sigma_{\mathrm{LATE},\mathrm{FS}}
 &=\E_w\!\left[
   \mu_{\mathrm{FS}}(S)
   \{\mu_{\mathrm{LATE}}(S)-\mathrm{LATE}\}
   \{\mu_{\mathrm{FS}}(S)-\mathrm{FS}\}
 \right]\\
 &=\Cov_w(\mu_{\mathrm{FS}}(S),\mu_{\mathrm{ITT}}(S))
   -\mathrm{LATE}\,
    \Var_w(\mu_{\mathrm{FS}}(S))\\
 &=\Var_w(\mu_{\mathrm{FS}}(S))
   \left(\beta_{\mathrm{ITT},\mathrm{FS}}
         -\mathrm{LATE}\right),
\end{align*}
which proves \eqref{eq:late-fs-identification}.

\subsection{Proof of Proposition~\ref{prop:late-prediction}}\label{subsec:proof-late-prediction}

For Point~1, for every $j,\ell\in\{1,\ldots,J\}$,
\begin{align*}
&\E_w\!\left[
 \{\mu_{\mathrm{ITT}_{d,j}}(S)-\mu_{\mathrm{FS}_d}(S)
 \mathrm{LATE}_{d,j,\mathrm{FS}_d^2}\}
 \{\mu_{\mathrm{ITT}_{d,\ell}}(S)-\mu_{\mathrm{FS}_d}(S)
 \mathrm{LATE}_{d,\ell,\mathrm{FS}_d^2}\}\right]\\
&\qquad=\E_w[\mu_{\mathrm{FS}_d}(S)^2]
 \Cov_{\mathrm{FS}_d^2}
 (\mu_{\mathrm{LATE}_{d,j}}(S),\mu_{\mathrm{LATE}_{d,\ell}}(S)).
\end{align*}
Stacking the identities for $j,\ell\in\{2,\ldots,J\}$ shows that the second-moment matrix of the transformed regressors is
\[
 \E_w[\mu_{\mathrm{FS}_d}(S)^2]
 \Var_{\mathrm{FS}_d^2}(\bm\mu_{\mathrm{LATE},d,-1}(S)),
\]
while their cross-moment vector with the transformed outcome is
\[
 \E_w[\mu_{\mathrm{FS}_d}(S)^2]
 \Cov_{\mathrm{FS}_d^2}
 (\bm\mu_{\mathrm{LATE},d,-1}(S),\mu_{\mathrm{LATE}_{d,1}}(S)).
\]
The coefficient from the no-intercept regression is the inverse of the first expression multiplied by the second. Therefore, it is equal to
\begin{align*}
 &\left\{\E_w[\mu_{\mathrm{FS}_d}(S)^2]
 \Var_{\mathrm{FS}_d^2}(\bm\mu_{\mathrm{LATE},d,-1}(S))\right\}^{-1}
 \E_w[\mu_{\mathrm{FS}_d}(S)^2]
 \Cov_{\mathrm{FS}_d^2}
 (\bm\mu_{\mathrm{LATE},d,-1}(S),\mu_{\mathrm{LATE}_{d,1}}(S))\\
 &\qquad=\Var_{\mathrm{FS}_d^2}(\bm\mu_{\mathrm{LATE},d,-1}(S))^{-1}
 \Cov_{\mathrm{FS}_d^2}
 (\bm\mu_{\mathrm{LATE},d,-1}(S),\mu_{\mathrm{LATE}_{d,1}}(S))
 =\beta^d.
\end{align*}

For Point~2, cancellation of the two first stages gives
\[
 \E_{\mathrm{FS}_1\mathrm{FS}_2}
 [\mu_{\mathrm{LATE}_{1,k}}(S)\mu_{\mathrm{LATE}_{2,k}}(S)]
 =\frac{\E_w[\mu_{\mathrm{ITT}_{1,k}}(S)\mu_{\mathrm{ITT}_{2,k}}(S)]}
 {\E_w[\mu_{\mathrm{FS}_1}(S)\mu_{\mathrm{FS}_2}(S)]}.
\]
Similarly,
\[
 \E_{\mathrm{FS}_1\mathrm{FS}_2}
 [\mu_{\mathrm{LATE}_{1,k}}(S)]
 =\frac{\E_w[\mu_{\mathrm{FS}_2}(S)\mu_{\mathrm{ITT}_{1,k}}(S)]}
 {\E_w[\mu_{\mathrm{FS}_1}(S)\mu_{\mathrm{FS}_2}(S)]},
 \qquad
 \E_{\mathrm{FS}_1\mathrm{FS}_2}
 [\mu_{\mathrm{LATE}_{2,k}}(S)]
 =\frac{\E_w[\mu_{\mathrm{FS}_1}(S)\mu_{\mathrm{ITT}_{2,k}}(S)]}
 {\E_w[\mu_{\mathrm{FS}_1}(S)\mu_{\mathrm{FS}_2}(S)]}.
\]
Under squared-first-stage-1 weights,
\[
 \E_{\mathrm{FS}_1^2}[\mu_{\mathrm{LATE}_{1,k}}(S)^2]
 =\frac{\E_w[\mu_{\mathrm{ITT}_{1,k}}(S)^2]}
 {\E_w[\mu_{\mathrm{FS}_1}(S)^2]},
 \qquad
 \E_{\mathrm{FS}_1^2}[\mu_{\mathrm{LATE}_{1,k}}(S)]
 =\frac{\E_w[\mu_{\mathrm{FS}_1}(S)\mu_{\mathrm{ITT}_{1,k}}(S)]}
 {\E_w[\mu_{\mathrm{FS}_1}(S)^2]}.
\]
Subtracting the products of the corresponding means and taking the covariance--variance ratio yields \eqref{eq:late-cross-treatment-identification}, proving Point~2.

\subsection{Proof of Proposition~\ref{prop:causal-mediation}}\label{subsec:proof-causal-mediation}

For brevity, write $A=\mu_{X}(S)$,
$I=\mathrm{IDE}(S)$, and $G=\mathrm{DE}(S)$. By
\eqref{eq:causal-mediation-decomposition},
$\mu_{Y}(S)=AI+G$. Assumption~\ref{ass:causal-mediation} therefore
implies
\begin{align*}
 \Cov_w(A,\mu_{Y}(S))
 &=\Cov_w(A,AI+G)\\
 &=\E_w[I]\{\E_w[A^2]-\E_w[A]^2\}\\
 &=\E_w[I]\Var_w(A).
\end{align*}
Dividing by $\Var_w(A)$ proves that
$\beta_{\mathrm{med}}=\E_w[I]$. The regression intercept is
\begin{align*}
 \alpha_{\mathrm{med}}
 &=\E_w[\mu_{Y}(S)]-\beta_{\mathrm{med}}\E_w[A]\\
 &=\E_w[A]\E_w[I]+\E_w[G]-\E_w[I]\E_w[A]
 =\E_w[G],
\end{align*}
which proves the result.

\subsection{Proof of Proposition~\ref{prop:providereffect}}\label{subsec:proof-provider-effect}

By \eqref{eq:model_providers_decomposition},
\[
 \mu_{\Delta_2}(S_i)-\mu_{\Delta_1}(S_i)
 =\epsilon_2^S(S_i)-\epsilon_1^S(S_i)
 +\epsilon_2^P(S_i,P_2(S_i))-\epsilon_1^P(S_i,P_1(S_i)),
\]
which proves Part~1.a). For Part~1.b), denote by $A_i$ and $B_i$ the first and the second difference on the right-hand side of the previous display, respectively. If $\Var_w(A_i+B_i)=0$ and $A_i$ and $B_i$ are not perfectly negatively correlated, then $\Var_w(A_i)=\Var_w(B_i)=0$. Moreover, if $\Var_w(B_i)=0$ and $\epsilon_2^P(S_i,P_2(S_i))$ and $\epsilon_1^P(S_i,P_1(S_i))$ are not perfectly positively correlated, then $\Var_w(\epsilon_2^P(S_i,P_2(S_i)))=\Var_w(\epsilon_1^P(S_i,P_1(S_i)))=0$. This proves necessity; sufficiency follows immediately by adding variables with zero weighted variance.

For Part 2, let $v=\Var_w(\mu_{\Delta_1}(S_i))=\Var_w(\mu_{\Delta_2}(S_i))>0$. Then
\[
 \sigma^2_{\Delta}=2v-2\Cov_w(\mu_{\Delta_1}(S_i),\mu_{\Delta_2}(S_i)),
 \qquad
 \beta_{21}=\frac{\Cov_w(\mu_{\Delta_1}(S_i),\mu_{\Delta_2}(S_i))}{v}.
\]
Consequently, $\sigma^2_{\Delta}=0$ if and only if $\Cov_w(\mu_{\Delta_1}(S_i),\mu_{\Delta_2}(S_i))=v$, which holds if and only if $\beta_{21}=1$.

\subsection{Proofs for Appendix~\ref{app:blp}: BLP comparisons}\label{app:blp-proofs}

\subsubsection{Proof of Lemma~\ref{lem:representation}}
\begin{proof}
Fix $k\in\{1,2\}$. For brevity, write $A_i=A_i^k$, $C_i=C_i^k$,
$\overline A_N=\overline A_N^k$, and $\overline C_N=\overline C_N^k$.
Let $a_i=A_i-\overline A_N$, $R_{u,i}=D_{u,i}-p_i$, and
$h_i=[p_i(1-p_i)]^{-1}$. The interaction regressor is
$Z_{u,i}=R_{u,i}a_i$. Within each validation subsample,
\begin{align*}
 \sum_{u\in\mathcal I_i^k}R_{u,i}&=0,\\
 \sum_{u\in\mathcal I_i^k}R_{u,i}^2
 &=N_i^k p_i(1-p_i),\\
 \sum_{u\in\mathcal I_i^k}R_{u,i}O_{u,i,1}
 &=N_i^k p_i(1-p_i)C_i.
\end{align*}
The first equality implies that $Z_{u,i}$ is weighted-orthogonal to every
site-constant baseline regressor, including $X_{1,i}^k$. Also,
\[
 \sum_i\sum_{u\in\mathcal I_i^k}h_i Z_{u,i}R_{u,i}
 =\sum_iN_i^k a_i=\frac12\sum_iN_i a_i=0.
\]
Therefore the interaction is weighted-orthogonal to every other
regressor in \eqref{eq:individual-blp}, and its coefficient is
\[
 \widehat\beta_2^k
 =\frac{\sum_{i,u\in\mathcal I_i^k}h_iZ_{u,i}O_{u,i,1}}
        {\sum_{i,u\in\mathcal I_i^k}h_iZ_{u,i}^2}
 =\frac{\sum_iN_i a_i C_i}{\sum_iN_i a_i^2}.
\]
Subtracting $\overline C_N$ in the numerator leaves it unchanged,
proving \eqref{eq:site-blp}.
\end{proof}

\subsubsection{Proof of Proposition~\ref{prop:power}}
\begin{proof}
We first derive the influence functions and their variances for a
generic $\theta\geq0$, and then apply them along the sequence
$\theta=\theta_n=\delta/\sqrt n$ in the proposition.
Under Assumption~\ref{ass:gaussian}, the two half-sample estimates admit
the representation
\[
 A_i=\mu+t_i+e_i,\qquad C_i=\mu+t_i+f_i,
\]
where $t_i\sim N(0,\theta)$ and $e_i,f_i\sim N(0,r)$ are mutually
independent. This follows from conditional independence of the
disjoint unit samples and from their site-invariant Gaussian error
distributions. Moreover,
$\widehat\mu_Y(S_i)=(A_i+C_i)/2$.

Set $a=\theta+r$,
$b=\theta/a$, $X_i=A_i-\mu$, and $Z_i=C_i-\mu$.
By Lemma~\ref{lem:representation},
$\widehat b=g(\widehat q,\widehat a_A,\widehat a_C)$, where
\[
 g(q,a_A,a_C)=\frac q2\left(\frac1{a_A}+\frac1{a_C}\right),
 \qquad
 \nabla g(\theta,a,a)=
 \left(\frac1a,-\frac{\theta}{2a^2},-\frac{\theta}{2a^2}\right).
\]
The influence functions of these three sample moments are respectively
$X_iZ_i-\theta$, $X_i^2-a$, and $Z_i^2-a$; replacing $\mu$ by the
sample means contributes only terms of order $O_{\Pp}(n^{-1})$.
The delta method therefore gives
\begin{align}
 \psi_{b,i}
 &=\frac{X_iZ_i-\theta}{a}
 -\frac{\theta}{2a^2}(X_i^2+Z_i^2-2a)\notag\\
 &=\frac1a\left[X_iZ_i-\frac b2(X_i^2+Z_i^2)\right].
 \label{eq:if-b}
\end{align}
To compute its variance, note that $(X_i,Z_i)$ is centered bivariate
normal with variances $a$ and covariance $\theta=ab$.
It therefore has the same distribution as
\[
 X_i=\sqrt a\,U_i,\qquad
 Z_i=\sqrt a\left(bU_i+\sqrt{1-b^2}\,V_i\right),
\]
where $U_i,V_i$ are independent standard normal variables. Their odd
moments vanish, $\E[U_i^2]=\E[V_i^2]=1$, and
$\E[U_i^4]=\E[V_i^4]=3$. Expanding the products and using independence,
\begin{align*}
 \E[X_i^2Z_i^2]
 &=a^2\left\{b^2\E[U_i^4]
       +2b\sqrt{1-b^2}\E[U_i^3V_i]
       +(1-b^2)\E[U_i^2V_i^2]\right\}\\
 &=a^2\{3b^2+(1-b^2)\}=a^2+2\theta^2,\\
 \E[X_i^3Z_i]
 &=a^2\left\{b\E[U_i^4]
       +\sqrt{1-b^2}\E[U_i^3V_i]\right\}
 =3a^2b=3a\theta,\\
 \E[X_i^4]&=a^2\E[U_i^4]=3a^2.
\end{align*}
Exchangeability of $X_i$ and $Z_i$ also gives
$\E[X_iZ_i^3]=3a\theta$ and $\E[Z_i^4]=3a^2$.
Because $\E[\psi_{b,i}]=a^{-1}(\theta-ba)=0$, squaring
\eqref{eq:if-b} and taking expectations yields
\begin{align*}
 \Var(\psi_{b,i})
 &=\frac1{a^2}\left\{
 \E[X_i^2Z_i^2]
 -b\big(\E[X_i^3Z_i]+\E[X_iZ_i^3]\big)
 +\frac{b^2}{4}\big(\E[X_i^4]+2\E[X_i^2Z_i^2]+\E[Z_i^4]\big)
 \right\}\\
 &=\frac1{a^2}\left\{
 a^2+2\theta^2-6ba\theta
 +\frac{b^2}{4}(8a^2+4\theta^2)\right\}\\
 &=1+2b^2-6b^2+2b^2+b^4
 =1-2b^2+b^4.
\end{align*}
Thus
\begin{equation}\label{eq:fixed-vb}
 V_b(\theta):=\Var(\psi_{b,i})=(1-b^2)^2.
\end{equation}
Now set $\theta=\theta_n=\delta/\sqrt n$ and
$b_n=\theta_n/(\theta_n+r)$. Then $V_b(\theta_n)\to1$ and
$\sqrt n\,b_n=\sqrt n\,\theta_n/(\theta_n+r)\to\delta/r$.
Gaussian moments are uniformly bounded along this local sequence,
so the triangular-array central limit theorem and delta method apply.
In particular,
\[
 \sqrt n(\widehat b-b_n)\Rightarrow N(0,1).
\]
The corresponding laws of large numbers and the plug-in formula
\eqref{eq:if-b-hat} show that $\widehat V_b\to_p1$.
Adding $\sqrt n\,b_n\to\delta/r$ and using $r=2v$ proves
\eqref{eq:limit-b} by Slutsky's theorem.

For our estimator, the Gaussian sample-mean/sample-variance
independence within each arm implies that $\widehat V_{Y,i}$ is
independent of $\widehat\mu_Y(S_i)$ conditional on $S_i$.
Its conditional distribution is site-invariant, and hence this
independence also holds unconditionally. The usual Gaussian
sample-variance formula gives $\E[\widehat V_{Y,i}]=v$ and
\eqref{eq:c}. The influence function is
\[
 \psi_{\sigma^2,i}
 =(\widehat\mu_Y(S_i)-\mu)^2-\widehat V_{Y,i}-\theta.
\]
Since $\widehat\mu_Y(S_i)-\mu\sim N(0,\theta+v)$,
\begin{equation}\label{eq:fixed-vtheta}
 V_{\sigma^2}(\theta)=2(\theta+v)^2+c.
\end{equation}
Along $\theta_n=\delta/\sqrt n$, uniform moment bounds and the central
limit theorem give
\[
 \sqrt n(\widehat\sigma^2_{\mathrm{ATE}}-\theta_n)
 \Rightarrow N(0,2v^2+c).
\]
Moreover, the empirical second moment of \eqref{eq:if-v-hat} converges
in probability to $2v^2+c$. Since $\sqrt n\,\theta_n=\delta$,
Slutsky's theorem proves \eqref{eq:limit-v}.
When $\delta=0$, both limiting distributions are standard normal,
which establishes asymptotic size $\alpha$. For $\delta>0$, taking
upper-tail probabilities at $z_{1-\alpha}$ gives the two power
functions stated in the proposition.

Write $a_d=\omega_d^2/N_d>0$, so $v=a_0+a_1$.
Because $N_d\geq4$,
\[
 c=2\sum_{d=0}^1\frac{a_d^2}{N_d-1}
 \leq\frac23(a_0^2+a_1^2)
 <\frac23v^2<2v^2.
\]
Consequently $\sqrt{2v^2+c}<2v$, so
$\delta/\sqrt{2v^2+c}>\delta/(2v)$ for every $\delta>0$.
Strict monotonicity of $\Phi$ proves
$\pi_{\sigma^2}(\delta)>\pi_b(\delta)$.

\end{proof}

\begin{landscape}
\section{Survey of Multi-Site RCTs}\label{sec:survey}

\begin{table}[H]
	\centering
	\caption{Multi-site RCTs in AEJ: Applied Economics 2014-2016} \label{table:exps}
	\resizebox{\paperwidth}{!}{
		\begin{tabular}{l c c c c}
			Title & Units of Observation & Units of Randomization & Sites & Randomization Units per Site \\
			\hline
			Keeping It Simple: Financial Literacy and Rules of Thumb & Individual clients & 1,193 individual clients & 107 barrios & 11.1\\
			Improving Educational Quality through Enhancing Community Participation: Results from a Randomized Field Experiment in Indonesia & Students & 520 schools & 44 subdistricts & 11.8\\
			The Demand for Medical Male Circumcision & Individuals & 1,634 individuals & 29 enumeration areas & 56.3\\
			Should Aid Reward Performance? Evidence from a Field Experiment on Health and Education in Indonesia & Individuals & 300 subdistricts (kecamatan) & 20 districts (kabupaten) & 15.0\\
			Private and Public Provision of Counseling to Job Seekers: Evidence from a Large Controlled Experiment & Individuals & 43,977 individuals & 216 employment offices & 203.6\\
			Estimating the Impact of Microcredit on Those Who Take It Up: Evidence from a Randomized Experiment in Morocco & Households & 162 villages (81 matched pairs) & 47 branches & 3.4\\
			Microcredit Impacts: Evidence from a Randomized Microcredit Program Placement Experiment by Compartamos Banco & Households & 250 geographic clusters & 45 superclusters & 5.6\\
			The Impacts of Microcredit: Evidence from Bosnia and Herzegovina & Individuals & 1,196 individuals & 282 city/towns or 14 branches & 85.4\\
			Social Networks and the Decision to Insure & Households & 5,335 households & 185 villages & 28.8\\
			Inputs in the Production of Early Childhood Human Capital: Evidence from Head Start & Individuals & 4,442 individuals & 353 Head Start centers & 12.6\\
			The Returns to Microenterprise Support among the Ultrapoor: A Field Experiment in Postwar Uganda & Individuals & 904 individuals & 60 villages & 15.1\\
			The Impact of High School Financial Education: Evidence from a Large-Scale Evaluation in Brazil & Students & 892 schools (matched pairs) & 6 states (municipalities not reported)& 148.7\\
			\hline
			Median & & & 46 & 15.0\\
			\hline
		\end{tabular}
	}
	\begin{minipage}{21.0cm}
		\scriptsize{\textit{Notes:} "The Returns to Microenterprise Support among the Ultrapoor: A Field Experiment in Postwar Uganda" corresponds to the Phase 2 experiment.
			"Social Networks and the Decision to Insure" corresponds to the household level randomization and analysis. In ``The Impact of High School Financial Education'', randomization was within matched pairs of schools stratified by municipality; the paper does not report the number of municipalities or pairs, so we count the six states in which the schools are located as sites. 
		}
	\end{minipage}
\end{table}
\end{landscape}

\end{document}